\documentclass[12pt]{article}
\pdfoutput=1
\usepackage[T1]{fontenc}
\usepackage[colorlinks,linkcolor=Blue,citecolor=Blue,bookmarks,bookmarksnumbered]{hyperref}
\usepackage[scaled=0.85]{helvet}
\usepackage{amsmath,amssymb,accents,mathrsfs,XoohmE}
\usepackage{graphicx,color}
\graphicspath{{Figures/}}
\usepackage{booktabs}
\usepackage{multirow}
\usepackage{placeins}
\usepackage{amsmath}
\usepackage{subfigure}
\usepackage{multirow}
\usepackage[table,x11names]{xcolor}
\usepackage{longtable}
\newtheorem{theorem}{Theorem}
\newtheorem{definition}[theorem]{Definition}

\definecolor{Green}  {rgb}{0.10,0.70,0.10} %  1
\definecolor{Orange} {rgb}{1.00,0.50,0.15} %  2
\definecolor{Red}    {rgb}{0.90,0.00,0.12} %  3
\definecolor{Purple} {rgb}{0.50,0.25,0.55} %  4
\definecolor{Turque} {rgb}{0.00,0.65,0.85} %  5
\definecolor{Blue}   {rgb}{0.00,0.00,1.00} %  6
\definecolor{Magenta}{rgb}{1.00,0.00,1.00} %  7
\definecolor{Gold}   {rgb}{1.00,0.75,0.25} %  8
\definecolor{Seaweed}{rgb}{0.01,0.24,0.09} %  9
\definecolor{Brown}  {rgb}{0.43,0.26,0.32} % 10
\definecolor{grey1}  {rgb}{0.20,0.20,0.20} % 11
\definecolor{grey2}  {rgb}{0.40,0.40,0.40} % 12
\definecolor{grey3}  {rgb}{0.60,0.60,0.60} % 13
\definecolor{grey4}  {rgb}{0.80,0.80,0.80} % 14
\definecolor{grey5}  {rgb}{0.90,0.90,0.90} % 15
\def\C#1#2{{\ifcase#1\or%Greg's color scheme
             \color{Green}\or \color{Orange}\or \color{Red}\or
              \color{Purple}\or \color{Turque}\or \color{Blue}\or
               \color{Magenta}\or \color{Gold}\or \color{Seaweed}\or
                \color{Brown}\or\color{grey1}\or\color{grey2}\or
                 \color{grey3}\else\color{grey4}\fi#2}}

\definecolor{Slate} {rgb}{0.00,0.45,0.55}

\usepackage{color} 
\usepackage{listings} 
\usepackage{setspace} 
 
\definecolor{Code}{rgb}{0,0,0} 
\definecolor{Decorators}{rgb}{0.5,0.5,0.5} 
\definecolor{Numbers}{rgb}{0.5,0,0} 
\definecolor{MatchingBrackets}{rgb}{0.25,0.5,0.5} 
\definecolor{Keywords}{rgb}{0,0,1} 
\definecolor{self}{rgb}{0,0,0} 
\definecolor{Strings}{rgb}{0,0.63,0} 
\definecolor{Comments}{rgb}{0,0.63,1} 
\definecolor{Backquotes}{rgb}{0,0,0} 
\definecolor{Classname}{rgb}{0,0,0} 
\definecolor{FunctionName}{rgb}{0,0,0} 
\definecolor{Operators}{rgb}{0,0,0} 
\definecolor{Background}{rgb}{0.98,0.98,0.98}

\lstdefinelanguage{Python}{ 
numbers=left, 
numberstyle=\footnotesize, 
numbersep=1em, 
xleftmargin=1em, 
framextopmargin=2em, 
framexbottommargin=2em, 
showspaces=false, 
showtabs=false, 
showstringspaces=false, 
frame=l, 
tabsize=4, 
basicstyle=\ttfamily\small\setstretch{1}, 
backgroundcolor=\color{Background}, 
commentstyle=\color{Comments}\slshape, 
stringstyle=\color{Strings}, 
morecomment=[s][\color{Strings}]{"""}{"""}, 
morecomment=[s][\color{Strings}]{'''}{'''}, 
morekeywords={import,from,class,def,for,while,if,is,in,elif,else,not,and,or,print,break,continue,return,True,False,None,access,as,,del,except,exec,finally,global,import,lambda,pass,print,raise,try,assert}, 
keywordstyle={\color{Keywords}\bfseries}, 
morekeywords={[2]@invariant,pylab,numpy,np,scipy}, 
keywordstyle={[2]\color{Decorators}\slshape}, 
emph={self}, 
emphstyle={\color{self}\slshape}, 
}

\def\rI{{\rm I}}
\def\rJ{{\rm J}}

\def\rL{{\rm L}}
\def\rR{{\rm R}}

\def\fracm#1#2{\hbox{\large{${\frac{{#1}}{{#2}}}$}}}

\def\be{\begin{equation}}
\def\ee{\end{equation}}
\newcommand{\bea}{\begin{eqnarray}}
\newcommand{\eea}{\end{eqnarray}}
\newcommand{\ena}{\end{eqnarray}}

\def\pp{{\mathchoice
          {%displaystyle
              \kern 1pt%
              \raise 1pt
              \vbox{\hrule width5pt height0.4pt depth0pt
                    \kern -2pt
                    \hbox{\kern 2.3pt
                          \vrule width0.4pt height6pt depth0pt
                          }
                    \kern -2pt
                    \hrule width5pt height0.4pt depth0pt}%
                    \kern 1pt
           }
            {%textstyle
              \kern 1pt%
              \raise 1pt
              \vbox{\hrule width4.3pt height0.4pt depth0pt
                    \kern -1.8pt
                    \hbox{\kern 1.95pt
                          \vrule width0.4pt height5.4pt depth0pt
                          }
                    \kern -1.8pt
                    \hrule width4.3pt height0.4pt depth0pt}%
                    \kern 1pt
            }
            {%scriptstyle
              \kern 0.5pt%
              \raise 1pt
              \vbox{\hrule width4.0pt height0.3pt depth0pt
                    \kern -1.9pt  %[e]=0.15pt
                    \hbox{\kern 1.85pt
                          \vrule width0.3pt height5.7pt depth0pt
                          }
                    \kern -1.9pt
                    \hrule width4.0pt height0.3pt depth0pt}%
                    \kern 0.5pt
            }
            {%scriptscriptstyle
              \kern 0.5pt%
              \raise 1pt
              \vbox{\hrule width3.6pt height0.3pt depth0pt
                    \kern -1.5pt
                    \hbox{\kern 1.65pt
                          \vrule width0.3pt height4.5pt depth0pt
                          }
                    \kern -1.5pt
                    \hrule width3.6pt height0.3pt depth0pt}%
                    \kern 0.5pt%}
            }
        }}

\def\mm{{\mathchoice
                       {%displaystyle
                             \kern 1pt
               \raise 1pt    \vbox{\hrule width5pt height0.4pt depth0pt
                                  \kern 2pt
                                  \hrule width5pt height0.4pt depth0pt}
                             \kern 1pt}
                       {%textstyle
                            \kern 1pt
               \raise 1pt \vbox{\hrule width4.3pt height0.4pt depth0pt
                                  \kern 1.8pt
                                  \hrule width4.3pt height0.4pt depth0pt}
                             \kern 1pt}
                       {%scriptstyle
                            \kern 0.5pt
               \raise 1pt
                            \vbox{\hrule width4.0pt height0.3pt depth0pt
                                  \kern 1.9pt
                                  \hrule width4.0pt height0.3pt depth0pt}
                            \kern 1pt}
                       {%scriptscriptstyle
                           \kern 0.5pt
             \raise 1pt  \vbox{\hrule width3.6pt height0.3pt depth0pt
                                  \kern 1.5pt
                                  \hrule width3.6pt height0.3pt depth0pt}
                           \kern 0.5pt}
                       }}

\def\ad{{\kern0.5pt
                   \alpha \kern-5.05pt \raise5.8pt\hbox{$\textstyle.$}\kern
0.5pt}}

\def\bd{{\kern0.5pt
                   \beta \kern-5.05pt \raise5.8pt\hbox{$\textstyle.$}\kern
0.5pt}}

\def\qd{{\kern0.5pt
                   q \kern-5.05pt \raise5.8pt\hbox{$\textstyle.$}\kern
0.5pt}}
\def\Dot#1{{\kern0.5pt
     {#1} \kern-5.05pt \raise5.8pt\hbox{$\textstyle.$}\kern
0.5pt}}

\catcode`@=11
\def\un#1{\relax\ifmmode\@@underline#1\else
        $\@@underline{\hbox{#1}}$\relax\fi}
\catcode`@=12

\def\a{\alpha}
\def\b{\beta}
\def\c{\chi}
\def\d{\delta}

\def\g{\gamma}

\def\i{\iota}
\def\j{\psi}
\def\k{\kappa}
\def\l{\lambda}

\def\o{\omega}

\def\r{\rho}

\def\t{\tau}

\def\L{\Lambda}
\def\O{\Omega}
\def\P{\Pi}

\def\S{\Sigma}

\def\ca{{\cal A}}

\def\cf{{\cal F}}

\def\dslash{\not{\hbox{\kern-2pt $\partial$}}}
\def\Dslash{\not{\hbox{\kern-4pt $D$}}}
\def\pslash{\not{\hbox{\kern-2.3pt $p$}}}
 \newtoks\slashfraction
 \slashfraction={.13}
 \def\slash#1{\setbox0\hbox{$ #1 $}
 \setbox0\hbox to \the\slashfraction\wd0{\hss \box0}/\box0 }
 
\def\kcr{{\hbox{\ro \char'170}}}                % right-handed rope
\def\ktl{{\hbox{\ro \char'170}}}        % top end for left-handed rope
\def\ktr{{\hbox{\ro \char'170}}}        % " right
\def\kbl{{\hbox{\ro \char'170}}}        % " bottom left
\def\kbr{{\hbox{\ro \char'170}}}        % " right
\def\plpl{\raise-2pt\hbox{$\raise3pt\hbox{$_+$}\hskip-6.67pt\raise0.0pt
\hbox{$^+$}\hskip 0.01pt$}}
\def\mimi{\raise-2pt\hbox{$\raise3pt\hbox{$_-$}\hskip-6.67pt\raise0.0pt
\hbox{$^-$}\hskip 0.01pt$}} 

\def\bo{{\raise.15ex\hbox{\large$\Box$}}}               % D'Alembertian
\def\pa{\partial}                                       % curly d
\def\TH{{\raise.2ex\hbox{$\displaystyle \bigodot$}\mskip-4.7mu \llap H \;}}
\def\face{{\raise.2ex\hbox{$\displaystyle \bigodot$}\mskip-2.2mu \llap {$\ddot
        \smile$}}}                                      % happy face
\def\dt#1{\on{\hbox{\bf .}}{#1}}                % (big) dot over
\def\Dot#1{\dt{#1}}

\def\Tilde#1{\widetilde{#1}}                    % big tilde
\def\Hat#1{\widehat{#1}}                        % big hat
\def\leftrightarrowfill{$\mathsurround=0pt \mathord\leftarrow \mkern-6mu
        \cleaders\hbox{$\mkern-2mu \mathord- \mkern-2mu$}\hfill
        \mkern-6mu \mathord\rightarrow$}
\def\dvec#1{\vbox{\ialign{##\crcr
        \leftrightarrowfill\crcr\noalign{\kern-1pt\nointerlineskip}
        $\hfil\displaystyle{#1}\hfil$\crcr}}}           % <--> accent
\def\dt#1{{\buildrel {\hbox{\LARGE .}} \over {#1}}}     % dot-over for sp/sb
\def\fracm#1#2{\hbox{\large{${\frac{{#1}}{{#2}}}$}}}
\def\sfrac#1#2{{\vphantom1\smash{\lower.5ex\hbox{\small$#1$}}\over
        \vphantom1\smash{\raise.4ex\hbox{\small$#2$}}}} % alternate fraction
\def\bfrac#1#2{{\vphantom1\smash{\lower.5ex\hbox{$#1$}}\over
        \vphantom1\smash{\raise.3ex\hbox{$#2$}}}}       % "
\def\afrac#1#2{{\vphantom1\smash{\lower.5ex\hbox{$#1$}}\over#2}}    % "
\def\perm#1{\langle #1 \rangle}

\def\pa{\partial}      
\let\bm\relax
\newcommand{\bm}[1]{{\boldsymbol{#1}}}

\def\ad{{\dot{\alpha}}}
\def\bd{{\dot{\beta}}}

 \font\rOpe=cmsy10                        % Ersatz for the non-standard rope font
 \def\ktl{{\hbox{\rOpe\char'170}}}        % top end for left-handed rope
 \def\kbl{{\hbox{\rOpe\char'170}}}        % bottom end for left-handed rope
 \def\kcr{{\reflectbox{\rOpe\char'170}}}        % right-handed rope
 \def\ktr{{\reflectbox{\rOpe\char'170}}}        % top end for right-handed rope
 \def\kbr{{\reflectbox{\rOpe\char'170}}}        % bottom end for right-handed rope
 \def\Border{\vbox{\hsize0pt% braided border
        \setlength{\unitlength}{1mm}
        \newcount\xco
        \newcount\yco
        \xco=-21
        \yco=12
        \begin{picture}(0,0)(-7.5,0)
        \put(\xco,\yco){$\ktl$}
        \advance\yco by-1
        {\loop
        \put(\xco,\yco){$\kcr$}
        \advance\yco by-2
        \ifnum\yco>-240
        \repeat
        \put(\xco,\yco){$\kbl$}}
        \xco=170
        \yco=12
        \put(\xco,\yco){$\ktr$}
        \advance\yco by-1
        {\loop
        \put(\xco,\yco){$\kcr$}
        \advance\yco by-2
        \ifnum\yco>-240
        \repeat
        \put(\xco,\yco){$\kbr$}}
        \put(-19.5,13){\scalebox{.6065}{%
         University of Maryland Center for String and Particle  Theory \&\ Physics Department%
        |University of Maryland Center for String and Particle  Theory \&\ Physics Department}}
        \put(-19.5,-241.5){\scalebox{.5835}{%
         ****University of Maryland * Center for String and
         Particle  Theory* Physics Department****University of Maryland *Center
        for String and Particle  Theory* Physics Department}}
        \end{picture}
        \par\vskip-8mm}}
\definecolor{UMred}{rgb}{.9,.05,.2}
\definecolor{HUblue}{rgb}{.0,.3,.7}

\definecolor{Red}    {rgb}{0.90,0.00,0.12} %  1
\definecolor{Blue}   {rgb}{0.00,0.00,1.00} %  2
\definecolor{Green}  {rgb}{0.10,0.70,0.10} %  3
\definecolor{Turque} {rgb}{0.00,0.65,0.85} %  4
\definecolor{Orange} {rgb}{1.00,0.50,0.15} %  5
\definecolor{Magenta}{rgb}{1.00,0.00,1.00} %  6
\definecolor{Gold}   {rgb}{1.00,0.75,0.25} %  7
\definecolor{Seaweed}{rgb}{0.01,0.24,0.09} %  8
\definecolor{Purple} {rgb}{0.50,0.25,0.55} %  9
\definecolor{Brown}  {rgb}{0.43,0.26,0.32} % 10
\definecolor{grey1}  {rgb}{0.20,0.20,0.20} % 11
\definecolor{grey2}  {rgb}{0.40,0.40,0.40} % 12
\definecolor{grey3}  {rgb}{0.60,0.60,0.60} % 13
\definecolor{grey4}  {rgb}{0.80,0.80,0.80} % 14
\definecolor{grey5}  {rgb}{0.90,0.90,0.90} % 15
\def\C#1#2{{\ifcase#1\or%TH color scheme
             \color{Red}\or \color{Green}\or \color{Blue}\or\
              \color{Turque}\or \color{Orange}\or \color{Magenta}\or 
               \color{Gold}\or \color{Seaweed}\or \color{Purple}\or
                \color{Brown}\or\color{grey1}\or\color{grey2}\or
                 \color{grey3}\else\color{grey4}\fi#2}}

\definecolor{Slate} {rgb}{0.00,0.45,0.55}

\newdimen\parshift\parshift=\parindent
\catcode`@=11
 \long\def\@footnotetext#1{\insert\footins{\reset@font\footnotesize
           \interlinepenalty\interfootnotelinepenalty\splittopskip%
            \footnotesep\splitmaxdepth\dp\strutbox\floatingpenalty\@MM%
             \hsize\columnwidth\addtolength{\hsize}{-2\parindent}
              \@parboxrestore\protected@edef\@currentlabel%
              {\csname p@footnote\endcsname\@thefnmark}%
                \color@begingroup%
                 \@makefntext{\rule\z@\footnotesep\ignorespaces#1%
                  \@finalstrut\strutbox}%
                \color@endgroup}}
 \long\def\@makefntext#1{\hglue\parshift%
           \vbox{\noindent\baselineskip=11pt plus.5pt minus.5pt\hb@xt@0em{\hss\@makefnmark\kern1pt}#1}}
\catcode`@=12

\newskip\humongous \humongous=0pt plus 1000pt minus 1000pt
\def\caja{\mathsurround=0pt}
\def\eqalign#1{\,\vcenter{\openup2\jot \caja
        \ialign{\strut \hfil$\displaystyle{##}$&$
        \displaystyle{{}##}$\hfil\crcr#1\crcr}}\,}
\newif\ifdtup

\makeatletter
\def\section{\@startsection{section}{1}{\z@}
        {3ex plus-1ex minus-.2ex}{1pt plus1pt}{\large\sf\bfseries\boldmath}}
\def\subsection{\@startsection{subsection}{2}{\z@}
         {1.5ex plus-1ex minus-.2ex}{0.01pt plus1pt}{\sf\slshape}}
\def\subsubsection{\@startsection{subsubsection}{3}{\z@}
          {1.5ex plus-1ex minus-.2ex}{0.01pt plus0.2pt}{\sf\boldmath}}
\def\paragraph{\@startsection{paragraph}{4}{\z@}
           {.75ex \@plus.5ex \@minus.2ex}{-2mm}{\sf\bfseries\boldmath}}
\makeatother

 \allowdisplaybreaks
 \seceq

\usepackage{lipsum}
\usepackage{listings}
\definecolor{MyDarkGreen}{rgb}{0.0,0.4,0.0} % This is the color used for comments
\usepackage[enableskew,vcentermath]{youngtab}

\definecolor{Hey}{rgb}{.9,.05,.4}
\definecolor{orange}{rgb}{1,.5,0}
\definecolor{plum}{rgb}{.4,0,.6}
\definecolor{R}{rgb}{1,0,0}
\definecolor{G}{rgb}{0.1,0.7,0}
\definecolor{B}{rgb}{0,0,1}
\begin{document}

\thispagestyle{empty}
\noindent{\small
%\today
\hfill{  \\ % un-comment-out and specify when done}  
%$~~~~~~~~~~~~~~~~~~~~~~~~~~~~~~~~~~~~~~~~~~~~~~~~~~~~~~~~~~~~~~~~~$
%$~~~~~~~~~~~~~~~~~~~~~~~~~~~~~~~~~~~~~~~~~~~~~~~~~~~~~~~~~~~~~~~~~$ {}
}}
\vspace*{0mm}
\begin{center}
{\large \bf
 \hskip0.05in
$\bm {\cal N}$ = 2 SUSY \&
 the Hexipentisteriruncicantitruncated 7-Simplex
 \\[2pt]
}   \vskip0.2in
{\large {
$~~~~~~~~$
Aleksander J.\ Cianciara\footnote{acianciara@princeton.edu}$^{a,b}$,
Zachary Coleman\footnote{zachary.coleman@trinity.ox.ac.uk}$^{,c,d,e}$,\newline
S.\ James Gates, Jr.\footnote{sylvester${}_-$gatess@umd.edu}$^{f}$, 
Youngik (Tom) Lee\footnote{youngik${}_-$lee@brown.edu}${}^{,g,h}$, and 
Ziyang Zhang\footnote{ziyang${}_-$zhang1@brown.edu}$^{,g, h}$
 $~~~~~~$
}}
\\*[4mm]
\emph{
\centering
$^{a}$Joseph Henry Laboratories, Princeton University, 
Princeton, NJ 08544, USA, \\
$^{b}$Institute for Advanced Study,
Princeton, NJ 08540, USA,\\ [12pt]
$^{c}$Pepperdine University, Natural Science Division,
%\\[12pt]
Malibu, CA 90263, USA,
\\
$^{d}$Mathematics Institute, University of Oxford, Oxford OX2 6GG, 
UK,\\
$^{e}$ Department of Physics, University of Oxford, Oxford OX1 3PU, UK,\\
[12pt]
$^{f} $Department of Physics, University of Maryland,
\\[1pt]
College Park, MD 20742-4111, USA,
\\[12pt]
$^{g}$Brown University, Department of Physics,
\\[1pt]
182 Hope Street, Barus \& Holley 545,
Providence, RI 02912, USA,
\\ [10pt] 
$^{h}$Brown Center for Theoretical Physics, \\[1pt] 
340 Brook Street, Barus Hall,
Providence, RI 02912, USA,
%\\[4pt] 
}
 \\*[4mm]
{ ABSTRACT}\\[4mm]
\parbox{142mm}{\parindent=2pc\indent\baselineskip=14pt plus1pt
We study algorithms for recursively creating arbitrary $N$-extended `supermultiplets' given minimal matrix representations of off-shell, $\cal N $ = 1 supermultiplet matrices.  We observe connections between the color vertex problems in graph theory and the different supermultiplet sets locations in the permutahedron by using the concepts of truncation and chromatic number. The concept of `hopping operators' is introduced, constructed, and then used to partition the $8!$ vertices of the permutahedron. We explicitly partition these into 5,040 octets constrained in locations on the permutahedron by a magic number rule.  Boolean factors in this recursive construction are shown to obey a doubly even binary flip rule.  Although these hopping operators do not generally constitute normal subgroups of the permutation group, we find that `ab-normal cosets' exist where the same left- and right-hoppers appear as unordered sets.  Finally, using computer simulations, we investigate the types of faces on higher-order permutahedron which may give rise to lower-order supermultiplets.

}
 \end{center}
\vfill
\vfill
\clearpage
%

%
%%%%%%%%%%%%%%%%%%%%%%%%%%%%%%%%%%%%%%%%%%%%%%%
%%%%%%%%%%%%%%%%%%%%%%%%%%%%%%%%%%%%%%%%%%%%%%%
%%%%%%%%%%%%%%%%%%%%%%%%%%%%%%%%%%%%%%%%%%%%%%%
\section{Introduction}\label{sec:1}
The question of how to create arbitrary 4D, $N$-extended supermultiplets
(represented by
matrices) starting from minimal 4D, $\cal N$ = 1 realizations began with \cite{GRana1,GRana2} and was one of the motivations for the DFGHILM collaboration.
Later it was realized \cite{FG} the matrices utilized could be interpreted as modified versions of adjacency matrices for a set of graphs given the names of ``adinkras.'' More recently \cite{pHEDRON}, it was shown that the permutahedron plays a role in the selection of supermultiplets, and that some faces of the permutahedron polytope related to a given value of $\cal N$ correspond to valid supermultiplets of a lower degree or extension \cite{Note}. In other areas, it's been shown that faces of polytopes may 
correspond to properties of effective field theories \cite{EFThedron1}. The present connection between representations of supersymmetry and permutation polytopes hints at an underlying polytopic representation theory that may be responsible for generating all valid representations of supersymmetry. Similar work has been carried out examining the structure of viruses \cite{viruses}, building off the idea of Watson and Crick that the  crystallographic symmetry of viruses could be explained by the repeated use of just a few protein subunits (in the present paper we call these hopping operators) \cite{watsoncrick}. 

The results in this work give an extension of the questions and results determined about
the recursive construction of higher $N$-extended 1D, adinkra matrices from lower $N$-extended 1D, adinkra matrices in the works of \cite{GRana1,GRana2}.  
In this work similar questions and results are determined for the extension of adinkra matrices associated with 4D, $\cal N$ = 1 models expanded to adinkra matrices associated with 4D, $\cal N$ = 2 models.

\subsection{A problem from the representation theory: at the foundation of 4-dimensional supersymmetry}\label{sec:11}

\begin{figure}[h]
\centering
\includegraphics[width=8.5cm,height=5cm]{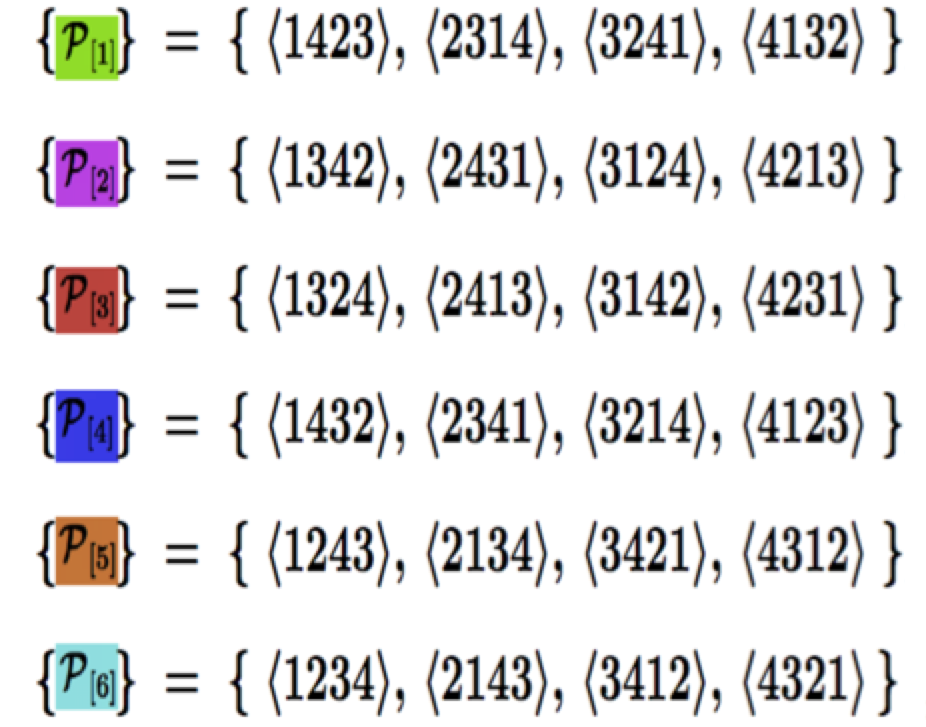}
\caption{Permutation matrix assignments to subsets}
\label{FigP1}
\end{figure}

\begin{figure}[h]
\centering
\begin{minipage}{.5\textwidth}
  \centering
  \includegraphics[width=.6\linewidth]{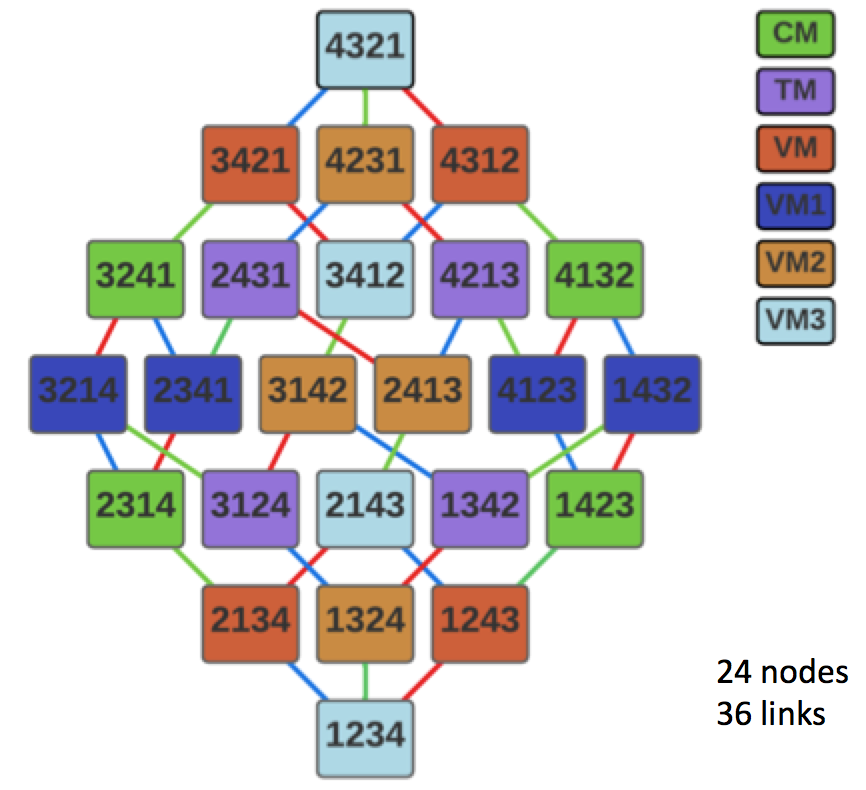}
  \caption{Permutahedron containing subset elements}
  \label{FigP2}
\end{minipage}%
\begin{minipage}{.5\textwidth}
  \centering
  \includegraphics[width=1.\linewidth]{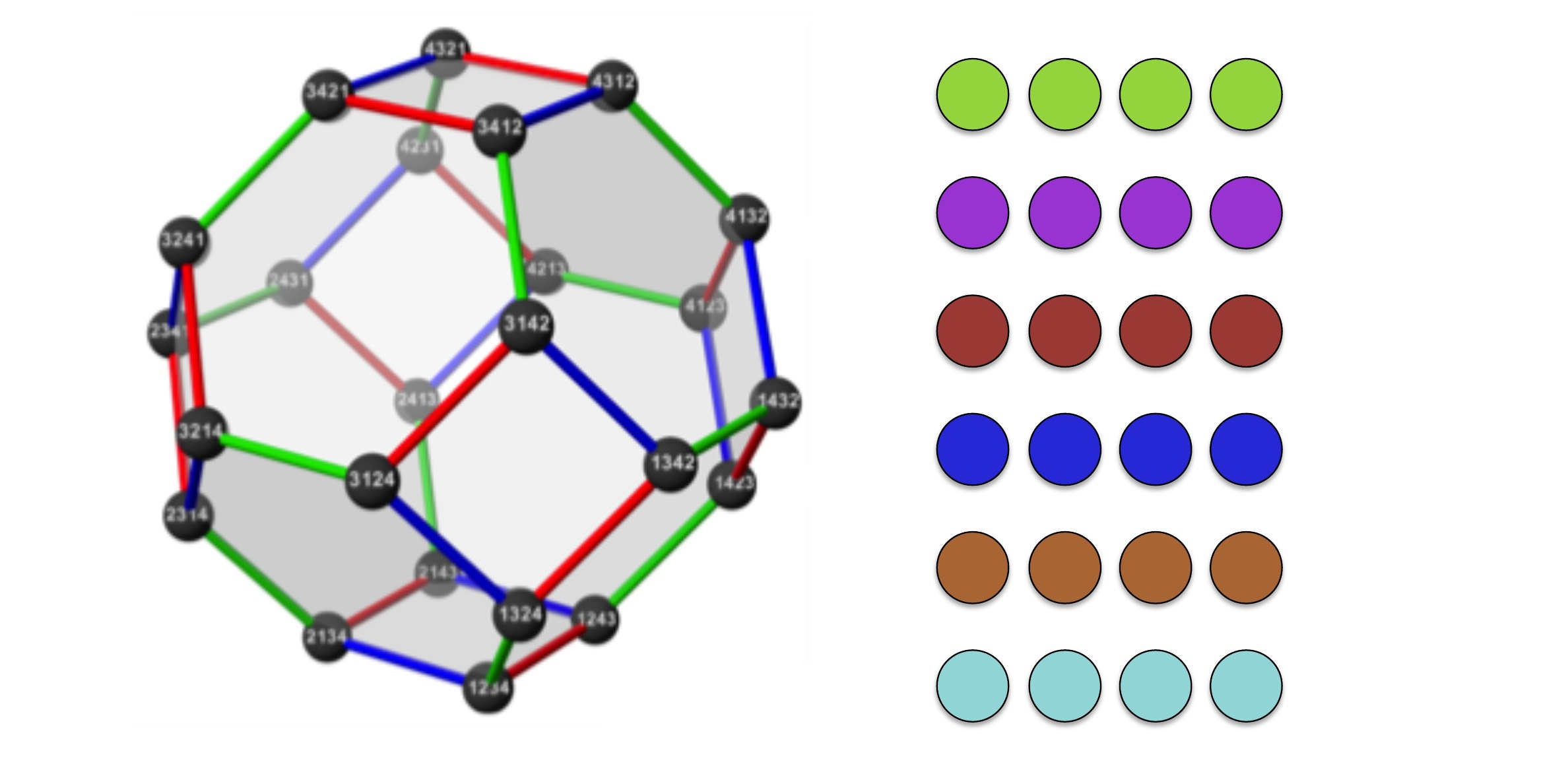}
  \caption{Permutahedron and subsets}
  \label{FigP3}
\end{minipage}
\end{figure}

In mathematics, the four color map conjecture dates back to 1852 \cite{Cit1}, and apparently its extension to a major theorem was the first proven by means of computers and information technology \cite{Cit2,Cit3}.  While at the time of the announcement in 1976 there was some controversy, by now there seems to be a general acceptance of the proof's validity.
%
%[Cit1]
%D.\ MacKenzie, Mechanizing Proof: Computing, Risk, %and Trust (MIT Press, 2004) p103
%
%[Cit1]
%G.\ Chartrand and Li.\ Lesniak, Graphs & Digraphs %(CRC Press, 2005) p.221
%
%[Cit3]
%Wilson (2014); Appel & Haken (1989); Thomas (1998, %pp. 852–853)
%
For a number of years a program of IT use has been under development \cite{X1,X2,X3,X4,X5} to explore problems at the mathematical foundations of supersymmetry.  In particular, the work of \cite{X1} showed the concept of ``off-shell'' 4D, $\cal N$ = 1 supersymmetry leading to the study of Coxeter Groups.  This work demonstrated that there exists a direct sum partitioning of the totality of BC(4) matrix representations into six disjoint subsets of matrices representing Clifford Algebras.  The permutation matrices are assigned as shown in Fig.~\ref{FigP1}
where each permutation matrix is indicated to belong to one of six (color-coded for convenience sake) subsets. Upon taking the absolute values of these BC(4) matrix representations, one is led to twenty-four matrices that provide a matrix representation of the permutation group of order four. The permutahedron is a well-known concept related to the study of permutation groups and all the members can properly be assigned using weak right Bruhat ordering \cite{weakbruhat} to the appropriate location as shown in Fig.~\ref{FigP2}.

This naturally leads to the question of whether there is an invariant associated with the 
permutation group of order four that determines the assignments of the members of the subsets to the distinct patterns of locations
in the permutahedron.  This answer is yes.  In the past \footnote{See the works in \cite{X3} and \cite{X5} for explicit definitions and discussions.}, we have used the symbols ${\cal G}_{(1)}$, and ${\cal G}_{(2)}$ to denote maps from adinkra graphs to the real numbers.
In this work, we define a new mapping operator that will be denoted by ${\cal G}_{(3)}$.  Unlike the previous two mapping operations, this new one is defined by the decoration of the nodes of the permutahedron by sets of colored spheres.

The permutahedron is shown in Fig.~\ref{FigP3}, and the six subsets are shown in Fig.~\ref{FigP2}.

% \begin{figure}[h]
% \centering
% \includegraphics[scale=.40]{FigP3.jpg}
% \caption{Permutahedron \& Subsets}
% \label{FigP3}
% \end{figure}

% \begin{figure}[h]
% \centering
% \includegraphics[scale=.30]{FigP2}
% \caption{Permutahedron Containing Subset Elements}
% \label{FigP2}
% \end{figure}

% In Fig. \ref{FigP2}, the elements of each subset are shown assigned to an appropriate location in the permutahedron.

\begin{figure}[h]
\centering
\includegraphics[scale=.45]{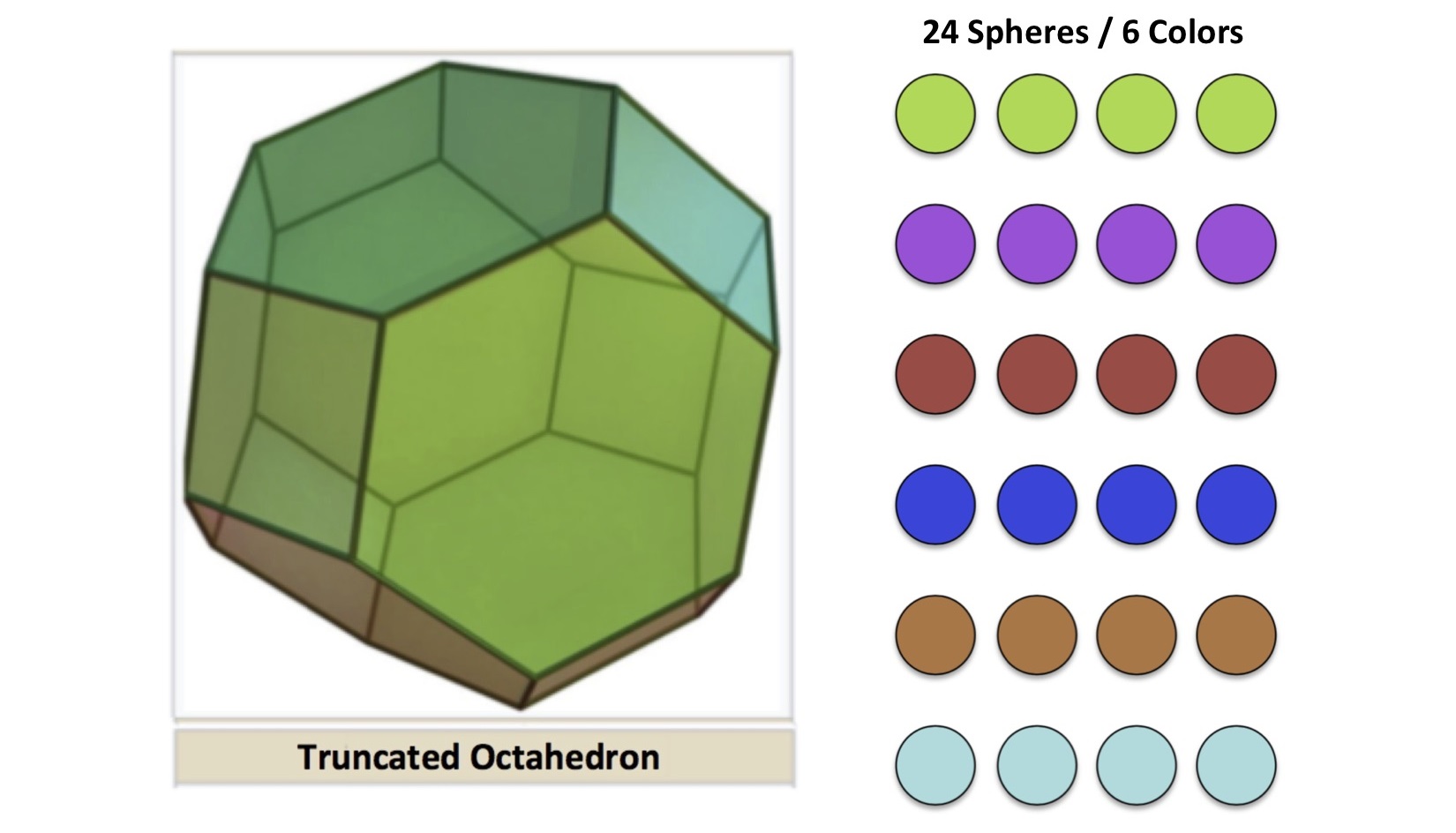}
\caption{Twenty-four spheres with six different colors next to a truncated octahedron}
\label{fig1}
\end{figure}

The truncated octahedron is a well known Archimedean solid that contains 24 vertices as shown in Fig.~\ref{fig1}.	
Let	us consider the case where 24 spheres (as shown to the right in Fig.~\ref{FigP3} are placed at vertices of the truncated octahedron, one sphere to each vertex to adorn the solid). 
However, the spheres are not all identical. Six disjoint sets each containing four spheres of the same color, are shown in Fig.~\ref{fig1}.	
After placing all of the spheres at vertices, a particular sphere can be specified, and then the minimum	number of edges traversed to go to the locations of the	remaining three	spheres of the same color can be noted and summed.	

A problem to solve is how to place the 24 spheres at the vertices so that the sum described above is the same independent of the choice of color.

As one example of such a solution, it can be noted that the truncated octahedron possesses six faces that are squares. So a solution is obtained by choosing four spheres of the same color and placing them at the vertices of one of the square faces.
Upon choosing one of these spheres as the fiducial one, there are two other spheres adjacent to it and connected by one edge. The remaining sphere is connected to this chosen sphere by two edges. Thus the sum of the numbers of edges connecting the chosen sphere to the other three is ${\cal G}{}_{(3)}$ = 1 + 1 + 2 = 4.
As all the square faces are equivalent, the choice of which color of spheres over which to carry out this calculation is irrelevant. 

An interesting question to entertain is whether there are other placements of the spheres which have the same property with respect to such an invariant and what is the numerical value for this invariant for these other placements.

There is a second such solution: it is the one obtained by using the image in Fig.~\ref{FigP2} as a guide to all the positioning of the colored spheres.  Carrying out the same calculation upon picking a fiducial sphere of any color leads to ${\cal G}{}_{(3)}$ = 2 + 4 + 6 = 12.  Furthermore, this result is independent of the color of the spheres that are chosen.
This choice of decoration is the one that is required to implement the disassociation of the Coxeter Group BC(4) matrices in six sets of matrices representing Clifford Algebras.  We call ${\cal G}{}_{(3)}$ = 12 the ``magic number'' and will return to this in the remainder of the discussion.

The problem described above can be generalized to permutations groups of higher order. For the permutation group of order $8$, there will be $8!$ nodes and the number of distinct colors will be $8$.  In this case, there are 191,520 faces, 141,120 edges, 40,320 nodes and 8 colors.

Moreover, the solution with ${\cal G}{}_{(3)}$ = 12 can be interpreted as a solution of a graph vertex coloring problem~\cite{Skiena}. The graph vertex coloring problem is defined as the process of coloring the nodes in such a way that no two nodes sharing a common edge of the graph have the same color. The smallest number of colors needed to color a graph $G$ is called the chromatic number, denoted as $\chi(G)$ \cite{Skiena}.  It can be seen that our first proposed solution with ${\cal G}{}_{(3)}$ = 4
is {\it {not}} a solution of the graph vertex coloring problem.

\begin{figure}[h]
\centering
\includegraphics[scale=0.92]{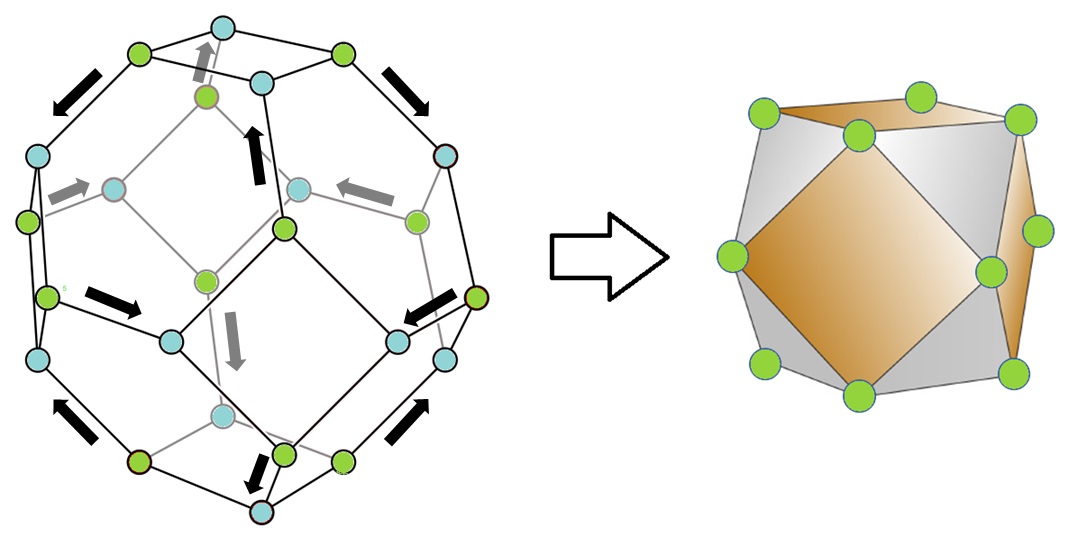}
\caption{Coloring a truncated octahedron with 2 colors}
\label{fig5}
\end{figure}

Then one question we can ask is whether any geometric method exists to construct the second coloring solution suggested above. 
In order to do this, first let us define the ``restricted chromatic number" under the stronger restrictions on graph coloring, as the smallest number of colors needed to color for an arbitrary graph $G$ that ${\cal G}{}_{(3)}$ has a constant value and independent of the choice of color. Then one can show that the restricted chromatic number of the permutahedron is 2 as in Fig.~\ref{fig5}.
Secondly, after coloring the vertices of the permutahedron with two colors (metallic gold and silver), we can transform the 12 metallic gold vertices of the permutahedron onto the cuboctahedron as on the right in Fig.~\ref{fig5}. (We can repeat the same process with all of the silver vertices.)

% \begin{figure}[H]
% \centering
% \begin{minipage}{.4\textwidth}
%   \centering
%   \includegraphics[scale=.45]{Colored_permutohedron_1.png}
%   \caption{Coloring a Truncated Octahedron with 2 colors}
%   \label{fig5}
% \end{minipage}%
% \hspace{0.6cm}
% \begin{minipage}{.4\textwidth}
%   \centering
%   \includegraphics[scale=.45]{Colored_permutohedron_2 .png}
%   \caption{Re-coloring a Fig. \ref{fig5} black nodes with 3 colors}
%   \label{fig6}
% \end{minipage}
% \end{figure}

\begin{figure}[h]
\centering
\includegraphics[scale=.37]{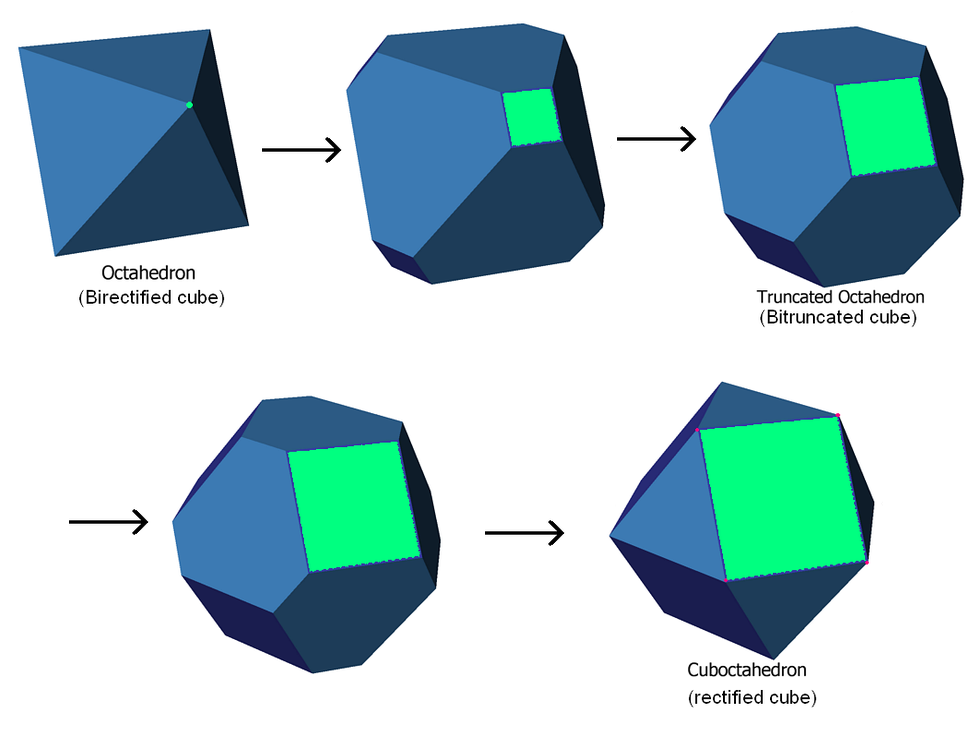}
\caption{The truncation process of an octahedron}
\label{trunc}
\end{figure}

Fig.~\ref{trunc} shows the relationship between the cuboctahedron and truncated octahedron. Starting from the octahedron, if we apply the uniform truncation, we obtain a truncated octahedron, which is also called a permutahedron (of order 4). If we continue to do deeper truncation, we can obtain the cuboctahedron as a result of rectified truncation.
This truncation process also can be interpreted as moving metallic gold nodes on the silver nodes as in Fig.~\ref{fig5}, which is a similar process of compressing the truncated octahedron from the eight cardinal directions to make a compact polytope (which is a cuboctahedron in this case) so that the metallic gold nodes are placed closest to each other.

\begin{figure}[h]
\centering
\includegraphics[scale=.41]{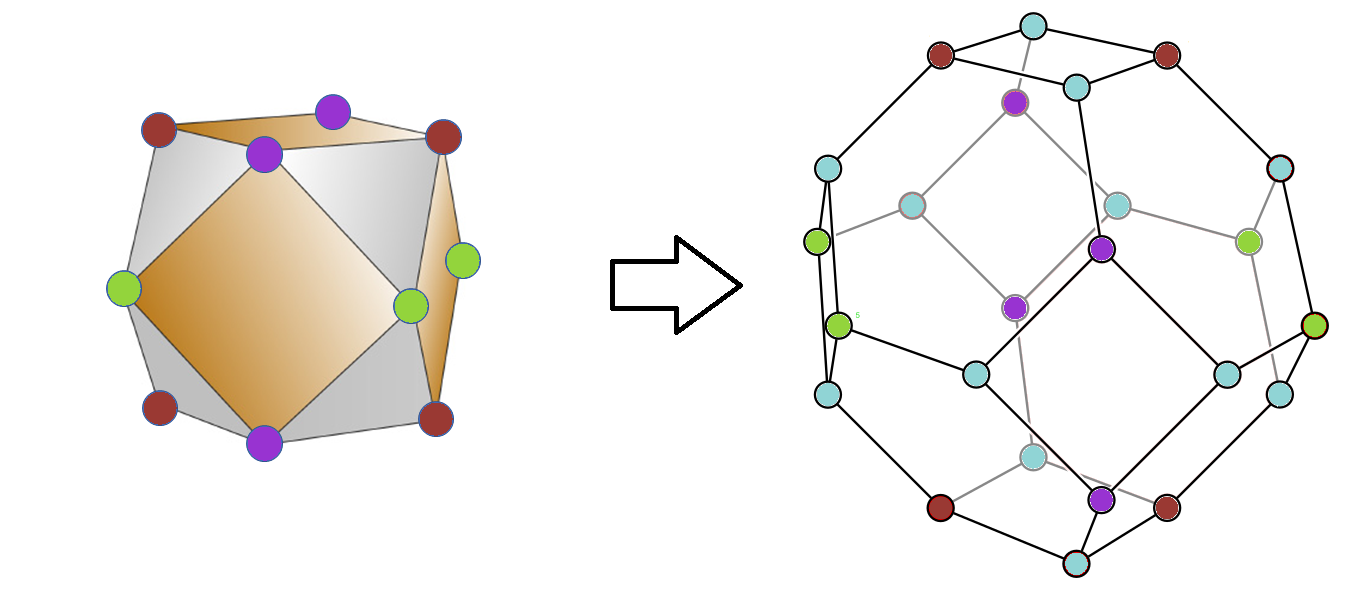}
\caption{Re-coloring metallic gold nodes in Fig.~\ref{fig5} with 3 colors (metallic gold, dark orchid, milano red)}
\label{fig6}
\end{figure}

\begin{figure}[h]
\centering
\includegraphics[scale=.41]{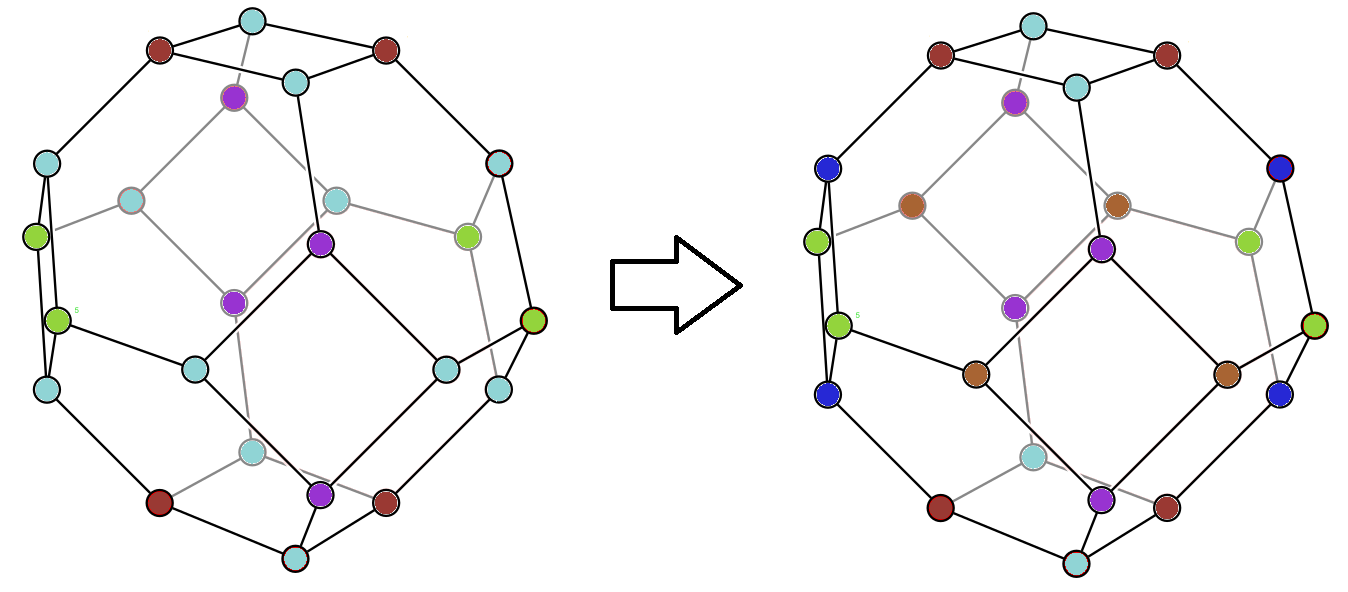}
\caption{Re-coloring silver nodes in the Fig.~\ref{fig6}  with 3 colors (silver, mai tai, persian blue)}
\label{fig8}
\end{figure}

Then, based on the fact that the restricted chromatic number for the cuboctahedron is 3 (as in Fig.~\ref{fig6}), we can re-color the metallic gold nodes with a different set of 3 colors (metallic gold, dark orchid, milano red) as on the right in Fig.~\ref{fig6}.
And if we process the same step on the silver vertices as Fig.~\ref{fig8}, we can get a new set of 6 $(=2\times 3)$ colors for permutahedron coloring with the appropriate coloring position that satisfies the restricted coloring condition as Fig.~\ref{FigP2} and maximizes the ${\cal G}{}_{(3)}$, which coincides with the second solution.

This approach is one of the available polytopic approach to find the positions of six sets of matrices representing Clifford Algebras on the permutahedron, and it can also be applied to the higher dimensional permutahedra.

Interestingly there are many different research activities based on developing the properties of the chromatic number $\chi(G)$. For example, Brook's Theorem states that the chromatic number of a graph is at most the maximum vertex degree, except when the graph is complete or an odd cycle.\cite{Brook}, and Grötzsch's theorem states that a triangle-free planar graph can be colored with only three colors. \cite{Grotzsch}

Thus, the importance of polytopic representation of the $N$-extended supermultiplets is the availability it provides for a bridge between various applications of mathematical fields of study such as graph theory, combinatorics, and supersymmetry. More detailed examples of SUSY representation will be presented in the remainder of the discussion.

\section{Recursive Construction of Supermultiplets}\label{sec:recursion}

Let us begin by discussing our use of the term ``supermultiplet'' in this work.
Throughout the following, the term supermultiplet refers to a collection of matrices.
The matrices in question are denoted by ${\bm {\rL}}{}_{{}_{\rI}}$ and
${\bm {\rR}}{}_{{}_{\rI}}$ where these are d $\times$ d matrices and the subscripted indices 
${{\rI}}$ take on values ${{\rI}}$ = 1,..., $N$ (for some integer $N $)\footnote{In this work, we primarily consider the case where $ N$ = d = 8.}.   These matrices can be extended to a 
set of $N$ (2d) $\times$ (2d) matrices (denoted by $ {\Hat {\bm \g}}{}_\rI$) by use of the formula
\be  \eqalign{
{\Hat {\bm \g}}{}_\rI  ~&=~ \fracm12 \, (\, {\bm  \sigma}^1 \,+\, i {\bm  \sigma}^2  \, ) \, \otimes \,  {\bm {\rL}}_\rI 
~+~ \fracm12 \, (\, {\bm  \sigma}^1 \,-\, i {\bm  \sigma}^2  \, ) \, \otimes \,  {\bm {\rR}}_\rI  \cr
%%%%%%%%%%%%%%%%%%%%%%%%%%%%%%%%%%%%%%%%%%%%%%
~&=~ \fracm1{\sqrt 2} \, (\, {\bm  \sigma}^+    \, ) \, \otimes \,  {\bm {\rL}}_\rI ~+~
\fracm1{\sqrt 2} \, (\, {\bm  \sigma}^-  \, ) \, \otimes \,  {\bm {\rR}}_\rI
~~~   } 
 \label{DefGmm}
\ee
for each collection ${\bm {\rL}}{}_{{}_{\rI}}$ and
${\bm {\rR}}{}_{{}_{\rI}}$.  Here, the symbol $\otimes$ denotes the outer product where
$ {\bm  \sigma}^1$ and $ {\bm  \sigma}^2$ are the usual first and second Pauli-matrices.

For our purposes, the collections ${\bm {\rL}}{}_{{}_{\rI}}$-${\bm {\rR}}{}_{{}_{\rI}}$ are distinguished into two types.  When the
collections satisfy (and thus form representations of Euclidean Clifford Algebras)
\begin{equation}
\left\{ \, {\Hat {\bm \g}}{}_\rI ~,~ {\Hat {\bm \g}}{}_\rJ  \, \right\} = 2\,  \delta_{\rm {IJ}} \, {\bm {\rm{I}}}{}_{2d} 
~~~,
\label{CLFF2}
\end{equation}
they are called an ``off-shell'' set.  This condition is realized when the
${\bm {\rL}}{}_{{}_{\rI}}$-${\bm {\rR}}{}_{{}_{\rI}}$ matrices satisfy the 
equations
\begin{align}
\begin{split}
{\bm {\rL}}{}_{{}_{\rI}} \, {\bm {\rR}}{}_{{}_{\rJ}} ~+~ {\bm {\rL}}{}_{{}_{\rJ}} \, {\bm {\rR}}{}_{{}_{\rI}}
& ~=~ 2\,  \d{}_{{}_{\rm {I \, J}}} \, {\bm {\rm{I}}}{}_{\rm d}  ~~~, ~~~
%%%%%%%%%%%%%%%%%%%%%%%%%%%%%%%%%%
{\bm {\rR}}{}_{{}_{\rI}} \, {\bm {\rL}}{}_{{}_{\rJ}} ~+~ {\bm {\rR}}{}_{{}_{\rJ}} \, {\bm {\rL}}{}_{{}_{\rI}}
~=~ 2\,  \d{}_{{}_{\rm {I \, J}}} \, {\bm {\rm{I}}}{}_{\rm d} ~~~.
\end{split}
\label{eq:GAlg1}
\end{align}
In previous works, these have been called the
${\cal {G R}}$(d, $N$) or ``garden algebra'' conditions recognizing the dependence on the parameters $\rm{d}$ and $N$.
 
However, some of the sets in the following
discussion satisfy a different condition of the form
\begin{equation}
\left\{ \, {\Hat {\bm \g}}{}_\rI ~,~ {\Hat {\bm \g}}{}_\rJ  \, \right\} = 2\,  \delta_{\rm {IJ}} \, 
{\bm {\rm{I}}}{}_{2\rm{d}}  ~+~ {\cal N}{}_{\rm {IJ}}{}^{\Hat \a} \, {\bm {\kappa}}{}_{\Hat \a} 
~~~,
\label{CLFF3}
\end{equation}
where the coefficients $ {\cal N}{}_{\rm {IJ}}{}^{\Hat \a \, ({\cal R})} $ and the sets of
$2\rm{d} $ $\times$ $ 2\rm{d}$ matrices ${\bm {\kappa}}{}_{\Hat \a}{}^{({\cal R})}$ are determined by the explicit forms of each ${\bm {\rL}}{}_{{}_{\rI}}$-${\bm {\rR}}{}_{{}_{\rI}}$ collection.
In this case above, the collection is called an ``on-shell'' set.
The coefficients ${\cal N}{}_{\rm {IJ}}{}^{\Hat \a \, ({\cal R})} $ may be interpreted as obstructions of an off-shell set to forming a Clifford algebra.

It is worth noting that matrix collections of ${\bm {\rL}}{}_{{}_{\rI}}$ and ${\bm {\rR}}{}_{{}_{\rI}}$ that are $\rm d$ $\times $  $\rm d $ matrices can be regarded as special cases of more general matrices.  
In this more general case, the $\rm d $ $\times $  $\rm d $ ${\bm {\rL}}{}_{{}_{\rI}}$ matrices are replaced by $ {\rm d}_L\times {\rm d}_R $ ${\bm {\rL}}{}_{{}_{\rI}}$ matrices where the integers $ {\rm d}_L$ and ${\rm d}_R $ satisfy $ {\rm d}_L$ $\ne $ $ {\rm d}_R$ .  
In a similar manner in the more general case, the $\rm d $ $\times $  $\rm d $ ${\bm {\rR}}{}_{{}_{\rI}}$ matrices are replaced by $ {\rm d}_L\times {\rm d}_R $ ${\bm {\rR}}{}_{{}_{\rI}}$ matrices.  
When this is done, the condition in (\ref{eq:GAlg1}) requires a replacement
of the form
\begin{align}
\begin{split}
{\bm {\rL}}{}_{{}_{\rI}} \, {\bm {\rR}}{}_{{}_{\rJ}} ~+~ {\bm {\rL}}{}_{{}_{\rJ}} \, {\bm {\rR}}{}_{{}_{\rI}}
& ~=~ 2\,  \d{}_{{}_{\rm {I \, J}}} \, {\bm {\rm{I}}}{}_{{\rm d}_L}  ~~~, ~~~
%%%%%%%%%%%%%%%%%%%%%%%%%%%%%%%%%%
{\bm {\rR}}{}_{{}_{\rI}} \, {\bm {\rL}}{}_{{}_{\rJ}} ~+~ {\bm {\rR}}{}_{{}_{\rJ}} \, {\bm {\rL}}{}_{{}_{\rI}}
~=~ 2\,  \d{}_{{}_{\rm {I \, J}}} \, {\bm {\rm{I}}}{}_{{\rm d}_R} ~~~.
\end{split}
\label{eq:GAlg2}
\end{align}
but there have never been uncovered realization where it is possible to simultaneously satisfy both of these conditions.  The situation ${\rm d}_L$ $\ne $ $ {\rm d}_R$. occurs often when a collection of fields only realize the supersymmetry algebra {\it {subject}} to some set of dynamical equations.
In the following, we will only consider the case were 
${\rm d}_L$ = $ {\rm d}_R$, but in distinction with
most previous studies this does not insure the realization of an ``off-shell'' system.

 It has been proposed that the structures above successfully ``mirror'' a situation that often occurs in discussions of supersymmetrical field theories when these are formulated in terms of the component fields of the representations. Adinkras are thus graphs defined over the component field space (but only gauge-invariant component fields) of SUSY representations.  The parameter
 $\rm d$ counts the number of
 {\it {gauge}}-{\it {invariant}}
 field components independent of dynamics while $N$ counts the number of supercharges.
 The case of 4D, $\cal N $ = 1 supergravity 
 can be used to illustrate this approach in an illuminating manner.  Let us illustrate this ``mirroring'' in an explicit example provided by
 four dimensional ``supergravity theory.''
 
 The original presentations of the four dimensional supergravity theory were completed by Freedman, van 
 Nieuwenhuizen, and Ferrara \cite{FFvN} and Deser, and 
 Zumino \cite{DZ}
 whose works were expressed in terms of the graviton $e{}_{\underline a }{}^{\underline m}(x)$ and the
 gravitino $\psi {}_{\underline a }{}^{\a}(x)$ fields.
 These works also contained the supersymmetry variations
$\d_Q e{}_{\underline a }{}^{\underline m}(x)$ and 
$\d_Q\psi {}_{\underline a }{}^{\a}(x)$.  Upon calculation of
 the commutator of two successive SUSY variations specified by
 fermionic Grassmann parameters $\epsilon_1$ and $\epsilon_2$ on either field variable one finds
\begin{align}
\left[ \,  \d_Q(\epsilon_1)   ~,~ \d_Q(\epsilon_2)   \, \right] ~=~ i \, 2
\left( \epsilon_1  \,  \g^{\underline a} \, \epsilon_2 \, \right)\, 
e{}_{\underline a }{}^{\underline m} \pa{}_{\underline m} ~+~
\d{}_{\rm {Gauge}}
~+~ { \d \, {\cal L}}{}_{{\rm field} }
\label{CLFF-FEQ}
\end{align}
where $ {\d{}_{\rm {Gauge }}}$ denotes
an appropriate gauge transformation for each field,
and
$ { \d \, {\cal L}}{}_{{\rm field} }$ denotes the
variation of a Lagrangian with respect to the 
fermionic field, but not the bosonic field,
on which the commutators are being evaluated.

Since adinkras are defined {\it {only}} over the gauge-invariant component fields, the first term on the RHS in (\ref{CLFF-FEQ}) has been proposed as the hologram of
the first term on the RHS of (\ref{CLFF3}).
Since adinkras only contain the gauge-invariant components, the second term on the RHS of (\ref{CLFF-FEQ}) vanishes identically.
Finally, the final term on the RHS of (\ref{CLFF-FEQ}) has been proposed as the hologram of the last term on the RHS of (\ref{CLFF3}).

Starting with the discussions\cite{permutadnk} 
and \cite{pHEDRON}, it has been found that in a number of cases the $ {\bm {\rR}}{}_{{}_{\rI}}$ matrices are the matrix transpose of the $ {\bm {\rL}}{}_{{}_{\rI}}$ matrices and furthermore the latter can be considered signed
permutations, i.e.~elements of Coxeter groups.  In the case where 
${\rI}$ and $\rm d$ take on values 1,...,4, the elements of the corresponding
Coxeter Group can be ``broken'' into six distinct sets 
\begin{align}
&\{C M\} = \{\langle 1\bar{4}2\bar{3}\rangle,\langle 23\bar{1}\bar{4}\rangle,\langle 3\bar{2}\bar{4}1\rangle,\langle 4132\rangle\}\label{4DN4C1}
\\
&\{T M\} = \{\langle 1\bar{3}\bar{4}\bar{2}\rangle,\langle 24\bar{3}1\rangle,\langle 312\bar{4}\rangle,\langle 4\bar{2}13\rangle\}
\label{4DN4T1}
\\
&\{V M\} = \{\langle 13\bar{2}\bar{4}\rangle,\langle 2\bar{4}1\bar{3}\rangle,\langle 3\bar{1}\bar{4}2\rangle,\langle 4231\rangle\}
\label{4DN4V1}
\\
&\left\{V M_{1}\right\} = \{\langle 1\bar{4}\bar{3}2\rangle,\langle \bar{2}\bar{3}41\rangle,\langle 3\bar{2}1\bar{4}\rangle,\langle 4123\rangle\} \\
&\left\{V M_{2}\right\} = \{\langle 12\bar{4}\bar{3}\rangle,\langle \bar{2}13\bar{4}\rangle,\langle 3421\rangle,\langle 4\bar{3}1\bar{2}\rangle\} \\
&\left\{V M_{3}\right\} = \{\langle 12\bar{3}\bar{4}\rangle,\langle \bar{2}1\bar{4}3\rangle,\langle 3412\rangle,\langle 4\bar{3}\bar{2}1\rangle\} 
\label{4DN4C6}
\end{align} 
(the notation and definitions of these can be found in the work of 
\cite{permutadnk}) whose union is the complete Coxeter group. 
Taking the absolute values of the Coxeter group elements in this discussion leads to the permutation group of order four.
The foundational tool used in \cite{permutadnk} to discover
the dissection of this Coxeter group, and its permutation subgroup, caused by
the condition in
(\ref{eq:GAlg1})  was a computer code used to
find all of its solutions.

In the end, this dissection process is equivalent to a sorting problem. The $4!$ elements of the permutation group of order four are sorted into six (i.e.~3!) disjointed subsets each containing four elements. The equivalent problem related
to the work completed in \cite{adnkKyeoh} must begin with the $8! = 40,320$ elements of the permutation group of order eight sorted into 5,040
(i.e.~7!) disjointed subsets each containing eight elements.

We begin noting some of the results for $\cal N$ = 2 supermultiplets\footnote{It should be noted the symbol $\cal N$ denotes the number of supercharges in a 
higher dimensional system while the symbol $N$ denotes the number of one dimensional supercharges realized in one dimensional systems obtained by reduction.} for  supersymmetry \cite{adnkKyeoh}, which for the chiral-chiral (CC), chiral-vector (CV) and chiral-tensor (CT) supermultiplets are given by 
\begin{equation}
\begin{aligned}[c]
&\mathbf{L}^{CC}_{1}=(170)_{b}\langle 14235867\rangle\\
&\mathbf{L}^{CC}_{2}=(204)_{b}\langle 23146758\rangle\\
&\mathbf{L}^{CC}_{3}=(102)_{b}\langle 32417685\rangle\\
&\mathbf{L}^{CC}_{4}=(0)_{b}\langle 41328576\rangle\\
&\mathbf{L}^{CC}_{5}=(15)_{b}\langle 58671423\rangle\\
&\mathbf{L}^{CC}_{6}=(105)_{b}\langle 67582314\rangle\\
&\mathbf{L}^{CC}_{7}=(195)_{b}\langle 76853241\rangle\\
&\mathbf{L}^{CC}_{8}=(165)_{b}\langle 85764132\rangle
\end{aligned}
\qquad \qquad
\begin{aligned}[c]
&\mathbf{L}^{CV}_{1} = (170)_{b}\langle 14236857\rangle \\
&\mathbf{L}^{CV}_{2} = (204)_{b}\langle 23145768\rangle \\
&\mathbf{L}^{CV}_{3} = (6)_{b}\langle 32418675\rangle \\
&\mathbf{L}^{CV}_{4} = (96)_{b}\langle 41327586\rangle \\
&\mathbf{L}^{CV}_{5} = (210)_{b}\langle 58672413\rangle \\
&\mathbf{L}^{CV}_{6} = (180)_{b}\langle 67581324\rangle \\
&\mathbf{L}^{CV}_{7} = (126)_{b}\langle 76854231\rangle \\
&\mathbf{L}^{CV}_{8} = (24)_{b}\langle 85763142\rangle
\end{aligned}
\qquad \qquad
\begin{aligned}[c]
&\mathbf{L}^{CT}_{1} = (234)_{b}\langle 14235786\rangle \\
&\mathbf{L}^{CT}_{2} = (76)_{b}\langle 23146875\rangle \\
&\mathbf{L}^{CT}_{3} = (134)_{b}\langle 32417568\rangle \\
&\mathbf{L}^{CT}_{4} = (32)_{b}\langle 41328657\rangle \\
&\mathbf{L}^{CT}_{5} = (11)_{b}\langle 58671342\rangle \\
&\mathbf{L}^{CT}_{6} = (173)_{b}\langle 67582431\rangle \\
&\mathbf{L}^{CT}_{7} = (103)_{b}\langle 76853124\rangle \\
&\mathbf{L}^{CT}_{8} = (193)_{b}\langle 85764213\rangle
\end{aligned}
\label{eq2.13}
\end{equation}
The elements of the garden algebra, being a subgroup of the Coxeter group, are composed of a signed matrix (the boolean factors in the above equations) times a permutation matrix, which together form the Coxeter group elements. Our method of generating higher $N$ supermultiplets will split the process into two parts. We generate the permutation matrices and the boolean factors separately, then combine the two. First we will examine the process of generating the analogous elements of higher $N$ permutation matrices. 

Finally, it is of note that the ${\bm {\rL}}{}_{{}_{\rI}}$-${\bm {\rR}}{}_{{}_{\rI}}$ matrices associated with the (CC) set satisfy the equations in (\ref{eq2.13}) while those associated with both the (CV) and (CT) sets satisfy the equations in (\ref{CLFF2}).

 This is in accord with the observation well known in the physics literature
 that there is no known off-shell formulation of the so-called 4D, ${\cal N}$ = 2 ``hypermultiplet'' described solely in terms of eight component fields.
 On the other hand, it is well known that there exist off-shell formulations of the so-called 4D, ${\cal N}$ = 2 vector supermultiplet
 and the 4D, ${\cal N}$ = 2 tensor supermultiplet in terms of  eight component fields.

\subsection{Generating higher SUSY permutation matrices from lower ones}
\label{sec:generating_Lmatrices}
We begin by demonstrating the process for recursively constructing the chiral-chiral $\cal N$=2 supermultiplets from the chiral $\cal N$=1 supermultiplet. Concatenating two copies of the first element of (\ref{4DN4C1}) (and ignoring the boolean factor represented by the overlined numbers for now), we have 

\begin{equation}
\begin{aligned}[c]
\langle 1423\rangle \oplus \langle 1423\rangle = \langle 14231423\rangle
\end{aligned}
\qquad \rightarrow \qquad
\begin{aligned}[c]
\langle 1423\rangle \oplus \langle 5867\rangle = \langle 14235867\rangle
\end{aligned}
\end{equation}
where we relabeled our second string so that our numbering would be from 1-8. We see that this indeed matches the permutation matrix component of $\mathbf{L}^{CC}_{1}$ above. For matrices $\mathbf{L}^{CC}_5-\mathbf{L}^{CC}_8$ we instead relabel the first string (adding 4 to the first four numbers instead of the last four). Similar manipulations can be carried out for all the remaining supermultiplets, notably even when we are combining different $\cal N$=1 supermultiplets. For example, to generate $|\mathbf{L}_1^{CT}|$, we combine
\begin{equation}
\begin{aligned}[c]
\langle 1423\rangle \oplus \langle 1342\rangle = \langle 14231342\rangle
\end{aligned}
\qquad \rightarrow \qquad
\begin{aligned}[c]
\langle 1423\rangle \oplus \langle 5786\rangle = \langle 14235786\rangle
\end{aligned}
\end{equation}
which matches with what was previously found for $\mathbf{L}_1^{CT}$ up to the boolean factor.

This process is also applied to the creation of the
$\mathbf{L}_2^{CT}$, $\mathbf{L}_3^{CT}$, and $\mathbf{L}_4^{CT}$ matrices.  By this means we have
gone from four 4 $\times$ 4 matrices to four 8 $\times$ 8 matrices.

The corresponding $|\mathbf{L}_5^{CT}|$ is found from relabeling the first string instead
\begin{equation}
\begin{aligned}[c]
\langle 1423\rangle \oplus \langle 1342\rangle = \langle 14231342\rangle
\end{aligned}
\qquad \rightarrow \qquad
\begin{aligned}[c]
\langle 5867\rangle \oplus \langle 1342\rangle = \langle 58671342\rangle
\end{aligned}
\end{equation}
which indeed matches the permutation element of the 4D, $\cal N$ = 2 chiral-tensor supermultiplet given above.
We apply this process to the creation of the
$\mathbf{L}_6^{CT}$, $\mathbf{L}_7^{CT}$, and $\mathbf{L}_8^{CT}$ matrices.  Now we have four more 8 $\times$ 8 matrices.

The discussion above was carried out on the basis of considering the matrices as realizations of permutation operators.  We can summarize 
these results in the forms of formulae applied to matrices, but {\it {without}} any reference to permutations.  The discussion above directly implies 
\begin{equation}
\begin{aligned}[c]
& 
| {\Hat {\bm L}}^{(multp1, multp2)}_\rI |  ~=~ \frac12 \, (\, {\bm  {\rm I}} \,+\,  {\bm  \sigma}^3  \, ) \, \otimes \,
| {\bm {\rL}}_\rI^{(multp1)} |
~+~
\frac12 \, (\, {\bm  {\rm I }} \,-\,  {\bm  \sigma}^3  \, )
\, \otimes \,
| {\bm {\rL}}_\rI^{(multp2)}    | ~~~~~~~~~,~~~ {\rm {for}}~ {\rm I} ~=~ 1,...,4 ~~~,
\\
& | {\Hat {\bm L}}^{(multp1, multp2)}_\rI | ~=~ 
\fracm1{\sqrt 2} \, (\, {\bm  \sigma}^-    \, )
 \, \otimes \,
| {\bm {\rL}}_\rI^{(multp1)} |
~+~
\fracm1{\sqrt 2} \, (\, {\bm  \sigma}^+    \, )
\, \otimes \,
| {\bm {\rL}}_\rI^{(multp2)}    | ~~~~~~~~~~~~~~~~,~~~ {\rm {for}}~ {\rm I} ~=~ 5,...,8 ~~~.
\end{aligned}\label{recursive_construction}
\end{equation}
Here ``multp1'', ``multp2'' denote the lower SUSY permutation matrices we use. At this point, it is also convenient to introduce a notational device that will be convenient later.  Upon making the definitions
\begin{equation}
{\bm {\rm P}}{}^{\pm} ~ \equiv ~ \frac12 \, (\, {\bm  {\rm I}} \,\pm\,  {\bm  \sigma}^3  \, )
\label{P+P-} 
\end{equation}
we can rewrite (\ref{recursive_construction}) in the form
\begin{equation}
\begin{aligned}[c]
& | {\Hat {\bm L}}^{(multp1, multp2)}_\rI | ~=~ {\bm {\rm P}}{}^{+}  \, \otimes \,
| {\bm {\rL}}_\rI^{(multp1)} |
~+~
{\bm {\rm P}}{}^{-} \otimes \,
| {\bm {\rL}}_\rI^{(multp2)}|
~~~~~~~~~~~~~~~~~~~~~~~~~,~~~ {\rm {for}}~ {\rm I} ~=~ 1,...,4 ~~~,
\\
& | {\Hat {\bm L}}^{(multp1, multp2)}_\rI | ~=~ \fracm1{\sqrt 2} \, (\, {\bm  \sigma}^-    \, ) \, \otimes \,
| {\bm {\rL}}_\rI^{(multp1)} |
~+~
\fracm1{\sqrt 2} \, (\, {\bm  \sigma}^+    \, )\otimes \,
| {\bm {\rL}}_\rI^{(multp2)}|.~~~~~~~~~,~~~ {\rm {for}}~ {\rm I} ~=~ 5,...,8 ~~~.
\end{aligned}
\label{ReW}
\end{equation}
where we have used the $ {\bm {\rm P}}$ notation to distinguish these matrices from elements in the permutation group such as appearing in (\ref{Prmx}) below.
The reader is reminded that these results \cite{adnkKyeoh} follow from two steps starting with 
the linear addition of pairs of actions for 4D, $\cal N$ = 1 off-shell supermultiplets.
First one obtains fermionic transformation laws that leave the pairs of actions invariant up 
to surface terms and then one reduces these transformation laws to one dimension to obtain 
the corresponding L-matrices.  Starting from the 4D, $\cal N$ = 1 off-shell chiral, vector, 
and tensor supermultiplets, this leads to six sets of L-matrices. We denote these matrices 
according to the pairings of the supermultiplets, hence (CC), (CV), (CT), (VV), (VT), (TT). 
We may think of these as labeling different representations. The work in \cite{GRana2} (that 
is based solely on considerations from purely one-dimensional SUSY) was the source of a seventh
set of matrices, and we can call this the (O) (for original) representation.  The absolute 
values of all seven sets lie within the permutation group of order eight. However, only the 
(O), (CV), (CT) sets satisfy the garden algebra condition that marks them as being `off-shell' 
representations.

\subsection{Generating the boolean factor for higher dimensional supermultiplets}\label{sec:generate}
While in Sec.~\ref{sec:generating_Lmatrices} we have recursively generated the unsigned L-matrices, the boolean factors are slightly more complicated to generate. We illustrate the general recursive construction for the Chiral-Tensor $\cal N$=2 supermultiplet. We first note that signed L-matrices can be decomposed into a sign factor matrix (boolean factor) times a permutation group matrix\cite{permutadnk}, i.e,
\begin{equation}\left(\mathbf{L}_{\mathrm{I}}\right)_{i}{}^{\hat{k}}=\left(\mathcal{S}^{(\mathrm{I})}\right)_{i}{}^{{\ell}}\left(\mathcal{P}_{(\mathrm{I})}\right)_{{\ell}}{}^{\hat{k}}
\label{Prmx}
\end{equation}

We generate the boolean factors from the sign factor matrices as follows:
\begin{equation}\left(\mathcal{S}^{(\mathrm{I})}\right)_{i}{}^{{\ell}}=\left[\begin{array}{cccc}
(-1)^{b_{1}} & 0 & 0 & \cdots \\
0 & (-1)^{b_{2}} & 0 & \cdots \\
0 & 0 & (-1)^{b_{2}} & \cdots \\
\vdots & \vdots & \vdots & \ddots
\end{array}\right] \quad \leftrightarrow \quad\left(\mathcal{R}_{\mathrm{I}}=\sum_{i=1}^{\mathrm{d}} b_{i} 2^{i-1}\right)_{b}=\begin{array}{c}
{\left[b_{1} b_{2} \cdots b_{\mathrm{d}}\right]_{2}},
\end{array}
\label{eq2.21}
\end{equation}

Here in (\ref{eq2.21}) $\mathcal{R}_\mathrm{I}$ represent the arbitrary real number, where we read binary words from left to right by convention (in contrast to the typical right to left construction). For example we can consider the decomposition of $\mathbf{L}_1$ for the $\cal N$=1 chiral multiplet, for which  we have 

\begin{equation}\left(\mathbf{L}_{1}\right)_{i}{}^{\hat{k}}=\left[\begin{array}{cccc}
1 & 0 & 0 & 0 \\
0 & 0 & 0 & -1 \\
0 & 1 & 0 & 0 \\
0 & 0 & -1 & 0
\end{array}\right]=\left[\begin{array}{cccc}
1 & 0 & 0 & 0 \\
0 & -1 & 0 & 0 \\
0 & 0 & 1 & 0 \\
0 & 0 & 0 & -1
\end{array}\right]\left[\begin{array}{cccc}
1 & 0 & 0 & 0 \\
0 & 0 & 0 & 1 \\
0 & 1 & 0 & 0 \\
0 & 0 & 1 & 0
\end{array}\right]=(10)_{b}\langle 1423\rangle=\langle 1 \overline{4} 2 \overline{3}\rangle.
\end{equation}

Turning now to the construction of $\mathbf{L}_1$ for the $\cal N$=2 chiral-tensor,
\begin{align}
    \begin{split}
    \mathbf{L}_1^{(CT)} &= \langle 1\bar{4}2\bar{3} \rangle \oplus \langle 1\bar{3}\bar{4}\bar{2} \rangle \\
        &= [01010111]_{b} \langle 14235786 \rangle
    \end{split}
\end{align}
where $[01010111]_{b}$ means the boolean factor [01010111]. Going forward we will bracket these factors and give them a subscript of b.

In the following, ${\bm {\cal B}}_{{\bm {\cal (R)}}}$ corresponds to a bitflip matrix (for given representation $\cal R$) obtained in the following way. To take the Chiral-Vector multiplet as an example, for elements $\mathbf{L}^{(CT)}_5-\mathbf{L}^{(CT)}_8$ we use the following rule: For the $i$-th element, take the $(i-4)$-th element's boolean factor and flip 4 neighboring bits (for example the fifth through the eighth bits. The choice of where to start is arbitrary). Do this to generate the boolean factors for $\mathbf{L}^{(CT)}_5-\mathbf{L}^{(CT)}_8$ and check if the garden algebra closes. If it does not close, cycle the choice of neighboring bits by one. For our example of choosing the fifth through the eighth bits, we would get a factor that wouldn't satisfy the garden algebra. Therefore we'd cycle to the next 4 neighboring bits, \{6, 7, 8, 1\}. In principle, there are 8 choices to cycle through (i.e \{1, 2, 3, 4\},\{2, 3, 4, 5\},\{3, 4, 5, 6\},\{4, 5, 6, 7\},\{5, 6, 7, 8\},\{6, 7, 8, 1\},\{7, 8, 1, 2\},\{8, 1, 2, 3\}) for $\cal N$=2. If none of these combinations work, then the supermultiplet does not have boolean factors that make it a garden algebra. The example calculation for $\mathbf{L}^{(CT)}_5$ ($\cal N$=2 chiral-tensor) is performed below, where we apriori know that the correct bits to flip are (6, 7, 8, 1). 

\begin{align}
    \begin{split}
    \mathbf{L}_5^{(CT)} &= \langle 1\bar{4}2\bar{3} \rangle \oplus \langle 1\bar{3}\bar{4}\bar{2} \rangle \\
        &= \text{Bitflip}_{(6,7,8,1)}([01010111]_{b}) \langle 58671342 \rangle \\
        &= [11010000]_{b} \langle 58671342 \rangle
    \end{split}
\end{align}

The corresponding bitflip matrices for the (CV) and (CT) sets take the respective forms

\begin{equation}
{\bm {\cal B}}_{(CV)} =  
\begin{bmatrix}
1 & 0 & 0 & 0 & 0 & 0 & 0 & 0 \\
0 & 1 & 0 & 0 & 0 & 0 & 0 & 0 \\
0 & 0 & 1 & 0 & 0 & 0 & 0 & 0 \\
0 & 0 & 0 & -1 & 0 & 0 & 0 & 0 \\
0 & 0 & 0 & 0 & -1 & 0 & 0 & 0 \\
0 & 0 & 0 & 0 & 0 & -1 & 0 & 0 \\
0 & 0 & 0 & 0 & 0 & 0 & -1 & 0 \\
0 & 0 & 0 & 0 & 0 & 0 & 0 & 1
\end{bmatrix} ~~~,~~~
{\bm {\cal B}}_{(CT)} =  
\begin{bmatrix}
-1 & 0 & 0 & 0 & 0 & 0 & 0 & 0 \\
0 & 1 & 0 & 0 & 0 & 0 & 0 & 0 \\
0 & 0 & 1 & 0 & 0 & 0 & 0 & 0 \\
0 & 0 & 0 & 1 & 0 & 0 & 0 & 0 \\
0 & 0 & 0 & 0 & 1 & 0 & 0 & 0 \\
0 & 0 & 0 & 0 & 0 & -1 & 0 & 0 \\
0 & 0 & 0 & 0 & 0 & 0 & -1 & 0 \\
0 & 0 & 0 & 0 & 0 & 0 & 0 & -1
\end{bmatrix}.
\end{equation}

\noindent The boolean factors of $\mathbf{L}^{(CT)}_5-\mathbf{L}^{(CT)}_8$ are dual to the boolean factors of $\mathbf{L}^{(CT)}_1-\mathbf{L}^{(CT)}_4$ by doubly even (4) bit flips on neighboring bits.

It is obvious that these matrices square to the identity element and as well are traceless.  Let us rewrite these
in the outer product notation in the forms
\begin{equation}
{\bm {\cal B}}_{(CV)} ~=~  
 (\,  {\bm {\rm P}}{}^{+}  \, ) \, \otimes \,
 {\bm { {\cal M}}}_{{\bm {\cal (CV)}}}  
~+~
 (\, {\bm {\rm P}}{}^{-}   \, )
\, \otimes \,
{\bm {\Tilde {\cal M}}}_{{\bm {\cal (CV)}}} 
 \end{equation}
 \begin{equation}
 {\bm {\cal B}}_{(CT)}
~=~ (\, {\bm {\rm P}}{}^{+}   \, ) \, \otimes \,
 {\bm {\cal M}}_{{\bm {\cal (CT)}}}
~+~
 (\, {\bm {\rm P}}{}^{-}  \, )
\, \otimes \,
{\bm {\Tilde {\cal M}}}_{{\bm {\cal (CT)}}} 
 \end{equation}
The quantities 
${\bm {\Tilde {\cal M}}}_{{\bm {\cal (CV)}}}$ and
${\bm {\Tilde {\cal M}}}_{{\bm {\cal (CT)}}}$ are
the bitflip duals of ${\bm { {\cal M}}}_{{\bm {\cal (CV)}}}$ and
${\bm { {\cal M}}}_{{\bm {\cal (CT)}}}$
and in these equations  ${\bm {{\cal M}}}_{{\bm {\cal (CV)}}}$ and ${\bm {{\cal M}}}_{{\bm {\cal (CT)}}}$
 are the 4 $\times$ 4 matrices given by
 \begin{equation}
  {\bm {\cal M}}_{{\bm {\cal (CV)}}} ~=~ 
  \begin{bmatrix}
1 & 0 & 0 & 0  \\
0 & 1 & 0 & 0  \\
0 & 0 & 1 & 0  \\
0 & 0 & 0 & -1 
\end{bmatrix}
~~~,~~~
{\bm {\cal M}}_{{\bm {\cal (CT)}}} ~=~ 
\begin{bmatrix}
-1 & 0 & 0 & 0  \\
0 & 1 & 0 & 0  \\
0 & 0 & 1 & 0  \\
0 & 0 & 0 & 1  
\end{bmatrix}
\end{equation}
These matrices make it clear that the form of the ${\bm {\cal M}}_{({\bm {{\cal R}}})}$
matrices determine the forms of the $ {\bm {\Tilde{\cal M}}}_{({\bm {\cal R}})}$ matrices (and thus the forms of the $ {\bm {\cal B}}_{({\cal R})}$ matrices)
for the representations $({\cal R})$ = $({CV})$ and $({CT})$.

In terms of matrix notation, we have
\begin{equation}
\begin{aligned}[c]
& 
{\Hat {\bm {\cal S}}}{}_\rI   ~=~  (\, {\bm {\rm P}}{}^{+}  \, ) \, \otimes \,
 {\bm {\cal S}}_\rI^{(multp1)} 
~+~
 (\, {\bm {\rm P}}{}^{-}  \, )
\, \otimes \,
 {\bm {\cal S}}_\rI^{(multp2)}     ~~~~~~~~~,~~~ {\rm {for}}~ {\rm I} ~=~ 1,...,4
\end{aligned}
\end{equation}

\begin{multline}
    {\Hat {\bm {\cal S}}}{}_\rI   ~=~  (\, {\bm {\rm P}}{}^{+}   \, ) \, \otimes \,
 {\bm {\cal M}}_{{\bm {\cal (R)}}} \cdot {\bm {\cal S}}_\rI^{(multp1)} 
~+~
 (\, {\bm {\rm P}}{}^{-}  \, )
\, \otimes \,
 {\bm {\cal M}}_{{\bm {\cal (R)}}} \cdot {\bm {\cal S}}_\rI^{(multp2)}     ~~~~~~~~~,~~~ \\{\rm {for}}~ {\rm I} ~=~ 5,...,8
\end{multline}

According to equations (3.3) and (3.6) in \cite{Note}, we obtain the HYMN value for each supermultiplet representation. This allows the following dissection:

\begin{eqnarray}
{\bm {\cal \hat C}}^{\bm(CT)}={\bm {\cal \hat C}}^{(CV)}
={\bm {\cal \hat C}}^{(O)}=\sigma_3\otimes {\bf{I}}_{8},\\
{\bm {\cal \hat C}}^{(CC)}={\bm {\cal \hat C}}^{(TT)}
={\bm {\cal \hat C}}^{(TV)}={\bm {\cal \hat C}}^{(VV)}={\bf{I}}_{16}
\end{eqnarray}

Here, the matrix denoted by $\bm{\cal \hat C}^{(\cal R)}$ is associated with each supermultiplet $\cal R$ and has eigenvalues known as HYMN values.\cite{Note}

The first class includes (CT), (CV), and (O), for which the HYMN matrix takes the form $\sigma_3\otimes {\bf{I}}_{8}$. The second class includes (CC), (TT), (TV), and (VV), for which the HYMN matrix takes the form ${\bf{I}}_{16}$. Based on the results presented in (\ref{CLFF2}), the available bitflip for (CT) was \{6, 7, 8, 1\}. Although this is not a mathematical proof, it is an observation that works for (CT), (CV), and (O), but not for any other combinations.

As part of our investigation, it is necessary to classify other supermultiplets such as CMN, O($2n$), and Harmonic $N$=2 systems using the Hexipentisteriruncicantitruncated 7-Simplex. We anticipate that this classification will align with the HYMN value classification and demonstrate a clear correlation between the two methods.

\subsection{A curious property maintained by the recursive construction}

There is a curious mathematical property maintained by the six supermultiplets mentioned from  (\ref{4DN4C1}) to (\ref{4DN4C6}). If we focus on the permutation matrices within a single representation, we can always pair the $n$-th matrix to $n$-th to last matrix so that the sum of the order of their ``weight'' (see Sec.~\ref{sec:0hopping}) is $d!+1$, where $d$ is the dimension of the permutation. Take $\left\{V M_{1}\right\}=\{\langle 1234\rangle ,\langle 2143\rangle ,\langle 3412\rangle ,\langle 4321\rangle \}$ for example. $\langle 1234\rangle $ is the smallest possible bracket notation for permutation of order $4$ and $\langle 4321\rangle $ is the largest. So after we arrange all permutations from smallest to largest, the orders of $\langle 1234\rangle $ and $\langle 4321\rangle $ are $1$ and $4!$ respectively. Their sum is $4!+1$. On the other hand, the orders of $\langle 2143\rangle $, $\langle 3412\rangle $ are $7$ and $18$ respectively. Still their sum is $4!+1$.

Another way of stating this property is that we can always pair the bracket notation so that the sum of digits on any position is $d+1$. For example, one pair within the VV supermultiplet \cite{Note} is $|\mathbf{L}_3^{(VV)}|=\langle 42318675\rangle $ and $|\mathbf{L}_6^{(VV)}|=\langle 57681324\rangle $ (ignoring the boolean factor). The first digit of 42318675 is 4 and the first digit of 57681324 is 5. The sum of their first digits is the dimension plus 1: $4+5=8+1$, as well as the sum of 6-th digit $6+3=8+1$. 

These two statements are equivalent. If we send every digit $i$ in braket notation to $d+1-i$, for instance $45728613$ maps to $54271386$, by symmetry the order of the preimage counting from small to large is exactly the order of the image counting from large to small.

This rule is preserved by our construction as well. Suppose we combine two $\cal N$=1 to obtain a $\cal N$=2 supermultiplet, and the two ``ingredients'' $\cal N$=1 satisfy the rule mentioned above. We can pair the $n$-th (for some $n\leq4$) L-matrix of the newly obtained supermultiplet with the $(8-i+1)$-th L-matrix (note that $d$=8 here). Since the `ingredients' satisfy the rule, by our construction method we know that the sum of the original digits at the same position would be $4+1$. On the other hand, the last half digits of the $n$-th L-matrix would be added $4$ from the original $\cal N$=1 matrix, whereas the first half digits of the $(8-i+1)$-th are added $4$. So the sum of digits at the same position of this pair would be $(4+1)+4=8+1$, maintaining the rule.

\section{Hopping Operators and a General Construction}
\label{sec:3}
\subsection{Introducing hopping operators}\label{sec:0hopping}

Within the domain of compact Lie algebras, the concepts of root and weight
spaces, the maximal torus, etc.\ have been recognized for decades as valuable tools to construct the representation theories of these algebras. 
The Jordan-Chevelley Decomposition of the generators of these Lie algebras
into elements of the maximal torus accompanied by an associated set of ``raising'' and ``lowering'' operators  continually show their value as
the foundations of Cartan's classification of all compact Lie algebras.

Starting in a lecture entitled ``Genomics, Networks, And Computational Concepts For Polytopic SUSY Representation Theory,'' delivered in 2021 \cite{X1}
it was proposed that although SUSY algebras are distinctly different from
compact Lie algebras, the concept of a ``weight space'' does likely exist,
even though (to the authors' best knowledge) the analog of the Jordan-Chevelley decomposition apparently does not exist for SUSY algebras.  Nevertheless, due to the relation of adinkras to Coxeter groups \cite{X2}, a structure has been identified for supersymmetry to play the role of root and weight spaces. An argument for such a structure goes as follows.

The elements of Coxeter groups may be regarded as signed permutations.  Thus the Coxeter group of length-$d$ elements---which we denote here by BC($d$)---consists of $ 2^d \, d!$ elements. Taking the absolute values of
the elements of BC($d$) yields the of the permutation group, which we denoted by $S_d$.  There is a well-known polytope called
the ``permutahedron'' \cite{pHR0n3} associated with $S_d$.  It was proposed in the works of \cite{Note,pHEDRON} that the permutahedron can act as a
weight space/root space for adinkras.  In order to achieve further progress
along these lines, it becomes a pressing matter to define quantities
that can play the role of ``raising and lowering operators''. What
defines the lowest or highest weight state of a representation?  In
particular, is there an intrinsic way to define the weight of a
permutation operator?

There are two natural ways to denote permutation operators: (a) cycle 
notation, and (b) ``bra-ket'' notation.  As an example, let us
consider the set of permutations given in cycle notation by
\be{ \eqalign{
\left\{ {~~}\mathcal{P}_{1} = () ~,~~
\mathcal{P}_{2} = (12)(34) ~,~~
\mathcal{P}_{3} = (13)(24)  ~,~~
\mathcal{P}_{4} = (14)(23)  {~~} \right\} ~~~,
} }
\label{eq:S1} \ee
which in ``bra-ket'' notation can be rewritten as
\be{ \eqalign{
\left\{ {~}\langle 1234\rangle ,~ \langle 2143\rangle , ~ \langle 3412\rangle , ~\langle 4321\rangle   {~} \right\}
~~~,} }
\label{eq:S3} 
\ee
respectively.
It should be apparent that the ``bra-ket'' notation provides a natural mapping between permutation operators and natural numbers, obtained by treating the list of digits in the brackets as a number in base-$d$.  We will refer to this map as the BK-map.  The BK-map thus induces a `weight' for 
each permutation, given by the order of their corresponding natural numbers.  Elements can be ordered according to their weight defined by the BK-map.  Inexorably, one is led to the concept that there is always a element of lightest weight that is natural and intrinsic in the considerations of permutations.

For example, application of the BK-map to the set in (\ref{eq:S1})
leads to the set of numbers given by
\begin{equation}
 \eqalign{
\left\{ {~}1234,~ 2143, ~ 3412, ~4321  {~} \right\}
~~~.}
\label{eq33} 
\end{equation}

So it is seen that the set of permutations given by~(\ref{eq:S1}) is already correctly ordered according by this weight.

This fact leads one to realize given any construction involving permutations, the elements can be ordered according to their weight defined by the BK-map.  Inexorably, one is led to the concept that there is always a element of lightest weight that is natural and intrinsic in the considerations of permutations.

Now we are in position to define ``hopping operators'' or ``hoppers.''  There are two sets of such operators: left hopping operators and right hopping operators.
We denote the A-th left ``hoppers'' by the symbol ${\mathcal H}{}^{(L)}_{\rm A}$ and the A-th right ``hoppers'' by the symbol ${\mathcal H}{}^{(R)}_{\rm A}$.  The hoppers are defined by the
equations (where $\mathcal{P}_{1}$ is the permutation operator of lightest weight in a given set),
\be{
\mathcal{P}_{\rm A} ~=~  {\mathcal H}{}^{(L)}_{\rm A} \, \mathcal{P}_{1}  ~~~~,~~~
\mathcal{P}_{\rm A} ~=~  \mathcal{P}_{1}  \, {\mathcal H}{}^{(R)}_{\rm A} ~~~,
} 
\label{eq:HpRs1}
\ee
so the hopping operators can be calculated from the permutation elements via the equations
\be{
\mathcal{P}_{\rm A} \, \left[ \mathcal{P}_{1}  \right]{}^{-1}~=~  {\mathcal H}{}^{(L)}_{\rm A} \,  ~~~,~~~
\left[ \mathcal{P}_{1}  \right]{}^{-1} \,
\mathcal{P}_{\rm A} ~=~  {\mathcal H}{}^{(R)}_{\rm A} 
~~~.
} 
\label{eq:HpRs2}
\ee

Before we leave this introduction to hoppers, it is of note that these operators have appeared in previous works on adinkras \cite{Hmy1,Hmy2,Hmy3,Hmy4,Hmy5,Hmy6}. In these older works the concepts of
``bosonic holoraumy matrices''
($ {\mathbf {\cal B}}{\,}_{{}_{\rm {I \, J}}}$) and 
``fermionic holoraumy matrices''
($ {\mathbf {\cal F}}{\,}_{{}_{\rm {I \, J}}}$) were introduced and defined by
\begin{align}
\begin{split}
{\bm {\rL}}{}_{{}_{\rI}} \, {\bm {\rR}}{}_{{}_{\rJ}} ~-~ {\bm {\rL}}{}_{{}_{\rJ}} \, {\bm {\rR}}{}_{{}_{\rI}}
& ~=~ {\mathbf {\cal B}}{\,}_{{}_{\rm {I \, J}}}   ~~~, ~~~
%%%%%%%%%%%%%%%%%%%%%%%%%%%%%%%%%%
{\bm {\rR}}{}_{{}_{\rI}} \, {\bm {\rL}}{}_{{}_{\rJ}} ~-~ {\bm {\rR}}{}_{{}_{\rJ}} \, {\bm {\rL}}{}_{{}_{\rI}}
~=~ {\mathbf {\cal F}}{\,}_{{}_{\rm {I \, J}}}  ~~~.
\end{split}
\label{eq:Hmy}
\end{align}
It is clear that $ {\mathbf {\cal B}}{\,}_{{}_{\rm {I \, J}}}$ and 
$ {\mathbf {\cal F}}{\,}_{{}_{\rm {I \, J}}}$ matrices have two parts.  To 
examine these two part we introduce ``left jumpers'' (${\mathcal J}{}^{(L)}_{\rm {A \, B}} $) and ``right jumpers''
(${\mathcal J}{}^{(R)}_{\rm {A \, B}} $) via the equations
\be{
\mathcal{P}_{A} ~=~  {\mathcal J}{}^{(L)}_{A \, B} \, \mathcal{P}_{B}  ~~~~,~~~
\mathcal{P}_{A} ~=~  \mathcal{P}_{B}  \, {\mathcal J}{}^{(R)}_{B \, A} ~~~,
} 
\label{eq:JmpRs1}
\ee
and thus we have the relations
\be{
\mathcal{P}_{A} \, [\mathcal{P}_{B} ]{}^{-1}~=~  {\mathcal J}{}^{(L)}_{A \, B}   ~~~~,~~~
[ \mathcal{P}_{B} ]{}^{-1}  \,
\mathcal{P}_{A} ~=~  {\mathcal J}{}^{(R)}_{B \, A} ~~~,
} 
\label{eq:JmpRs2}
\ee
and hence jumpers take an arbitrary permutation described by $\mathcal{P}_{\rm A}$ and ``jumps'' it to the permutation described by $\mathcal{P}_{\rm B}$.
The holoraumy matrices are therefore given by
\begin{align}
\begin{split}
{\mathbf {\cal B}}{\,}_{{}_{\rm {A \, B}}} ~=~  {\mathcal J}{}^{(L)}_{A \, B} 
~-~  {\mathcal J}{}^{(L)}_{B \, A} 
~~~,~~~
{\mathbf {\cal F}}{\,}_{{}_{\rm {A \, B}}} ~=~
  {\mathcal J}{}^{(R)}_{A \, B} 
~-~  {\mathcal J}{}^{(R)}_{B \, A} 
~~~.
\end{split}
\label{eq:BFmtrXs}
\end{align}
Thus, the hoppers are only some of the components of the larger ${\mathbf {\cal B}}{\,}_{{}_{\rm {A \, B}}} $
and ${\mathbf {\cal F}}{\,}_{{}_{\rm {A \, B}}} $ matrices.
In particular, we see
\be{
{\mathcal J}{}^{(L)}_{A \, 1} ~=~ {\mathcal H}{}^{(L)}_{A}  ~~~~,~~~
{\mathcal J}{}^{(R)}_{1 \, A} ~=~ {\mathcal H}{}^{(R)}_{A}
~~~.
} 
\label{eq:HppR}
\ee

As noted in the work of \cite{Hmy2}, the holoraumy matrices themselves satisfy
the equations
\be{  \eqalign{
\big[ \,  {\mathbf {\cal B}}{\,}_{{}_{\rm {I \, J}}} ~,~ 
{\mathbf {\cal B}}{\,}_{{}_{\rm {K \, L}}} \, \big] ~&=~  
2\d_{IL}\, {\mathbf {\cal B}}_{\rm {JK}} -2\d_{\rm {IK}}\,{\mathbf {\cal B}}_{JL} +2\d_{\rm {JK}}\,{\mathbf {\cal B}}_{\rm {IL}} -2\d_{\rm {JL}}\, {\mathbf {\cal B}}_{\rm {IK}} ~~~,
}}
\label{eq:2Spnsb}
\ee
\be{  \eqalign{
\big[ \,  {\mathbf {\cal F}}{\,}_{{}_{\rm {I \, J}}} ~,~ 
{\mathbf {\cal F}}{\,}_{{}_{\rm {K \, L}}} \, \big] ~&=~  
2\d_{IL}\, {\mathbf {\cal F}}_{\rm {JK}} -2\d_{\rm {IK}}\,{\mathbf {\cal F}}_{JL} +2\d_{\rm {JK}}\,{\mathbf {\cal F}}_{\rm {IL}} -2\d_{\rm {JL}}\, {\mathbf {\cal F}}_{\rm {IK}} ~~~.
}}
\label{eq:2Spnsf}
\ee
Once the existence of weights and hoppers is accepted,
then one can recall some structures from the theory of compact Lie algebras.  In this latter case, the Jordan-Chevalley decomposition implies that all of the generators of a compact Lie algebra can be partitioned into elements of either an abelian ``torus'' algebra, raising operators, or lowering operators.  The eigenvalues of the generators belonging to the abelian torus algebra define a space of weights for a set of states.  The raising operators act on the state space which possesses a state of lowest weight.  When a raising operator is applied to this lowest state, this produces a state of higher weight in a given representation.

In close analogy, for the permutations the application of the hoppers to the state with the lowest weight yields a higher weight state. So effectively hoppers are `raising operators' from the lowest weight state in the set of permutations.

\subsection{Using Hopping Operators}\label{sec:hopping}
The sets of permutations presented at the beginning of Sec.~\ref{sec:recursion} can be viewed as sets of hopping operators.  Given one element of a supermultiplet, the hopping operators allow us to generate the rest of the supermultiplet. In $\cal N$=1, the Klein group elements are the hopping operators. In $\cal N$=2, the Rana octet $\mathcal{R}_8$ plays the role of the hopping operators. In general, the hopping operators will be given by a group of permutations isomorphic to $(Z_2)^{\oplus \mathcal{N}+1}$.  See Sec.~\ref{sec:partition}. For example, for $\cal N$=1, the hopping operators (listed in cycle notation) are given by  

\begin{equation}
\begin{aligned}
&\mathcal{P}_{1} = () \\
&\mathcal{P}_{2} = (12)(34) \\
&\mathcal{P}_{3} = (13)(24) \\
&\mathcal{P}_{4} = (14)(23) 
\end{aligned}
\end{equation}

\begin{figure}[h]
\centering
\includegraphics[scale=.60]{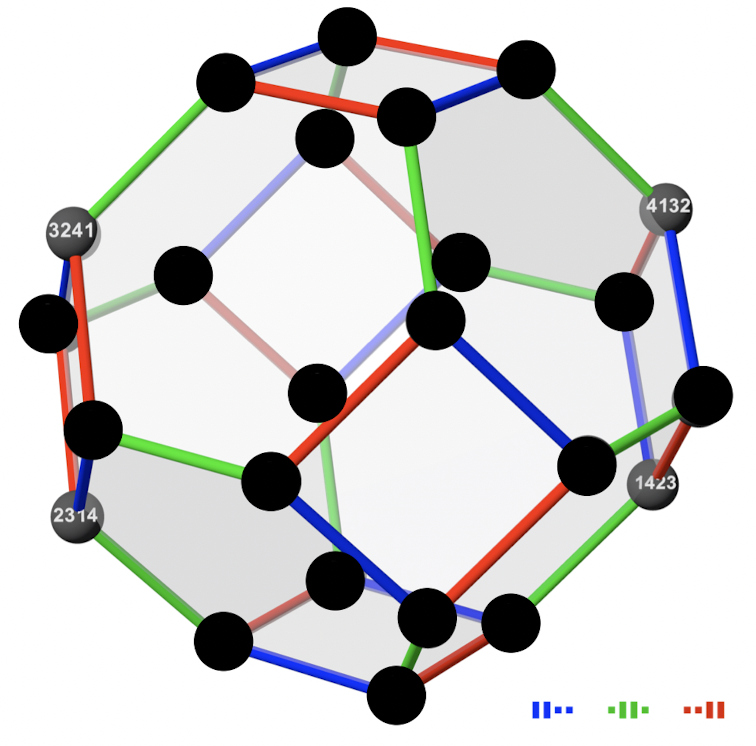}
\caption{Elements of the Chiral multiplet. Starting at 1423, the hopping operators will generate the three other elements.}
\end{figure}

Similarly, for 4D $\cal N$=2 (dimensionally reduced to D=1, $\cal N$=8), we can generate all the elements of a supermultiplet given a single element and the hopping operators. For example, for the chiral-vector supermultiplet, starting with $\mathbf{L}^{CV}_{1} = (170)_{b}\langle 14236857\rangle$ (and focusing only on the permutation matrix component), we can act with the hopping operators 

\begin{equation}
\begin{array}{llcl}
\mathcal{P}_{1}=\mathbf{I}_{2} \otimes \mathbf{I}_{2} \otimes \mathbf{I}_{2}= & () & = & \langle 12345678\rangle \\
\mathcal{P}_{2}=\mathbf{I}_{2} \otimes \mathbf{I}_{2} \otimes \sigma^{1}= & (12)(34)(56)(78) & = & \langle 21436587\rangle \\
\mathcal{P}_{3}=\mathbf{I}_{2} \otimes \sigma^{1} \otimes \mathbf{I}_{2}= & (13)(24)(57)(68) & = & \langle 34127856\rangle \\
\mathcal{P}_{4}=\mathbf{I}_{2} \otimes \sigma^{1} \otimes \sigma^{1}= & (14)(23)(58)(67) & = & \langle 43218765\rangle \\
\mathcal{P}_{5}=\sigma^{1} \otimes \mathbf{I}_{2} \otimes \mathbf{I}_{2}= & (15)(26)(37)(48) & = & \langle 56781234\rangle \\
\mathcal{P}_{6}=\sigma^{1} \otimes \mathbf{I}_{2} \otimes \sigma^{1}= & (16)(25)(38)(47) & = & \langle 65872143\rangle \\
\mathcal{P}_{7}=\sigma^{1} \otimes \sigma^{1} \otimes \mathbf{I}_{2}= & (17)(28)(35)(46) & = & \langle 78563412\rangle \\
\mathcal{P}_{8}=\sigma^{1} \otimes \sigma^{1} \otimes \sigma^{1}= & (18)(27)(36)(45) & = & \langle 87654321\rangle
\end{array}
\end{equation}
in which case we can reproduce all elements of the chiral-vector supermultiplet as follows.

\begin{equation}
\begin{aligned}[c]
&\mathcal{P}_{1} |\mathbf{L}^{CV}_{1}| = ()                \langle 14236857\rangle = \langle 14236857\rangle = |\mathbf{L}^{CV}_{1}| \\
&\mathcal{P}_{2} |\mathbf{L}^{CV}_{1}| = (12)(34)(56)(78) \langle 14236857\rangle = \langle 23145768\rangle = |\mathbf{L}^{CV}_{2}|  \\
&\mathcal{P}_{3} |\mathbf{L}^{CV}_{1}| = (13)(24)(57)(68) \langle 14236857\rangle = \langle 32418675\rangle = |\mathbf{L}^{CV}_{3}|  \\
&\mathcal{P}_{4} |\mathbf{L}^{CV}_{1}| = (14)(23)(58)(67) \langle 14236857\rangle = \langle 41327586\rangle = |\mathbf{L}^{CV}_{4}| \\
&\mathcal{P}_{5} |\mathbf{L}^{CV}_{1}| = (15)(26)(37)(48) \langle 14236857\rangle = \langle 58672413\rangle = |\mathbf{L}^{CV}_{5}|  \\
&\mathcal{P}_{6} |\mathbf{L}^{CV}_{1}| = (16)(25)(38)(47) \langle 14236857\rangle = \langle 67581324\rangle = |\mathbf{L}^{CV}_{6}|  \\
&\mathcal{P}_{7} |\mathbf{L}^{CV}_{1}| = (17)(28)(35)(46) \langle 14236857\rangle = \langle 76854231\rangle = |\mathbf{L}^{CV}_{7}| \\
&\mathcal{P}_{8} |\mathbf{L}^{CV}_{1}| = (18)(27)(36)(45) \langle 14236857\rangle = \langle 85763142\rangle = |\mathbf{L}^{CV}_{8}|  
\end{aligned}
\end{equation}

\subsection{Partitioning the Permutahedron via Cosets}\label{sec:partition}
It was found in \cite{Cosets} that the partitioning of $S_4$ can be understood in terms of cosets of Klein's Vierergruppe, i.e.~the set

\begin{equation}
    V=\{\perm{1234},\perm{2143},\perm{3412},\perm{4321}\}.
\end{equation}

The Vierergruppe is a normal subgroup of $S_4$, and it is isomorphic to the group $(Z_2)^{\oplus2}$.  

Here, we define cosets and normal subgroups as follows:
\begin{definition}
Let $\mathcal H=\{h_i\}$ be a subgroup of $S_n$ and $g\in S_n$ be any group element. We call the set $g\mathcal H=\{gh_i\}$ the {\bf left coset} by $g$, and the set $\mathcal Hg=\{h_i g\}$ the {\bf right coset} by $g$.
\label{def4-1}
\end{definition}

\begin{definition}
A subgroup $\mathcal{X}$ of the group $G$ is a \textbf{normal subgroup} of $G$ if and only if $g^{-1}xg\in \mathcal{X}$ for every $x\in \mathcal{X}$ and $g\in G$. 
\label{def4-2}
\end{definition}
A consequence of this definition is that, if $\mathcal{X}$ is normal in $G$, left and right cosets of $\mathcal{X}$ are equivalent.  That is, $\mathcal{X}g=g\mathcal{X},$ $\forall g\in G$.

In particular, each of the supermultiplets of $\cal N$=1 supersymmetry can be written as right cosets (or equivalently, left cosets, since $V$ is normal in $S_4$) of $V$:\begin{align}
    VM_3 &= V\\
    VM_2 &= V\perm{2134}\\
    VM_1 &= V\perm{3214}\\
    VM   &= V\perm{1324}\\
    CM   &= V\perm{2314}\\
    TM   &= V\perm{3124}.
\end{align}

When applied to the permutahedron, we recall findings regarding the magic number rules within and between these sets \cite{pHEDRON}.  With a suitable choice of sign prefactors, the Vierergruppe and its cosets form valid garden algebras.

Returning to \cite{Cosets}, the authors note that using the analogue of the Vierergruppe in $S_8$, the same partitioning can be accomplished.  We can construct this analogue, which we call the \textit{Rana subgroup~$\mathcal R_8$}, by following the method outlined in Sec.~\ref{sec:recursion}, starting with the Vierergruppe.  The result is
\begin{equation}
    \begin{aligned}[c]
    \mathcal{R}_8=
    \{& \perm{12345678},\perm{21436587},\perm{34127856},\perm{43218765},\\
    & \perm{56781234},\perm{65872143},\perm{78563412},\perm{87654321}\}.
    \end{aligned}
\end{equation}
This $\cal N$=2 supermultiplet is also identified in \cite{GRana1} and is isomorphic to the group $(Z_2)^{\oplus3}$.  As $\mathcal{R}_8$ forms a subgroup of $S_8$, its cosets will partition $S_8$ into octets.  However, in contrast with the Vierergruppe in $S_4$, $\mathcal{R}_8$ is not normal in $S_8$, so taking left and right cosets will not necessarily yield the same partitions.  Indeed, there are 30 conjugate subgroups of $\mathcal{R}_8$ that, when cosets are taken, will each yield a separate partition of $S_8$ \cite{Cosets}. The chiral-chiral, chiral-vector, and chiral-tensor supermultiplets are right cosets of $\mathcal{R}_8$. See Sec.~\ref{sec:hopping} for an explicit construction of the chiral-vector supermultiplet by this method.  Again, with suitable choices of sign prefactors, $\mathcal{R}_8$, its conjugates, and their cosets for valid garden algebras.

%mention the octet list that we generated? how can we have this list play into the paper?

We can further employ this method to generate a partition of $S_{16}$ via a recursive construction of the Rana subgroup (here, we must delineate terms of a permutation with commas to avoid ambiguity):
\begin{equation}
    \begin{aligned}[c]
    \mathcal{R}_{16}=
    \{& \perm{1, 2, 3, 4, 5, 6, 7, 8, 9, 10, 11, 12, 13, 14, 15, 16},
        \perm{2, 1, 4, 3, 6, 5, 8, 7, 10, 9, 12, 11, 14, 13, 16, 15},\\
      & \perm{3, 4, 1, 2, 7, 8, 5, 6, 11, 12, 9, 10, 15, 16, 13, 14},
        \perm{4, 3, 2, 1, 8, 7, 6, 5, 12, 11, 10, 9, 16, 15, 14, 13},\\
      & \perm{5, 6, 7, 8, 1, 2, 3, 4, 13, 14, 15, 16, 9, 10, 11, 12},
        \perm{6, 5, 8, 7, 2, 1, 4, 3, 14, 13, 16, 15, 10, 9, 12, 11},\\
      & \perm{7, 8, 5, 6, 3, 4, 1, 2, 15, 16, 13, 14, 11, 12, 9, 10},
        \perm{8, 7, 6, 5, 4, 3, 2, 1, 16, 15, 14, 13, 12, 11, 10, 9},\\
      & \perm{9, 10, 11, 12, 13, 14, 15, 16, 1, 2, 3, 4, 5, 6, 7, 8},
        \perm{10, 9, 12, 11, 14, 13, 16, 15, 2, 1, 4, 3, 6, 5, 8, 7},\\
      & \perm{11, 12, 9, 10, 15, 16, 13, 14, 3, 4, 1, 2, 7, 8, 5, 6},
        \perm{12, 11, 10, 9, 16, 15, 14, 13, 4, 3, 2, 1, 8, 7, 6, 5},\\
      & \perm{13, 14, 15, 16, 9, 10, 11, 12, 5, 6, 7, 8, 1, 2, 3, 4},
        \perm{14, 13, 16, 15, 10, 9, 12, 11, 6, 5, 8, 7, 2, 1, 4, 3},\\
      & \perm{15, 16, 13, 14, 11, 12, 9, 10, 7, 8, 5, 6, 3, 4, 1, 2},
        \perm{16, 15, 14, 13, 12, 11, 10, 9, 8, 7, 6, 5, 4, 3, 2, 1}    \}.
    \end{aligned}
\end{equation}

We once again note that $\mathcal{R}_{16}$ is isomorphic to $(Z_2)^{\oplus4}$, and it is a subgroup of $S_{16}$ (though not a normal subgroup).  Thus, the set forms two separate partitions of $S_{16}$: that formed by left cosets, and that formed by right cosets.  However, in contrast with previous cases, we find by direct search that there is no suitable choice of signs to convert $\mathcal{R}_{16}$ into a garden algebra.  The results in \cite{GRana1} indicate that we may have to go as far as $S_{128}$ to find a valid $N=16$ garden algebra.  The reason for this breach of pattern we leave for future exploration.

\subsection{The magic number}
In \cite{pHEDRON}, the idea of a number (hereafter called ``the magic number'') that constrains the ``positions'' of elements within a supermultiplet relative to the rest of the permutahedron was introduced.  The magic number intriguingly refers to two different but equivalent notions: the \textit{intra}-tet correlator, and the \textit{inter}-tet correlator. (These were originally dubbed ``intra-quartet'' and ``inter-quartet'' correlators, as they were studied in the context of $\cal N$=4 supersymmetry.) First, the intra-tet correlator of a given supermultiplet $\mathcal{A}$ and element $\pi\in \mathcal{A}$, labeled $\rm{Intra}_\pi(\mathcal{A})$, is the sum of the distances between $\pi$ and each other element of $\mathcal{A}$; that is,\begin{equation}
    \rm{Intra}_\pi(\mathcal{A})=\sum_{\alpha\in \mathcal{A}}d(\pi,\alpha),\label{eq321}
\end{equation} where $\rm d(\cdot,\cdot)$ is the hamming or Bruhat distance function detailed in \cite{pHEDRON}.
Second, the inter-tet correlator of a given permutation $\pi\in \mathcal{A}$ to a supermultiplet $\mathcal{B}\neq \mathcal{A}$, labeled $\rm{Inter}_\pi(\mathcal{A},\mathcal{B})$ is the sum of the distances between $\pi$ and each element of $\mathcal{B}$; that is,\begin{equation}
    \rm{Inter}_\pi(\mathcal{A},\mathcal{B})=\sum_{\beta\in \mathcal{B}}d(\pi,\beta).\label{eq322}
\end{equation}
What is remarkable about these quantities, and what in fact merits their collective description as the magic number, is that, in both $S_4$ and $S_8$, for any supermultiplets $\mathcal{A},\mathcal{B}$ and any elements $\pi,\sigma\in \mathcal{A}$,\begin{equation}\label{eq:magicnumber}
    \rm{Intra}_\pi(\mathcal{A})=\rm{Intra}_\sigma(\mathcal{A})=\rm{Inter}_\pi(\mathcal{A},\mathcal{B})=\rm{Inter}_\sigma(\mathcal{A},\mathcal{B}).
\end{equation}
In $S_4$ and $S_8$, this has been confirmed by brute force calculation, the magic number in $S_4$ being 12 and in $S_8$ being 112.  In $S_{16}$, $\rm{Intra}_{\perm{1,2,3,4,5,6,7,8,9,10,11,12,13,14,15,16}}(\mathcal{R}_{16})=960$, though the sheer size of $S_{16}$ has precluded a full validation of the identity~(\ref{eq:magicnumber}).  We conjecture that this magic number rule will hold for partitions in arbitrary $S_{2^n}$, the magic number being ${\cal G}_{(3)}=$$2^{2n-2}(2^n-1)$.

Notice that for a permutahedron of dimension $D$, the average distance between two vertices would be $(D-1)D/4$, which can be proved by induction. When $D=2^n$ and there are $D$ vertices in a multiplet, we can then approximate the magic number by the average distance\begin{equation}
    (D-1)D/4\times D=(2^n-1)2^{2n}/4=2^{2n-2}(2^n-1)\label{eq3.24}
\end{equation}
 which exactly matches with our conjectured magic number.

Future work may focus on the theoretical grounding for this magic number rule, while the rule itself may function as a crucial technique for the validation of supermultiplets.

\section{The Appearance of Ab-normal Cosets in BC(8)}
\label{sec:4}
We have seen in Sec.~\ref{sec:partition} that, in the 4D $\mathcal{N}=1$ case, left and right cosets are equivalent for purposes of partitioning the permutahedron, but in the 4D $\mathcal{N}=2$ case, they are not.  Here, we investigate some consequences of this fact.
%Based on the definitions of Cosets, we can define two kinds of cosets, left- and right-cosets. And as shortly mentioned in Section \ref{sec:partition}, in 4D $\mathcal{N}=1$ case, there is no difference between left- and right-cosets, but in 4D $\mathcal{N}=2$ case we can see the discrepancies between them.

\subsection{4D $\mathcal{N}=1$ ab-normal coset}
Let us first explicitly calculate left and right cosets for the 4D $\mathcal{N}=1$ case using the Vierergruppe $VM_3$ as the set of hopping operators.  This set has also been dubbed the diadem set $O$~\cite{pHEDRON}.
Also the diadem $O$ is the same as the object called Rana subgroup $\mathcal{R}_n$ in Sec.~\ref{sec:partition}. 
We calculate the left- and right-cosets of 4D $\mathcal{N}=1$ case by $O$ with every element of $S_4$.
%{To be specific first by using Def. \ref{def4-2} we can get the normal subgroup of the $S_4$ is Klein four-subgroup which is same as a diadem, $O$.  And by using Def. \ref{def4-1}, we can calculate left- and right-cosets of $S_4$.}
Using Def. \ref{def4-1}, we calculate the left and right cosets of $O$ by elements in $S_4$.
As expected, we find six distinct cosets.
Table. \ref{tab01} and \ref{tab02} shows the left- and right-coset list for 4D $\mathcal{N}=1$ case.

\begin{longtable}{|l|l|l|l|l|l|l|l|}
\hline
No. & Element 1 & Element 2 & Element 3 &  Element 4 & Type\\ \hline
1 & 1234 & 2143 & 3412 & 4321 & $O$/$VM_3$ \\ \hline
2 & 1243 & 2134 & 3421 & 4312 & $VM_2$\\ \hline
3 & 1324 & 2413 & 3142 & 4231 & $VM$\\ \hline
4 & 1342 & 2431 & 3124 & 4213 & $TM$\\ \hline
5 & 1423 & 2314 & 3241 & 4132 & $CM$\\ \hline
6 & 1432 & 2341 & 3214 & 4123 & $\rm VM_1$\\ \hline
\caption{$S_4$ left-cosets list}
\label{tab01}
\end{longtable}

\begin{longtable}{|l|l|l|l|l|l|l|l|}
\hline
No. & Element 1 & Element 2 & Element 3 &  Element 4 & Type\\ \hline
1 & 1234 & 2143 & 3412 & 4321 & $O$/$VM_3$\\ \hline
2 & 1243 & 2134 & 4312 & 3421 & $ VM_2$\\ \hline
3 & 1324 & 3142 & 2413 & 4231 & $VM$\\ \hline
4 & 1342 & 3124 & 4213 & 2431 & $TM$\\ \hline
5 & 1423 & 4132 & 2314 & 3241 & $CM$\\ \hline
6 & 1432 & 4123 & 3214 & 2341 & $ VM_1$\\ \hline
\caption{$S_4$ right-cosets list}
\label{tab02}
\end{longtable}

Here, we write 1234 to mean the permutation $\langle 1234 \rangle$ for convinience.
Also, we call each row of Table~\ref{tab01} and Table~\ref{tab02} a Kevin's Pizza slice.
Kevin's Pizza refers to a partitioning of the permutation group of order four~\cite{pHEDRON} (as in Fig.~\ref{FigP1}). 
 Table~\ref{tab01} and Table~\ref{tab02} show that although the Vierergruppe generates identical left and right cosets, the order in which these elements occur changes. For example,  the third row of Table~\ref{tab01} and Table~\ref{tab02} contain the same set elements but have different orders.
To describe this particular mathematical property, we  define

% \begin{definition}
% When O is a diadem of the $S_n$, and if any $i$-th of n-th Extended Kevin's Pizza slice ${\rm KP_i^{(n)}}$ satisfies $\forall g\in {\rm KP_i^{(n)}}$, $Og=gO$ then we call ${\rm KP_i^{(n)}}$ an {\bf ab-normal coset}.
% \label{def03}
% \end{definition}

\begin{definition}
Let $O$ be a diadem of $S_n$, and let ${\rm KP}_i^{(n)}$ be the $i$-th slice of Kevin's Pizza on $S_n$.  If $Og=gO$ $\forall g\in {\rm KP}_i^{(n)}$, then we call ${\rm KP}_i^{(n)}$ an {\bf ab-normal coset}.
\label{def03}
\end{definition}

Here the $i$-th slice of Kevin's Pizza on $S_n$ is defined as 
\begin{equation}
{\rm KP}_i^{(n)}=Og_i,
\label{eq4.8}
\end{equation}
where $O$ is a diadem of $S_n$.

where the diadem $O$ is Rana subgroup $\mathcal{R}_n$ defined in Sec.~\ref{sec:hopping}-\ref{sec:partition} and $g_i$ is the $i$-th element in $S_n$ (weighted by the BK-map).

Based on Def. \ref{def03}, in the $S_4$ case, we can say that all of the Kevin's pizza slices are ab-normal cosets.

\subsection{Left versus right hoppers}
In $S_8$, using the set $\mathcal{R}_8$, explicit construction of the left- and right-hoppers yields
\begin{scriptsize}
\begin{longtable}{|l|l|l|l|l|l|l|l|l|l|}
    \hline
    Left Hoppers & Element1& Element2& Element3& Element4& Element5& Element6& Element7& Element8 \\ \hline
    $CMCM$  & 12345678 & 21436587 & 34127856 & 43218765 & 56781234 & 65872143 & 78563412 & 87654321 \\ \hline
    $CMTM$  & 12345678 & 21436587 & 34127856 & 43218765 & 56781234 & 65872143 & 78563412 & 87654321 \\ \hline
    $CMVM$  & 12345678 & 21436587 & 34127856 & 43218765 & 56781234 & 65872143 & 78563412 & 87654321 \\ \hline
    $TMVM$  & 12345678 & 21436587 & 34127856 & 43218765 & 56781234 & 65872143 & 78563412 & 87654321  \\ \hline
    $TMTM$  & 12345678 & 21436587 & 34127856 & 43218765 & 56781234 & 65872143 & 78563412 & 87654321 \\ \hline
    $VMVM$  & 12345678 & 21436587 & 34127856 & 43218765 & 56781234 & 65872143 & 78563412 & 87654321 \\ \hline
    $O$   & 12345678 & 21436587 & 34127856 & 43218765 & 56781234 & 65872143 & 78563412 & 87654321  \\ \hline
\caption{Left Hoppers in 4D $\mathcal{N}=2$}
\label{tab:lefthoppers}
\end{longtable}
\end{scriptsize}

\begin{scriptsize}
\begin{longtable}{|l|l|l|l|l|l|l|l|l|l|}
    \hline
    Right Hoppers & Element1& Element2& Element3& Element4& Element5& Element6& Element7& Element8 \\ \hline
    $CMCM$ &  12345678 & 34127856 & 43218765 & 21436587 & 56781234 & 78563412 & 87654321 & 65872143 \\ \hline
    $CMTM$ & 12345678 & 34128765 & 43216587 & 21437856 & 57861423 & 86573241 & 68754132 & 75682314 \\ \hline
    $CMVM$ & 12345678 & 34127856 & 43216587 & 21438765 & 76583214 & 58761432 & 85672341 & 67854123 \\ \hline
    $TMVM$ & 12345678 & 43217856 & 21436587 & 34128765 & 78654312 & 56871243 & 87563421 & 65782134 \\ \hline
    $TMTM$ & 12345678 & 43218765 & 21436587 & 34127856 & 56781234 & 87654321 & 65872143 & 78563412 \\ \hline
    $VMVM$ & 12345678 & 34127856 & 21436587 & 43218765 & 56781234 & 78563412 & 65872143 & 87654321  \\ \hline
    $O$ & 12345678 & 21436587 & 34127856 & 43218765 & 56781234 & 65872143 & 78563412 & 87654321  \\ \hline    
\caption{Right Hoppers in 4D $\mathcal{N}=2$}
\label{tab:righthoppers}
\end{longtable}
\end{scriptsize}

Here $CM$, $TM$, $VM$ are chiral-, tensor-, vector-multiplets defined from (\ref{4DN4C1}), (\ref{4DN4T1}), (\ref{4DN4V1}) and $CMCM$ is same as $(CC)$ which is the chiral-chiral $\mathcal{N}=2$ supermultiplet generated from concatenating two copies of the $\mathcal{N}=1$ supermultiplet in Sec.~\ref{sec:generating_Lmatrices}.
And one can observe that in general the left and right hoppers of $CMCM$, $VMVM$, $TMTM$, and $O$ have a different ordering of elements, however, when taken as a set they are the same. This shows that there is an ab-normal coset in which the left and right cosets are the same as a set at 4D $\mathcal{N}=2$ case, which we can do a 0-brane reduction on 8D $\mathcal{N}=1$ (Eight color) case.

\subsection{Eight color ab-normal coset}
In general for the group $S_n$, we have \cite{groupth}:

\begin{theorem}
Let n $\geq$ 5.  Then the only normal subgroups of $S_n$ are \{()\}, $A_n$, and $S_n$.
\end{theorem}

Here $A_n$ is the alternating group: the group of even permutations on a set of length $n$.
Therefore, from $n \geq 5$, the diadem $O$ is not a normal subgroup of $S_n$.

In 4D $\mathcal{N}=2$ case, which is equivalent to 8D $\mathcal{N}=1$ after a 0-brane reduction we constructed the 8-th Extended Kevin's Pizza using a diadem $O$ (Rana subgroup $\mathcal{R}_8$). Thus the total number of 8D $\mathcal{N}=1$ 8-th Extended Kevin's Pizza slices are 7!=5040 because one 8-th Extended Kevin's Pizza slice has 8 elements.

Then for every $g\in {\rm{KP}}_i^{(n)}$, $n=8$, we have

\begin{equation}
g{\mathcal H^{(R)}={\rm KP}_i^{(n)}},\quad {\mathcal H^{(L)}}g={\rm KP}_i^{(n)},
\label{eq4.1}
\end{equation}

\begin{equation}
{\rm \mathcal H^{(R)}=\{\mathcal H_1^{(R)}, \mathcal H_2^{(R)}, ..., \mathcal H_8^{(R)}}\},
\quad 
{\rm \mathcal H^{(L)}}=\{\mathcal H_1^{(L)}, \mathcal H_2^{(L)}, ..., \mathcal H_8^{(L)}\}.
\label{eq4.10}
\end{equation}
Here $\rm \mathcal H_A^{(L)}$, $\rm \mathcal H_A^{(R)}$ are defined in (\ref{eq:HpRs1}).

And when $\mathcal H^{(L)}$ and $\mathcal H^{(R)}$ are the same, we find
\begin{equation}
{\rm \mathcal H^{(L)}}g=g{\rm \mathcal H^{(L)}}={\rm KP}_i^{(n)}.
\label{eq4.11}
\end{equation}

By using (\ref{eq4.8}) and (\ref{eq4.11}),

\begin{equation}
{\rm \mathcal H^{(L)}}={O}
\label{eq4.2}
\end{equation}

Therefore, if there is any ${\rm KP}_i^{(n)}$ which has the same left- and right-hoppers simultaneously, then from (\ref{eq4.11}), (\ref{eq4.2}) and Def. \ref{def03} we can call this kind of $n$-th extended Kevin's Pizza slice an ab-normal coset.

In the appendix, Table.~\ref{tab5} shows the ab-normal coset list on eight color case with the lexicographic order (BK-map) from $\langle12345678\rangle$ to $\langle18726345\rangle$.
As a result, one can see there are 168 sets in which the set of left- and right-hoppers are the same, which means each 8-th extended Kevin's Pizza slices on the list is an ab-normal coset.
And the highlighted rows' elements of Table.~\ref{tab5} have (4D $\mathcal{N}=1$ supermultiplet)+(4D $\mathcal{N}=1$ supermultiplet) decomposable structures, such as $VMVM$. And from row 72, the highlighted row patterns are based on the repetition of the first 24 rows, but it does not have particular type naming compared to 4D $\mathcal{N}=1$ supermultiplet.

Thus, in further investigation, one need is to clarify why the highlighted supermultiplet type consists of the same copies of 4D $\mathcal{N}=1$ supermultiplet such as (CC), (VV), (TT) rather than different combinations, eg., (CT), (CV), (VT), and why there is particular number (168) of ab-normal cosets which exists in 4D $\mathcal{N}=2$.

\section{Towards a Polytopic Representation Theory for Supersymmetry}\label{sec:5}
\begin{figure}[h]
    \includegraphics[width=.37\textwidth]{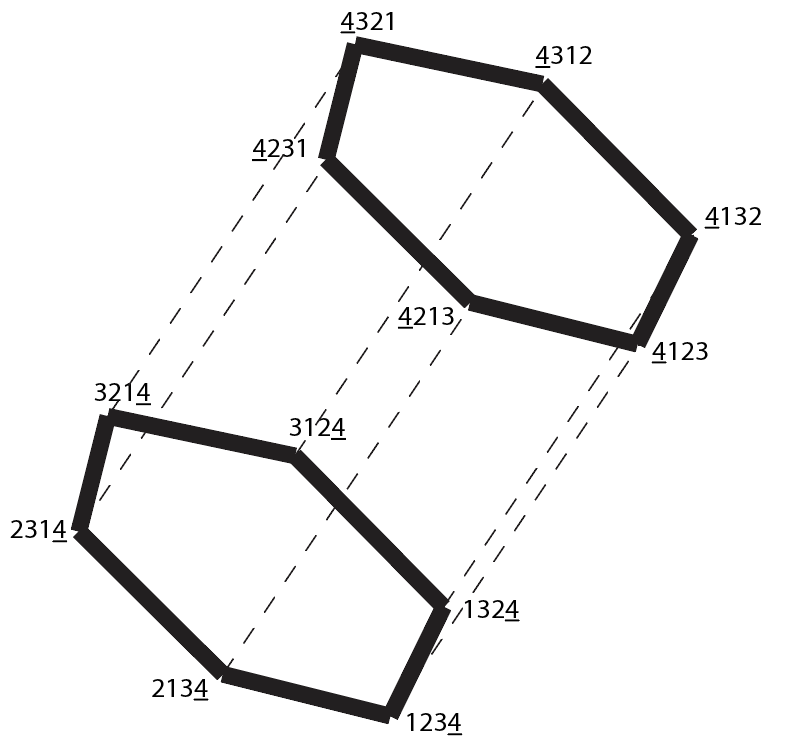}\hfill
    \includegraphics[width=.37\textwidth]{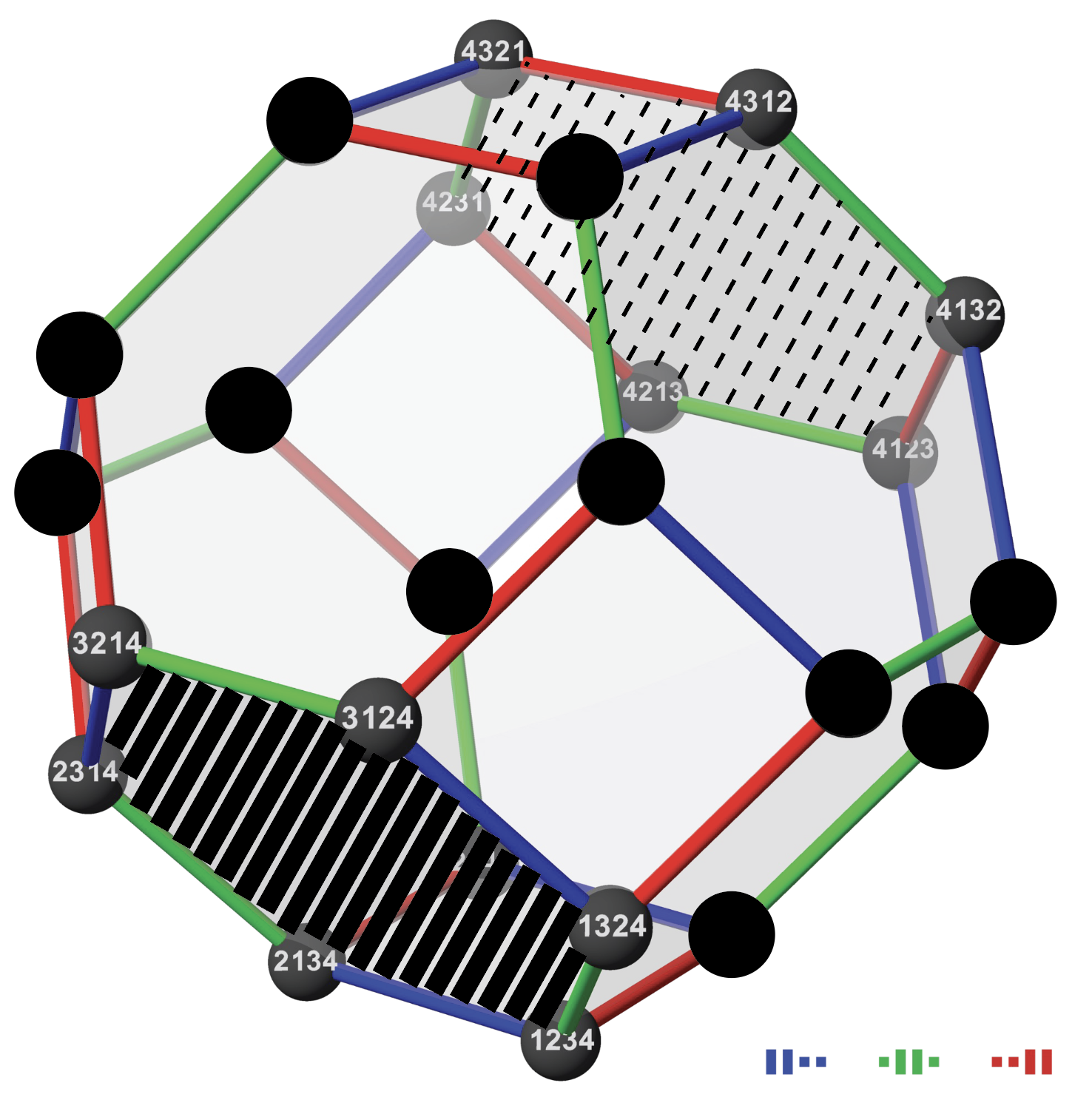}\hfill
    \caption{Illustration of the hexagon pair and its position on the 3rd order permutahedron}
    \label{fig10}
\end{figure}

\begin{figure}[h]
    \includegraphics[width=.37\textwidth]{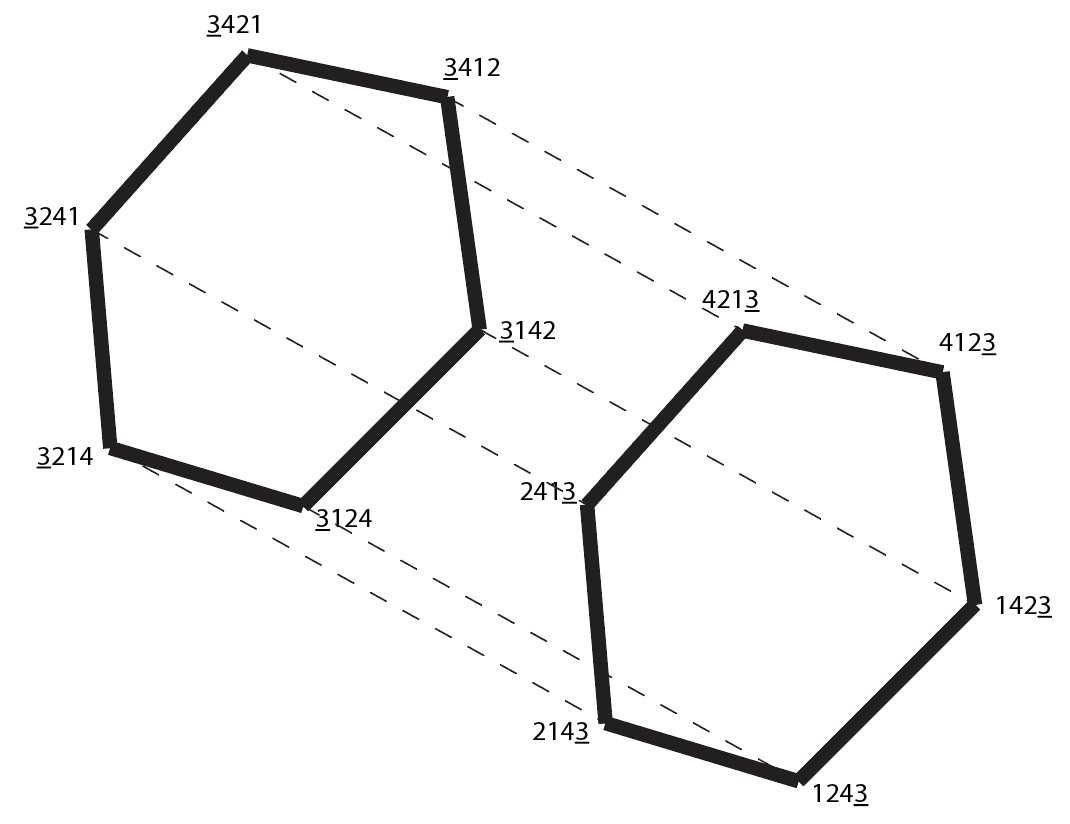}\hfill
    \includegraphics[width=.37\textwidth]{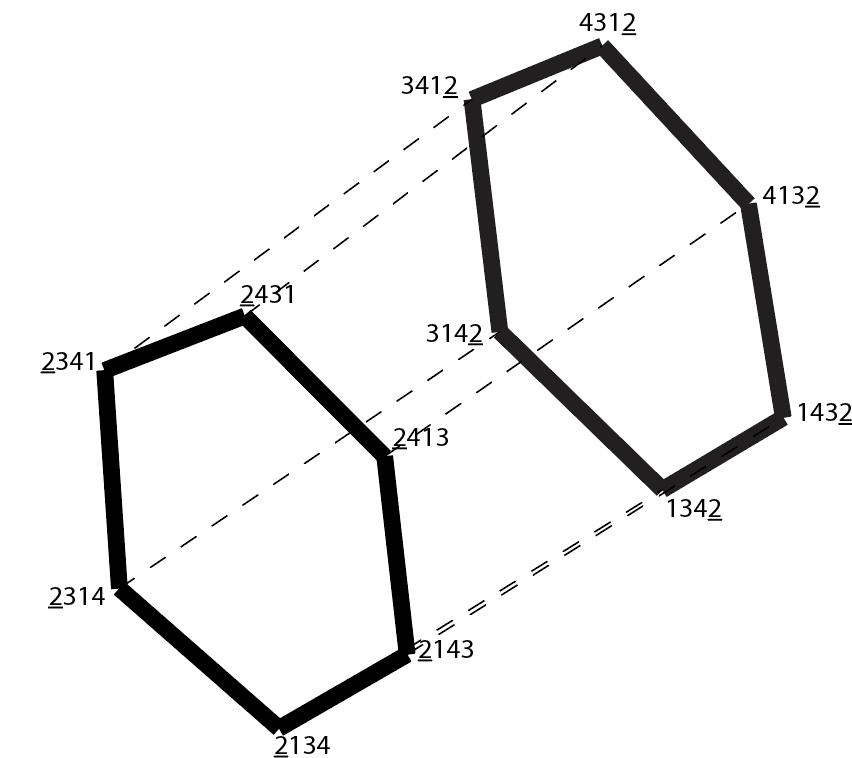}\hfill
    \includegraphics[width=.37\textwidth]{H3.png}\hfill
    \includegraphics[width=.37\textwidth]{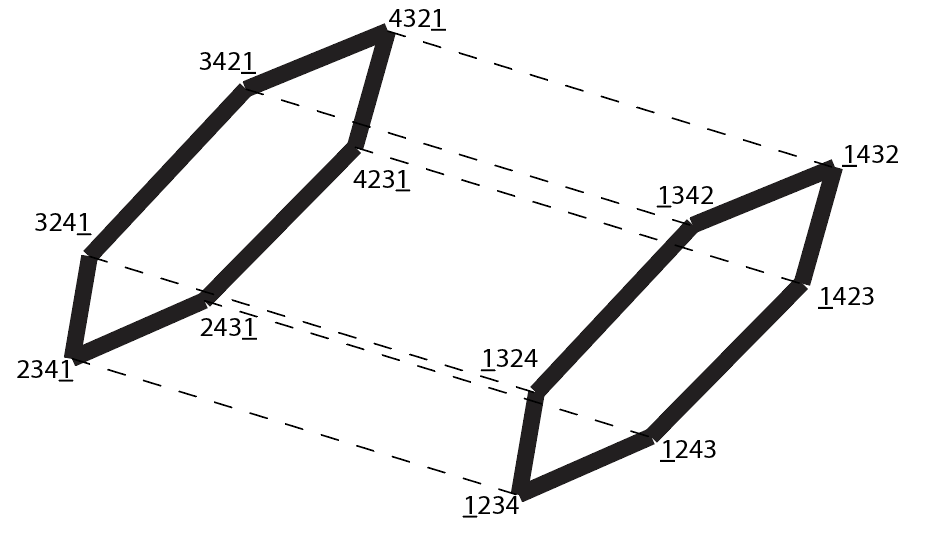}
    \caption{Illustration of the four hexagon pairs that generate the 3rd order permutahedron}
    \label{fig11}
\end{figure}

In our previous work \cite{Note}, we discussed the potential to view faces of the permutahedron as being composed of valid supermultiplets of a lower degree. In fact, we can realize this by examining how $\cal N$=3 permutahedra (hexagons) conspire to create the $\cal N$=4 permutahedron (truncated octahedron). In Fig. \ref{fig10} and Fig. \ref{fig11}, the construction of an $\cal N$=4 permutahedron out of $\cal N$=3 permutahedra is demonstrated. We start by taking the three element string ``123'' and using bruhat weak ordering to create two hexagons. Then, we add a ``4'' to the left hand side of all elements of one hexagon, and a ``4'' to the right hand side of all elements of the other hexagon. Fig. \ref{fig10} shows how this pair of hexagons fits on the completed $\cal N$=4 permutahedron. By following this construction for three more pairs of hexagons, where the respective added element is a ``1'', ``2'', and ``3'' (as depicted in Fig. \ref{fig11}), we can complete the construction of the $\cal N$=4 permutahedron. The square faces (which in our previous work \cite{Note} were identified as supermultiplets of a lower degree or extension) are seen to arise as a consequence of creating the pairs of hexagonal faces. 


\begin{figure}[h]
    \centering
    \includegraphics[width=.5\textwidth]{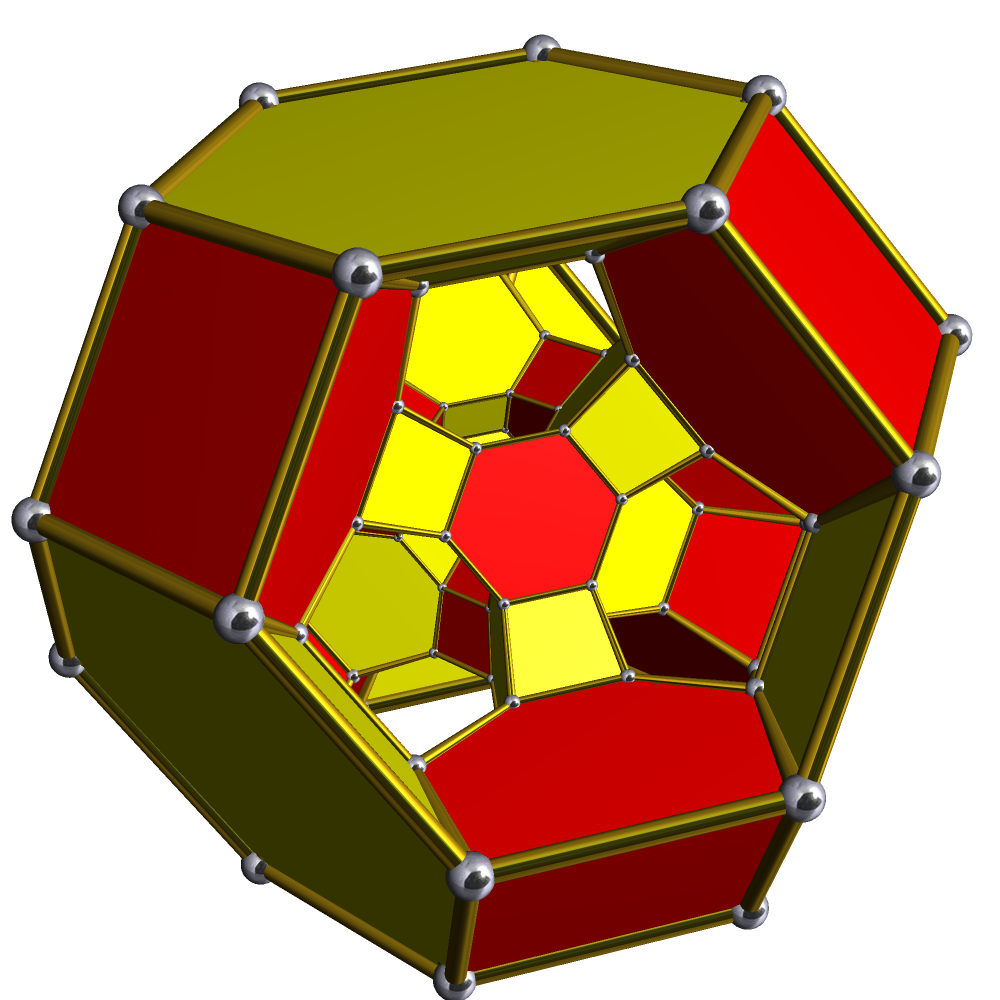}\hfill
    \includegraphics[width=.3\textwidth]{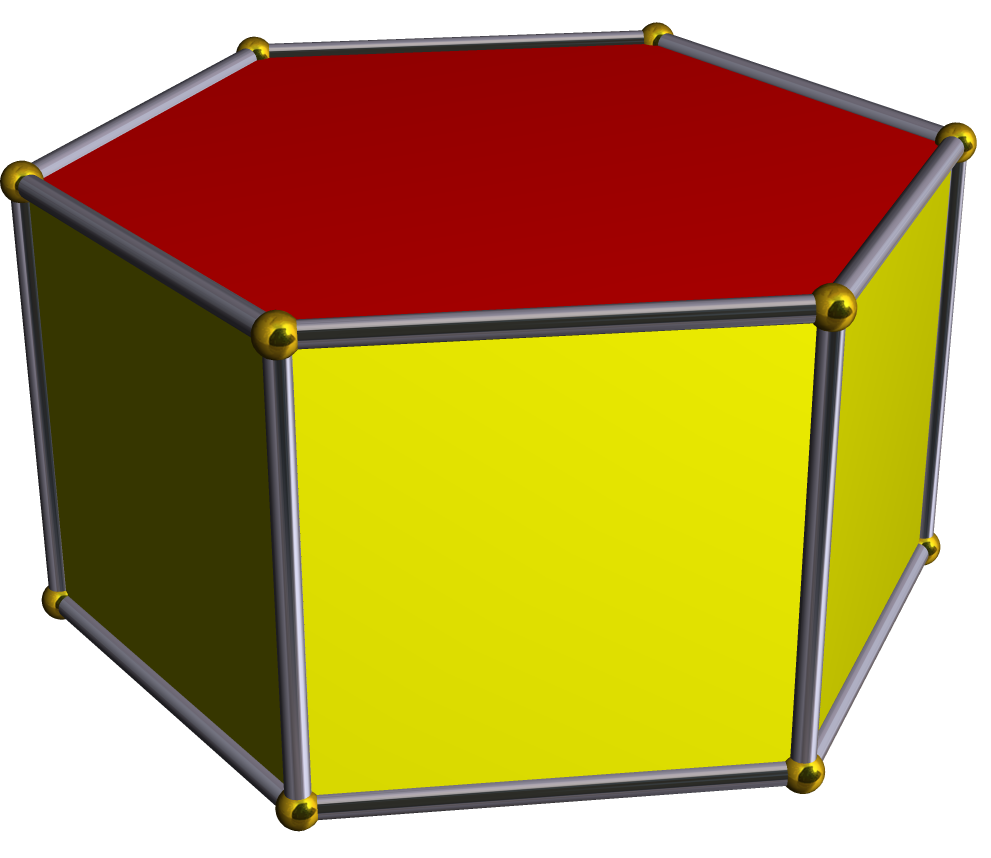}\hfill
    \includegraphics[width=.3\textwidth]{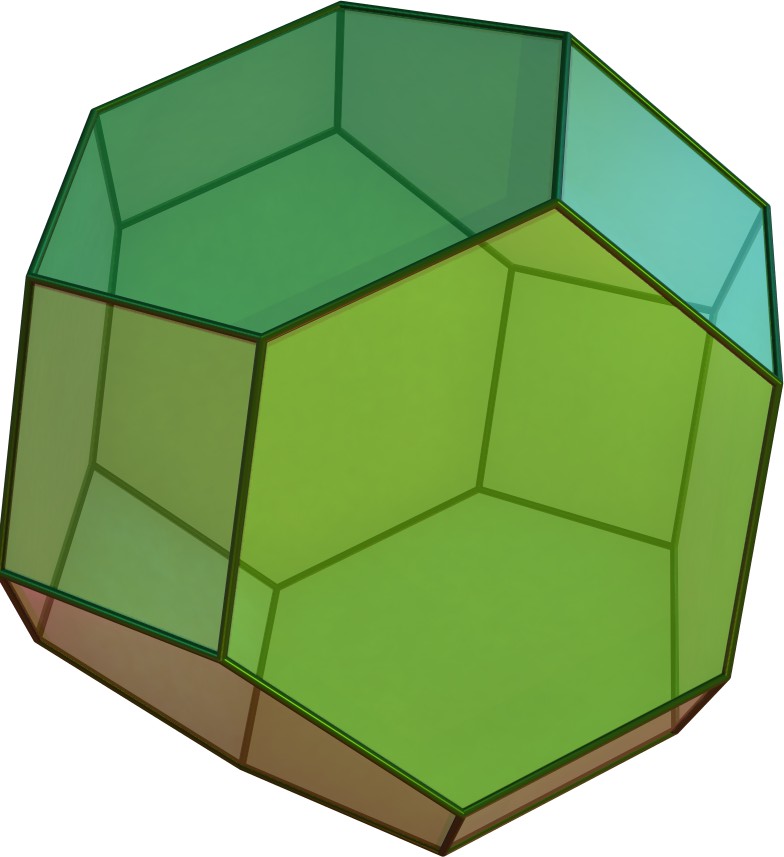}\hfill
    \caption{The omnitruncated 5-cell, constructed from truncated octahedra and hexagonal prisms}
    \label{fig12}
\end{figure}

A similar construction may be carried out for higher $\cal N$ permutahedra. For example, the permutahedra of order 5, the omnitruncated 5-cell, depicted in Fig. \ref{fig12}. This permutahedra is composed of 10 truncated octahedra, and 20 hexagonal prisms. If we were to carry out the analogous construction as for the $\cal N$=4 permutahedra, we would find that the truncated octahedra being put together would make the hexagonal prisms similarly emergent as the square faces were in the $\cal N$=4 case. This is strongly suggestive that our approach of creating higher degree supermultiplets out of lower degree supermultiplets is mathematically robust. Creating higher order supermultiplets out of lower order supermultiplets can be reduced to a problem of identifying vertices on a permutahedron of higher degree. 

It's notable that the types of 2-faces remain the same even as $\cal N$ scales. This isn't true for any of the other types of faces we examined (up to 5-faces).

\begin{tabular}{ |p{2cm}||p{2cm}|p{2cm}|p{2cm}|  }
 \hline
 \multicolumn{3}{|c|}{Types of 2-faces} \\
 \hline
 $\cal{N}$ & \{4\} & \{6\}\\
 \hline
 3&0 &1\\
 4&6 &8\\
 5&90 &60\\
 6&1080 & 480\\
 7&12600  & 4200\\
 8&151200  & 40320   \\
 \hline
\end{tabular}

\begin{tabular}{ |p{2cm}||p{2cm}|p{2cm}|p{2cm}|  }
 \hline
 \multicolumn{4}{|c|}{Types of 3-faces} \\
 \hline
 $\cal{N}$ & \{8\} & \{12\} & \{24\}\\
 \hline
 4&0 &0&1\\
 5&0 &20&10\\
 6&90 &360&90\\
 7&2520 & 5040&840\\
 8&50400  & 67200&8400\\
 \hline
\end{tabular}

\begin{tabular}{ |p{2cm}||p{2cm}|p{2cm}|p{2cm}|p{2cm}|p{2cm}|  }
 \hline
 \multicolumn{6}{|c|}{Types of 4-faces} \\
 \hline
 $\cal{N}$ & \{16\} & \{24\} & \{36\} & \{48\} & \{120\}\\
 \hline
 5&0 &0&0&0&1\\
 6&0 &0&20&30&12\\
 7&0 &630&420&630&126\\
 8&2520 & 20160&6720&10080&1344\\

 \hline
\end{tabular}

\begin{tabular}{ |p{2cm}||p{2cm}|p{2cm}|p{2cm}|p{2cm}|p{2cm}|  }
 \hline
 \multicolumn{6}{|c|}{Types of 5-faces} \\
 \hline
 $\cal{N}$ & \{72\} & \{96\} & \{144\} & \{240\} & \{720\}\\
 \hline
 6&0 &0&0&0&1\\
 7&0 &0&70&42&14\\
 8&1680 & 1260&1680&1008&168\\
 \hline
\end{tabular}

\begin{tabular}{ |p{2cm}||p{2cm}|p{2cm}|p{2cm}|p{2cm}|p{2cm}|  }
 \hline
 \multicolumn{5}{|c|}{Types of 6-faces} \\
 \hline
 $\cal{N}$ & \{72\} & \{96\} & \{144\} & \{240\} \\
 \hline
 7&0 &0&0&1\\
 8&70 & 112&56&16\\
 \hline
\end{tabular}

\section{Conclusions and Summary}

In this work, we have begun the exploration of the permutahedron associated with the omnitruncated 7-simplex.  With regard to supersymmetry, this is the appropriate setting for models where
eight independent supercharges occur.  

In Sec.~\ref{sec:1}, we reviewed past works and brought a new perspective to them.  We began our presentation by reviewing some ``traditional'' 
color problems of mathematics such as the four color map problem and the Graph Vertex Coloring Problem. The latter can be considered in 
the case where the polytope is the truncated octahedron. This polytope occurs in the structure of the permutahedron associated with the 
permutation group of order four, $ \{ {\mathbb{S}}{}_{4} \}$.  We also noted in Fig. \ref{FigP2} that Clifford Algebra matrices are 
associated with the color-coded permutation matrices.  While we did not mention it in Sec.~\ref{sec:11}, that pattern is also a solution 
to the Graph Coloring Vertex Problem!  

An examination of that image shows that no edge is connected to two vertices associated with the same color.  This is true of the 
square faces, and it is also {\it {true}} for the hexagonal faces!  So adinkras, which have adjacency matrices that form Clifford Algebras, 
in the cases under study can be associated with solutions of the `traditional' Graph Vertex Coloring Problem.  Should this hold in general,
the emergence of SUSY via graph theory, would be intimately connected with the ``traditional'' Graph Vertex Coloring Problem.  This is one of 
the most suggestive, and perhaps important, realization presented in the current work!

In Sec.~\ref{sec:recursion}, the focus was on how previously uncovered adinkras and associated Clifford Algebras in the case of four colors, can be as a
starting to construct the similar adinkras and associated Clifford Algebras in the case of eight colors.  We demonstrated a set of constructive rules that accomplish this goal.

In Sec.~\ref{sec:3}, the challenge of uncovering operators that play the role of raising operators in compact Lie Algebras (or alternating as
``hopping operator'' in systems of lattices in physics problems) for the space of adinkras was considered.  Although the Jordan-Chevelley Decomposition of the generators in Lie algebras appear not to apply,
in any obvious fashion, due to the appearance of permutation matrices,
a concept that induces a weighted ordering can be obtained from another
viewpoint.  As noted in equations (\ref{eq:S3}) and (\ref{eq33}), there is a natural notion of the weight of permutation matrices that is induced by a map from the action of the permutation matrices on their carrier spaces into the real numbers.  This notion of ordering is actually implicit in the
results that follow from Bruhat ordering.  Using these concepts, we 
constructed elementary left hopping and right hopping operators which have the effect of generating translation on the space of the permutahedron.   These were given for both the cases of four-color
and eight-color systems.

This same section also formally introduced the concept of the ``magic number.''  In the first chapter, a verbal description of it was given
and as well the symbol ${\cal G}{}_{(3)}$ was introduced for this
quantity.  In equations (\ref{eq321}) and (\ref{eq322}) two formal mathematical definitions have been presented to precisely define the values of
${\cal G}{}_{(3)}$.  For the cases of $ \{ {\mathbb{S}}{}_{4} \}$,
$ \{ {\mathbb{S}}{}_{8} \}$, and $ \{ {\mathbb{S}}{}_{16} \}$, our
studies indicate the results shown in Table. \ref{tab06}
\begin{longtable}{|l|l|}
\hline
${\cal G}{}_{(3)}$-Value & Permutation Group \\ \hline
{~~~~~}12 &  ${~~~~~~~~~~~}\{ {\mathbb{S}}{}_{4} \}$   \\ \hline
{~~~\,~}112 &  ${~~~~~~~~~~~}\{ {\mathbb{S}}{}_{8} \}$   \\ \hline
{~~~\,~}960 &  ${~~~~~~~~~~~}\{ {\mathbb{S}}{}_{16} \}$   \\ \hline
\caption{$S_4$ left-coset}
\label{tab06}
\end{longtable}  \noindent
and we conjecture the formula in equation (\ref{eq3.24}) will hold for the
special cases of $ \{ {\mathbb{S}}{}_{2^r} \}$.

In Sec.~\ref{sec:4} we report on a new phenomenon.  For the case
of $ \{ {\mathbb{S}}{}_{4} \}$ the left hoppers and the right
hoppers are one and the same when considered as unordered sets.  This is due to the fact that the quartets that appear are cosets of $ \{ {\mathbb{S}}{}_{4} \}$ with respect to the Klein Group and the latter is a normal subgroup in forming these cosets.  In the case of $ \{ {\mathbb{S}}{}_{8} \}$ we are no longer dealing with a normal subgroup.  But remarkably enough when one looks at sets instead of individual elements, there are one-hundred and sixty-eight subsets that act as if they are normal groups.
We have given the name `ab-normal' to these sets.
In further investigations, we can do more analysis on how many ab-normal cosets appear in more high-order dimensions. 

Finally, in Sec.~\ref{sec:5} we investigated the $k$-face analysis of higher-order permutahedron. We constructed a method to build 4-order permutahedron from 2D hexagon pairs. The $k$-face analysis yields the polytopic way to construct higher-degree supermultiplets out of lower-degree supermultiplets. By using the computer simulation, we obtain the partial $k$-face spectrum up to $\mathcal{N}=8$ and $k=5$. In further investigations, we can do more analysis on how many numbers of $k$-face type exists on particular $\mathcal{N}$-order permutahedron.

Let us end by making a comment about a new possible restriction that is revealed by the current work. We can
refer to this as the `restriction of decomposability.'

In the case of the four color case, there were found to be six subsets of the permutation group
$\{ {\mathbb{S}}{}_{4}\} $, embedded within the Coxeter Group, that led to 1D, $N$ = 4
supermultiplets, i.e. provide the basis for Clfford Algebras.  Thus, if one begins with these 
six subsets as a starting point to construct 1D, $N$ = 8 off-shell supermultiplets in pairs,
then the experience of the work in \cite{Note} informs us that each member of the pairs must
be chosen to be distinct.  So there must be 6 $\times$ 5 = 30 such pairings as the basis to 
construct 1D, $N$ = 8 off-shell supermultiplets that are decomposable!  It is notable that 
the number 30 has also appeared in one of our previous investigations \cite{Cosets}. 
Whether these occurrences are simply the result of accidents is a topic for future
study.  This observation has the potential to imposed drastic reductions in the space of acceptable
${\bm {\rL}}{}_{{}_{\rI}}$-${\bm {\rR}}{}_{{}_{\rI}}$ collections.

\vspace{.05in}
 \begin{center}
\parbox{4in}{{\it ``Behind it all is surely an idea so simple, so beautiful, $~\,~$ that when we grasp it - in a decade, a century, or a $~~$ millennium - we will all say to each other, how could it have been 
 otherwise? How could we have been so stupid?'' \\ ${~}$
\\ ${~}$ }\,\,-\,\, John Archibald Wheeler}
 \parbox{4in}{
 $~~$} 
 \end{center}
 \noindent
{\bf {Acknowledgements}}\\[.1in] \indent

We wish to acknowledge conversation with Dr.\ D.\ E.\ A.\ Gates for useful comments on this work.
The research of S.\ J.\ G., A.\ C.\, and Z.\ Z.\ was supported during a portion of the time it was carried out by the endowment of the Ford Foundation Professorship of Physics at Brown 
University.  All also gratefully acknowledge the support of the Brown Theoretical Physics Center.   In addition, the research of S.\ J.\ G. is supported by the
Clark Leadership Chair in Science endowment at the University of Maryland - College Park.

\newpage
\section{Appendix}

\subsection{Eight color ab-normal cosets}

% The available number of 8D $\mathcal{N}=1$ Kevin's Pizza is 7!=5040.
% Let $\rm KP_n$ is a n-th Kevin Pizza of a group, then for $g\in \rm KP_n$ the Eq. (\ref{eq7.1}) is hold.

% \begin{equation}
% g{\rm H_L=KP_n},\quad {\rm H_R}g={\rm KP_n}
% \label{eq7.1}
% \end{equation}

% And when $H_L$ and $H_R$ is same, we can get,
% \begin{equation}
% {\rm H_L}g=g{\rm H_L}={\rm KP_n},\quad {\rm H_L}={\rm O}
% \end{equation}

% Therefore if there is any $\rm KP_n$ which has same left- and right-hoppers simultaneously, then from Def. \ref{def1} and Def. \ref{def3} we can call this kind of Kevin's Pizza as ab-normal coset.

% The Table. \ref{tab5} shows the ab-normal coset list on 8D case with The lexicographic order from $\langle12345678\rangle$ to $\langle18765432\rangle$.
% As a result one can see there are 168 sets which left and right-coset is same, which means the each coset is an ab-normal coset.
% Here, for example type $\rm VM_3VM_3$ means, the 8-string has $\rm VM_3$+$\rm VM_3$ structure.
% To be specific, if we have element 2143, which is element of $\rm VM_3$, then $\rm VM_3VM_3$ contains 21436587, where 6587 also have 2143 order with $1\rightarrow5,2\rightarrow6,
% 3\rightarrow7,4\rightarrow8$.
% And here highlighted rows' elements have 4D decomposable structure, such as VMVM etc. And from row 72, the highlighted patterns are based on the repetition of first 24 rows.

\begin{scriptsize}
\begin{longtable}{|l|l|l|l|l|l|l|l|l|l|}
    \hline
    No. & Element1& Element2& Element3& Element4& Element5& Element6& Element7& Element8& Type \\ \hline
    \rowcolor{lightgray} 1& 12345678 & 21436587 & 34127856 & 43218765 & 56781234 & 65872143 & 78563412 & 87654321 & O/$\rm VM_3VM_3$ \\ \hline
    2& 12346587 & 21435678 & 34128765 & 43217856 & 56782143 & 65871234 & 78564321 & 87653412 & ~ \\ \hline
    3& 12347856 & 21438765 & 34125678 & 43216587 & 56783412 & 65874321 & 78561234 & 87652143 & ~ \\ \hline
    4& 12348765 & 21437856 & 34126587 & 43215678 & 56784321 & 65873412 & 78562143 & 87651234 & ~ \\ \hline
    \rowcolor{lightgray}5& 12435687 & 21346578 & 34217865 & 43128756 & 56871243 & 65782134 & 78653421 & 87564312 &  $\rm VM_2VM_2$ \\ \hline
    6& 12436578 & 21345687 & 34218756 & 43127865 & 56872134 & 65781243 & 78654312 & 87563421 & \\ \hline
    7& 12437865 & 21348756 & 34215687 & 43126578 & 56873421 & 65784312 & 78651243 & 87562134 & ~ \\ \hline
    8   & 12438756 & 21347865 & 34216578 & 43125687 & 56874312 & 65783421 & 78652134 & 87561243 & ~ \\ \hline
    9   & 12563478 & 21654387 & 34781256 & 43872165 & 56127834 & 65218743 & 78345612 & 87436521 & ~ \\ \hline
    10  & 12564387 & 21653478 & 34782165 & 43871256 & 56128743 & 65217834 & 78346521 & 87435612 & ~ \\ \hline
    11  & 12567834 & 21658743 & 34785612 & 43876521 & 56123478 & 65214387 & 78341256 & 87432165 & ~ \\ \hline
    12  & 12568743 & 21657834 & 34786521 & 43875612 & 56124387 & 65213478 & 78342165 & 87431256 & ~ \\ \hline
    13  & 12653487 & 21564378 & 34871265 & 43782156 & 56217843 & 65128734 & 78435621 & 87346512 & ~ \\ \hline
    14  & 12654378 & 21563487 & 34872156 & 43781265 & 56218734 & 65127843 & 78436512 & 87345621 & ~ \\ \hline
    15  & 12657843 & 21568734 & 34875621 & 43786512 & 56213487 & 65124378 & 78431265 & 87342156 & ~ \\ \hline
    16  & 12658734 & 21567843 & 34876512 & 43785621 & 56214378 & 65123487 & 78432156 & 87341265 & ~ \\ \hline
    17   & 12783456 & 21874365 & 34561278 & 43652187 & 56347812 & 65438721 & 78125634 & 87216543 & ~ \\ \hline
    18   & 12784365 & 21873456 & 34562187 & 43651278 & 56348721 & 65437812 & 78126543 & 87215634 & ~ \\ \hline
    19   & 12785634 & 21876543 & 34567812 & 43658721 & 56341278 & 65432187 & 78123456 & 87214365 & ~ \\ \hline
    20   & 12786543 & 21875634 & 34568721 & 43657812 & 56342187 & 65431278 & 78124365 & 87213456 & ~ \\ \hline
    21   & 12873465 & 21784356 & 34651287 & 43562178 & 56437821 & 65348712 & 78215643 & 87126534 & ~ \\ \hline
    22   & 12874356 & 21783465 & 34652178 & 43561287 & 56438712 & 65347821 & 78216534 & 87125643 & ~ \\ \hline
    23   & 12875643 & 21786534 & 34657821 & 43568712 & 56431287 & 65342178 & 78213465 & 87124356 & ~ \\ \hline
    24   & 12876534 & 21785643 & 34658712 & 43567821 & 56432178 & 65341287 & 78214356 & 87123465 & ~ \\ \hline
    \rowcolor{lightgray}25   & 13245768 & 24136857 & 31427586 & 42318675 & 57681324 & 68572413 & 75863142 & 86754231 & $\rm VMVM$ \\ \hline
    26   & 13246857 & 24135768 & 31428675 & 42317586 & 57682413 & 68571324 & 75864231 & 86753142 & ~ \\ \hline
    27   & 13247586 & 24138675 & 31425768 & 42316857 & 57683142 & 68574231 & 75861324 & 86752413 & ~ \\ \hline
    28   & 13248675 & 24137586 & 31426857 & 42315768 & 57684231 & 68573142 & 75862413 & 86751324 & ~ \\ \hline
    \rowcolor{lightgray}29   & 13425786 & 24316875 & 31247568 & 42138657 & 57861342 & 68752431 & 75683124 & 86574213 & $\rm TMTM$ \\ \hline
    30   & 13426875 & 24315786 & 31248657 & 42137568 & 57862431 & 68751342 & 75684213 & 86573124 & ~ \\ \hline
    31   & 13427568 & 24318657 & 31245786 & 42136875 & 57863124 & 68754213 & 75681342 & 86572431 & ~ \\ \hline
    32   & 13428657 & 24317568 & 31246875 & 42135786 & 57864213 & 68753124 & 75682431 & 86571342 & ~ \\ \hline
    33  & 13572468 & 24681357 & 31754286 & 42863175 & 57136824 & 68245713 & 75318642 & 86427531 & ~ \\ \hline
    34  & 13574286 & 24683175 & 31752468 & 42861357 & 57138642 & 68247531 & 75316824 & 86425713 & ~ \\ \hline
    35  & 13576824 & 24685713 & 31758642 & 42867531 & 57132468 & 68241357 & 75314286 & 86423175 & ~ \\ \hline
    36  & 13578642 & 24687531 & 31756824 & 42865713 & 57134286 & 68243175 & 75312468 & 86421357 & ~ \\ \hline
    37  & 13682457 & 24571368 & 31864275 & 42753186 & 57246813 & 68135724 & 75428631 & 86317542 & ~ \\ \hline
    38  & 13684275 & 24573186 & 31862457 & 42751368 & 57248631 & 68137542 & 75426813 & 86315724 & ~ \\ \hline
    39  & 13685724 & 24576813 & 31867542 & 42758631 & 57241368 & 68132457 & 75423186 & 86314275 & ~ \\ \hline
    40  & 13687542 & 24578631 & 31865724 & 42756813 & 57243186 & 68134275 & 75421368 & 86312457 & ~ \\ \hline
    41  & 13752486 & 24861375 & 31574268 & 42683157 & 57316842 & 68425731 & 75138624 & 86247513 & ~ \\ \hline
    42  & 13754268 & 24863157 & 31572486 & 42681375 & 57318624 & 68427513 & 75136842 & 86245731 & ~ \\ \hline
    43  & 13756842 & 24865731 & 31578624 & 42687513 & 57312486 & 68421375 & 75134268 & 86243157 & ~ \\ \hline
    44  & 13758624 & 24867513 & 31576842 & 42685731 & 57314268 & 68423157 & 75132486 & 86241375 & ~ \\ \hline
    45  & 13862475 & 24751386 & 31684257 & 42573168 & 57426831 & 68315742 & 75248613 & 86137524 & ~ \\ \hline
    46  & 13864257 & 24753168 & 31682475 & 42571386 & 57428613 & 68317524 & 75246831 & 86135742 & ~ \\ \hline
    47  & 13865742 & 24756831 & 31687524 & 42578613 & 57421386 & 68312475 & 75243168 & 86134257 & ~ \\ \hline
    48  & 13867524 & 24758613 & 31685742 & 42576831 & 57423168 & 68314257 & 75241386 & 86132475 & ~ \\ \hline
    \rowcolor{lightgray}49  & 14235867 & 23146758 & 32417685 & 41328576 & 58671423 & 67582314 & 76853241 & 85764132 & $\rm CMCM$ \\ \hline
    50  & 14236758 & 23145867 & 32418576 & 41327685 & 58672314 & 67581423 & 76854132 & 85763241 & ~ \\ \hline
    51  & 14237685 & 23148576 & 32415867 & 41326758 & 58673241 & 67584132 & 76851423 & 85762314 & ~ \\ \hline
    52  & 14238576 & 23147685 & 32416758 & 41325867 & 58674132 & 67583241 & 76852314 & 85761423 & ~ \\ \hline
    \rowcolor{lightgray}53  & 14325876 & 23416785 & 32147658 & 41238567 & 58761432 & 67852341 & 76583214 & 85674123 & $\rm VM_1VM_1$ \\ \hline
    54  & 14326785 & 23415876 & 32148567 & 41237658 & 58762341 & 67851432 & 76584123 & 85673214 & ~ \\ \hline
    55  & 14327658 & 23418567 & 32145876 & 41236785 & 58763214 & 67854123 & 76581432 & 85672341 & ~ \\ \hline
    56  & 14328567 & 23417658 & 32146785 & 41235876 & 58764123 & 67853214 & 76582341 & 85671432 & ~ \\ \hline
    57  & 14582367 & 23671458 & 32764185 & 41853276 & 58146723 & 67235814 & 76328541 & 85417632 & ~ \\ \hline
    58  & 14583276 & 23674185 & 32761458 & 41852367 & 58147632 & 67238541 & 76325814 & 85416723 & ~ \\ \hline
    59  & 14586723 & 23675814 & 32768541 & 41857632 & 58142367 & 67231458 & 76324185 & 85413276 & ~ \\ \hline
    60  & 14587632 & 23678541 & 32765814 & 41856723 & 58143276 & 67234185 & 76321458 & 85412367 & ~ \\ \hline
    61  & 14672358 & 23581467 & 32854176 & 41763285 & 58236714 & 67145823 & 76418532 & 85327641 & ~ \\ \hline
    62  & 14673285 & 23584176 & 32851467 & 41762358 & 58237641 & 67148532 & 76415823 & 85326714 & ~ \\ \hline
    63  & 14675823 & 23586714 & 32857641 & 41768532 & 58231467 & 67142358 & 76413285 & 85324176 & ~ \\ \hline
    64  & 14678532 & 23587641 & 32856714 & 41765823 & 58234176 & 67143285 & 76412358 & 85321467 & ~ \\ \hline
    65  & 14762385 & 23851476 & 32584167 & 41673258 & 58326741 & 67415832 & 76148523 & 85237614 & ~ \\ \hline
    66  & 14763258 & 23854167 & 32581476 & 41672385 & 58327614 & 67418523 & 76145832 & 85236741 & ~ \\ \hline
    67  & 14765832 & 23856741 & 32587614 & 41678523 & 58321476 & 67412385 & 76143258 & 85234167 & ~ \\ \hline
    68  & 14768523 & 23857614 & 32586741 & 41675832 & 58324167 & 67413258 & 76142385 & 85231476 & ~ \\ \hline
    69  & 14852376 & 23761485 & 32674158 & 41583267 & 58416732 & 67325841 & 76238514 & 85147623 & ~ \\ \hline
    70  & 14853267 & 23764158 & 32671485 & 41582376 & 58417623 & 67328514 & 76235841 & 85146732 & ~ \\ \hline
    71  & 14856732 & 23765841 & 32678514 & 41587623 & 58412376 & 67321485 & 76234158 & 85143267 & ~ \\ \hline
    72  & 14857623 & 23768514 & 32675841 & 41586732 & 58413267 & 67324158 & 76231485 & 85142376 & ~ \\ \hline
    \rowcolor{lightgray}73  & 15263748 & 26154837 & 37481526 & 48372615 & 51627384 & 62518473 & 73845162 & 84736251 & ~ \\ \hline
    74  & 15264837 & 26153748 & 37482615 & 48371526 & 51628473 & 62517384 & 73846251 & 84735162 & ~ \\ \hline
    75  & 15267384 & 26158473 & 37485162 & 48376251 & 51623748 & 62514837 & 73841526 & 84732615 & ~ \\ \hline
    76  & 15268473 & 26157384 & 37486251 & 48375162 & 51624837 & 62513748 & 73842615 & 84731526 & ~ \\ \hline
    \rowcolor{lightgray}77  & 15372648 & 26481537 & 37154826 & 48263715 & 51736284 & 62845173 & 73518462 & 84627351 & ~ \\ \hline
    78  & 15374826 & 26483715 & 37152648 & 48261537 & 51738462 & 62847351 & 73516284 & 84625173 & ~ \\ \hline
    79  & 15376284 & 26485173 & 37158462 & 48267351 & 51732648 & 62841537 & 73514826 & 84623715 & ~ \\ \hline
    80  & 15378462 & 26487351 & 37156284 & 48265173 & 51734826 & 62843715 & 73512648 & 84621537 & ~ \\ \hline
    81  & 15482637 & 26371548 & 37264815 & 48153726 & 51846273 & 62735184 & 73628451 & 84517362 & ~ \\ \hline
    82  & 15483726 & 26374815 & 37261548 & 48152637 & 51847362 & 62738451 & 73625184 & 84516273 & ~ \\ \hline
    83  & 15486273 & 26375184 & 37268451 & 48157362 & 51842637 & 62731548 & 73624815 & 84513726 & ~ \\ \hline
    84  & 15487362 & 26378451 & 37265184 & 48156273 & 51843726 & 62734815 & 73621548 & 84512637 & ~ \\ \hline
    85  & 15623784 & 26514873 & 37841562 & 48732651 & 51267348 & 62158437 & 73485126 & 84376215 & ~ \\ \hline
    86  & 15624873 & 26513784 & 37842651 & 48731562 & 51268437 & 62157348 & 73486215 & 84375126 & ~ \\ \hline
    87  & 15627348 & 26518437 & 37845126 & 48736215 & 51263784 & 62154873 & 73481562 & 84372651 & ~ \\ \hline
    88  & 15628437 & 26517348 & 37846215 & 48735126 & 51264873 & 62153784 & 73482651 & 84371562 & ~ \\ \hline
    89  & 15732684 & 26841573 & 37514862 & 48623751 & 51376248 & 62485137 & 73158426 & 84267315 & ~ \\ \hline
    90  & 15734862 & 26843751 & 37512684 & 48621573 & 51378426 & 62487315 & 73156248 & 84265137 & ~ \\ \hline
    91  & 15736248 & 26845137 & 37518426 & 48627315 & 51372684 & 62481573 & 73154862 & 84263751 & ~ \\ \hline
    92  & 15738426 & 26847315 & 37516248 & 48625137 & 51374862 & 62483751 & 73152684 & 84261573 & ~ \\ \hline
    93  & 15842673 & 26731584 & 37624851 & 48513762 & 51486237 & 62375148 & 73268415 & 84157326 & ~ \\ \hline
    94  & 15843762 & 26734851 & 37621584 & 48512673 & 51487326 & 62378415 & 73265148 & 84156237 & ~ \\ \hline
    95  & 15846237 & 26735148 & 37628415 & 48517326 & 51482673 & 62371584 & 73264851 & 84153762 & ~ \\ \hline
    96  & 15847326 & 26738415 & 37625148 & 48516237 & 51483762 & 62374851 & 73261584 & 84152673 & ~ \\ \hline
    \rowcolor{lightgray}97  & 16253847 & 25164738 & 38471625 & 47382516 & 52617483 & 61528374 & 74835261 & 83746152 & ~ \\ \hline
    98  & 16254738 & 25163847 & 38472516 & 47381625 & 52618374 & 61527483 & 74836152 & 83745261 & ~ \\ \hline
    99  & 16257483 & 25168374 & 38475261 & 47386152 & 52613847 & 61524738 & 74831625 & 83742516 & ~ \\ \hline
    100  & 16258374 & 25167483 & 38476152 & 47385261 & 52614738 & 61523847 & 74832516 & 83741625 & ~ \\ \hline
    \rowcolor{lightgray}101  & 16382547 & 25471638 & 38164725 & 47253816 & 52746183 & 61835274 & 74528361 & 83617452 & ~ \\ \hline
    102  & 16384725 & 25473816 & 38162547 & 47251638 & 52748361 & 61837452 & 74526183 & 83615274 & ~ \\ \hline
    103  & 16385274 & 25476183 & 38167452 & 47258361 & 52741638 & 61832547 & 74523816 & 83614725 & ~ \\ \hline
    104  & 16387452 & 25478361 & 38165274 & 47256183 & 52743816 & 61834725 & 74521638 & 83612547 & ~ \\ \hline
    105  & 16472538 & 25381647 & 38254716 & 47163825 & 52836174 & 61745283 & 74618352 & 83527461 & ~ \\ \hline
    106  & 16473825 & 25384716 & 38251647 & 47162538 & 52837461 & 61748352 & 74615283 & 83526174 & ~ \\ \hline
    107  & 16475283 & 25386174 & 38257461 & 47168352 & 52831647 & 61742538 & 74613825 & 83524716 & ~ \\ \hline
    108  & 16478352 & 25387461 & 38256174 & 47165283 & 52834716 & 61743825 & 74612538 & 83521647 & ~ \\ \hline
    109  & 16523874 & 25614783 & 38741652 & 47832561 & 52167438 & 61258347 & 74385216 & 83476125 & ~ \\ \hline
    110  & 16524783 & 25613874 & 38742561 & 47831652 & 52168347 & 61257438 & 74386125 & 83475216 & ~ \\ \hline
    111  & 16527438 & 25618347 & 38745216 & 47836125 & 52163874 & 61254783 & 74381652 & 83472561 & ~ \\ \hline
    112  & 16528347 & 25617438 & 38746125 & 47835216 & 52164783 & 61253874 & 74382561 & 83471652 & ~ \\ \hline
    113  & 16742583 & 25831674 & 38524761 & 47613852 & 52386147 & 61475238 & 74168325 & 83257416 & ~ \\ \hline
    114 & 16743852 & 25834761 & 38521674 & 47612583 & 52387416 & 61478325 & 74165238 & 83256147 & ~ \\ \hline
    115 & 16745238 & 25836147 & 38527416 & 47618325 & 52381674 & 61472583 & 74163852 & 83254761 & ~ \\ \hline
    116 & 16748325 & 25837416 & 38526147 & 47615238 & 52384761 & 61473852 & 74162583 & 83251674 & ~ \\ \hline
    117 & 16832574 & 25741683 & 38614752 & 47523861 & 52476138 & 61385247 & 74258316 & 83167425 & ~ \\ \hline
    118 & 16834752 & 25743861 & 38612574 & 47521683 & 52478316 & 61387425 & 74256138 & 83165247 & ~ \\ \hline
    119 & 16835247 & 25746138 & 38617425 & 47528316 & 52471683 & 61382574 & 74253861 & 83164752 & ~ \\ \hline
    120 & 16837425 & 25748316 & 38615247 & 47526138 & 52473861 & 61384752 & 74251683 & 83162574 & ~ \\ \hline
    \rowcolor{lightgray}121 & 17283546 & 28174635 & 35461728 & 46352817 & 53647182 & 64538271 & 71825364 & 82716453 & ~ \\ \hline
    122 & 17284635 & 28173546 & 35462817 & 46351728 & 53648271 & 64537182 & 71826453 & 82715364 & ~ \\ \hline
    123 & 17285364 & 28176453 & 35467182 & 46358271 & 53641728 & 64532817 & 71823546 & 82714635 & ~ \\ \hline
    124 & 17286453 & 28175364 & 35468271 & 46357182 & 53642817 & 64531728 & 71824635 & 82713546 & ~ \\ \hline
    \rowcolor{lightgray}125 & 17352846 & 28461735 & 35174628 & 46283517 & 53716482 & 64825371 & 71538264 & 82647153 & ~ \\ \hline
    126 & 17354628 & 28463517 & 35172846 & 46281735 & 53718264 & 64827153 & 71536482 & 82645371 & ~ \\ \hline
    127 & 17356482 & 28465371 & 35178264 & 46287153 & 53712846 & 64821735 & 71534628 & 82643517 & ~ \\ \hline
    128 & 17358264 & 28467153 & 35176482 & 46285371 & 53714628 & 64823517 & 71532846 & 82641735 & ~ \\ \hline
    129 & 17462835 & 28351746 & 35284617 & 46173528 & 53826471 & 64715382 & 71648253 & 82537164 & ~ \\ \hline
    130 & 17463528 & 28354617 & 35281746 & 46172835 & 53827164 & 64718253 & 71645382 & 82536471 & ~ \\ \hline
    131 & 17465382 & 28356471 & 35287164 & 46178253 & 53821746 & 64712835 & 71643528 & 82534617 & ~ \\ \hline
    132 & 17468253 & 28357164 & 35286471 & 46175382 & 53824617 & 64713528 & 71642835 & 82531746 & ~ \\ \hline
    133 & 17532864 & 28641753 & 35714682 & 46823571 & 53176428 & 64285317 & 71358246 & 82467135 & ~ \\ \hline
    134 & 17534682 & 28643571 & 35712864 & 46821753 & 53178246 & 64287135 & 71356428 & 82465317 & ~ \\ \hline
    135 & 17536428 & 28645317 & 35718246 & 46827135 & 53172864 & 64281753 & 71354682 & 82463571 & ~ \\ \hline
    136 & 17538246 & 28647135 & 35716428 & 46825317 & 53174682 & 64283571 & 71352864 & 82461753 & ~ \\ \hline
    137 & 17642853 & 28531764 & 35824671 & 46713582 & 53286417 & 64175328 & 71468235 & 82357146 & ~ \\ \hline
    138 & 17643582 & 28534671 & 35821764 & 46712853 & 53287146 & 64178235 & 71465328 & 82356417 & ~ \\ \hline
    139 & 17645328 & 28536417 & 35827146 & 46718235 & 53281764 & 64172853 & 71463582 & 82354671 & ~ \\ \hline
    140 & 17648235 & 28537146 & 35826417 & 46715328 & 53284671 & 64173582 & 71462853 & 82351764 & ~ \\ \hline
    141 & 17823564 & 28714653 & 35641782 & 46532871 & 53467128 & 64358217 & 71285346 & 82176435 & ~ \\ \hline
    142 & 17824653 & 28713564 & 35642871 & 46531782 & 53468217 & 64357128 & 71286435 & 82175346 & ~ \\ \hline
    143  & 17825346 & 28716435 & 35647128 & 46538217 & 53461782 & 64352871 & 71283564 & 82174653 & ~ \\ \hline
    144  & 17826435 & 28715346 & 35648217 & 46537128 & 53462871 & 64351782 & 71284653 & 82173564 & ~ \\ \hline
    \rowcolor{lightgray}145 & 18273645 & 27184536 & 36451827 & 45362718 & 54637281 & 63548172 & 72815463 & 81726354 & ~ \\ \hline
    146 & 18274536 & 27183645 & 36452718 & 45361827 & 54638172 & 63547281 & 72816354 & 81725463 & ~ \\ \hline
    147 & 18275463 & 27186354 & 36457281 & 45368172 & 54631827 & 63542718 & 72813645 & 81724536 & ~ \\ \hline
    148 & 18276354 & 27185463 & 36458172 & 45367281 & 54632718 & 63541827 & 72814536 & 81723645 & ~ \\ \hline
    \rowcolor{lightgray}149 & 18362745 & 27451836 & 36184527 & 45273618 & 54726381 & 63815472 & 72548163 & 81637254 & ~ \\ \hline
    150 & 18364527 & 27453618 & 36182745 & 45271836 & 54728163 & 63817254 & 72546381 & 81635472 & ~ \\ \hline
    151 & 18365472 & 27456381 & 36187254 & 45278163 & 54721836 & 63812745 & 72543618 & 81634527 & ~ \\ \hline
    152 & 18367254 & 27458163 & 36185472 & 45276381 & 54723618 & 63814527 & 72541836 & 81632745 & ~ \\ \hline
    153 & 18452736 & 27361845 & 36274518 & 45183627 & 54816372 & 63725481 & 72638154 & 81547263 & ~ \\ \hline
    154 & 18453627 & 27364518 & 36271845 & 45182736 & 54817263 & 63728154 & 72635481 & 81546372 & ~ \\ \hline
    155 & 18456372 & 27365481 & 36278154 & 45187263 & 54812736 & 63721845 & 72634518 & 81543627 & ~ \\ \hline
    156 & 18457263 & 27368154 & 36275481 & 45186372 & 54813627 & 63724518 & 72631845 & 81542736 & ~ \\ \hline
    157 & 18542763 & 27631854 & 36724581 & 45813672 & 54186327 & 63275418 & 72368145 & 81457236 & ~ \\ \hline
    158 & 18543672 & 27634581 & 36721854 & 45812763 & 54187236 & 63278145 & 72365418 & 81456327 & ~ \\ \hline
    159 & 18546327 & 27635418 & 36728145 & 45817236 & 54182763 & 63271854 & 72364581 & 81453672 & ~ \\ \hline
    160 & 18547236 & 27638145 & 36725418 & 45816327 & 54183672 & 63274581 & 72361854 & 81452763 & ~ \\ \hline
    161 & 18632754 & 27541863 & 36814572 & 45723681 & 54276318 & 63185427 & 72458136 & 81367245 & ~ \\ \hline
    162 & 18634572 & 27543681 & 36812754 & 45721863 & 54278136 & 63187245 & 72456318 & 81365427 & ~ \\ \hline
    163 & 18635427 & 27546318 & 36817245 & 45728136 & 54271863 & 63182754 & 72453681 & 81364572 & ~ \\ \hline
    164 & 18637245 & 27548136 & 36815427 & 45726318 & 54273681 & 63184572 & 72451863 & 81362754 & ~ \\ \hline
    165 & 18723654 & 27814563 & 36541872 & 45632781 & 54367218 & 63458127 & 72185436 & 81276345 & ~ \\ \hline
    166 & 18724563 & 27813654 & 36542781 & 45631872 & 54368127 & 63457218 & 72186345 & 81275436 & ~ \\ \hline
    167 & 18725436 & 27816345 & 36547218 & 45638127 & 54361872 & 63452781 & 72183654 & 81274563 & ~ \\ \hline
    168 & 18726345 & 27815436 & 36548127 & 45637218 & 54362781 & 63451872 & 72184563 & 81273654 & ~ \\ \hline
\caption{Eight color ab-normal cosets list}
\label{tab5}
\end{longtable}
\end{scriptsize}

Here column `type' means the type of the supermultiplet. For example $\rm VM_3VM_3$ means, the 8-string has $\rm VM_3$+$\rm VM_3$ structure.
To be specific, if we have element 2143, which is an element of $\rm VM_3$, then $\rm VM_3VM_3$ should contains 21436587, where 6587 also has 2143 order with $1\rightarrow5,\;2\rightarrow6,\;3\rightarrow7,\;4\rightarrow8$ mapping.


\begin{thebibliography}{99}
\small\frenchspacing\raggedright

\bibitem{GRana1}
S.\ J.\ Gates, Jr., and L.\ Rana,  ``A Theory of Spinning Particles for Large 
N-extended Supersymmetry (I),'' {\bf {Phys.\ Lett.\  B352}} (1995) 50;
DOI: 10.1016/0370-2693(95)00474-Y, arXiv [hep-th:9504025].

\bibitem{GRana2}
S.\ J.\ Gates Jr., and L.\ Rana, ``A Theory of Spinning 
Particles for Large N-extended Supersymmetry (II),'' ibid.\ {\bf {Phys.\ Lett.\  
B369}} (1996) 262; DOI: 10.1016/0370-2693(95)01542-6, arXiv [hep-th:9510151].

\bibitem{FG}
M.\ Faux, and S.\ J.\ Gates, Jr., ``Adinkras: A Graphical technology for supersymmetric representation theory,''
{\bf {Phys.\ Rev.\ D 71}} (2005) 065002; DOI: 10.1103/PhysRevD.71.065002, arXiv [hep-th:0408004].

\bibitem{permutadnk} 
I.\ Chappell, II, S.\ J.\ Gates, Jr., and T.\ H\" ubsch, ``Adinkra (In)Equivalence 
From Coxeter Group Representations: A Case Study,''  {\bf { Int.\ J.\ Mod.\ Phys.\  
 A29}} (2014) 06, 1450029;  
DOI: 10.1142/S0217751X14500298, arXiv [hep-th: 1210.0478]. 

\bibitem{pHEDRON}
A.\ J.\ Cianciara, S.\ J.\ Gates, Jr., Y.\ Hu, R.\ Kirk, 
``The 300 ``Correlators" Suggests 4D, $\cal N$ = 1 SUSY Is a Solution to a Set
of Sudoku Puzzles," 
e-Print: arXiv [hep-th: 2012.13308].

\bibitem{Note}
D.\ D.\ Bristow, J.\ H.\ Caporaletti, A.\ J.\ Cianciara, S.\ J.\ Gates, D.\ Levine, and G.\ Yerger, 
``A Note On Exemplary Off-Shell Constructions Of 4D, N = 2 Supersymmetry Representations,''  
\textbf{J. High Energ. Phys. 2022, 104} (2022);
DOI: 10.1007/JHEP04(2022)104,
arXiv [hep-th: 2012.14015].

\bibitem{EFThedron1}
N.\ Arkani-Hamed, TC.\ Huang, and YT.\ Huang, 
``The EFT-hedron,'' 
\textbf{J. High Energ. Phys. 2021, 259} (2021);
DOI: 10.1007/JHEP05(2021)259.

\bibitem{viruses}
R.\ Kaplan, J.\ Klobusicky, S.\ Pandey, D.\ H.\ Gracias, and G.\ Menon, 
``Building polyhedra by self-assembly: theory and experiment,''
\textbf{Artif Life. 2014 Fall; 20(4):409-39};
DOI: 10.1162/ARTL\_a\_00144. Epub 2014 Aug 22. PMID: 25148546.

\bibitem{watsoncrick}
F.\ Crick and J.\ Watson, Structure of Small Viruses. Nature 177, 473Ð475 (1956). doi:10.1038/177473a0

\bibitem{Cit1}
D.\ MacKenzie, Mechanizing Proof: Computing, Risk, and Trust (MIT Press, 2004) p103

\bibitem{Cit2}
G.\ Chartrand and Li.\ Lesniak, Graphs \& Digraphs (CRC Press, 2005) p.221

\bibitem{Cit3}
Wilson (2014); Appel \& Haken (1989); Thomas (1998, pp. 852–853)

\bibitem{X1}
I.\ Chappell,  S.\  J.\ Gates, Jr., and T.\ H\" ubsch,  ``Adinkra (in)equivalence from Coxeter group
representations: A case study,''     Int.\ J.\ Mod.\ Phys.\ A 29 (2014) 06, 1450029, DOI:
10.1142/S0217751X14500298, 1210.0478 [hep-th].

\bibitem{X2}
M.\ Calkins, D.\ E.\ A.\ Gates, S.\  J.\ Gates, Jr., and W.\ M.\  Golding, ``Think Different: Applying
the Old Macintosh Mantra to the Computability of the SUSY Auxiliary Field Problem,''
JHEP 04 (2015) 056, DOI: 10.1007/JHEP04(2015)056, 1502.04164 [hep-th].
 
\bibitem{X3}
S.\  J.\ Gates, Jr.,  F.\ Guyton,  S.\ Harmalkar, D.\ Kessler, and V. Korotkikh, ``Adinkras from
ordered quartets of BC4 Coxeter group elements and regarding 1,358,954,496 matrix elements
of the Gadget,''
 JHEP 06 (2017) 006, DOI: 10.1007/JHEP06(2017)006,  1701.00304 [hep-th].  
 
\bibitem{X4}
S.\  J.\ Gates, Jr., K.\ Iga, L.\ Kang, V. Korotkikh, and K.\ Stiffler, ``
Generating all 36,864 Four-Color Adinkras via Signed Permutations and Organizing into $\ell$-
and $\tilde \ell$-Equivalence Classes,'' Symmetry 11 (2019) 1, 120, DOI:10.3390/sym11010120,
1712.07826 [hep-th].
 
\bibitem{X5}
S.\  J.\ Gates, Jr., L.\ Kang, D.\ Kessler, and V. Korotkikh,
``Adinkras from ordered quartets of BC4 Coxeter group elements and regarding another Gadget’s
1,358,954,496 matrix elements,''   Int.\ J.\ Mod.\ Phys.\ A 33 (2018) 12, 1850066, DOI:
10.1142/S0217751X18500665, 1802.02890 [hep-th]

\bibitem{weakbruhat}
Björner, Anders (1984), "Orderings of Coxeter groups", in Greene, Curtis (ed.), Combinatorics and algebra (Boulder, Colo., 1983), Contemp. Math., vol. 34, Providence, R.I.: American Mathematical Society, ISBN 978-0-8218-5029-9, MR 0777701

\bibitem{Skiena}
S.\ Skiena, ``The Algorithm Design Manual", Springer (2008), 2nd ed., 544-548 pp

\bibitem{Heawood}
Grünbaum, Branko; Szilassi, Lajos (2009), ``Geometric Realizations of Special Toroidal Complexes", Contributions to Discrete Mathematics, 4 (1): 21–39, doi:10.11575/cdm.v4i1.61986, ISSN 1715-0868

\bibitem{Brook}
Brooks, R. L. ``On Coloring the Nodes of a Network." Proc. Cambridge Philos. Soc. 37, 194-197, 1941.

\bibitem{Grotzsch}
Steinberg, Richard; Younger, D. H. (1989), ``Grötzsch's Theorem for the projective plane", Ars Combinatoria, 28: 15–31 pp.

\bibitem{FFvN}
D.\ Z.\ Freedman, P.\ van Nieuwenhuizen, and S.\ Ferrara, ``Progress Toward a Theory of Supergravity,'' 
{\bf {Phys.\ Rev.\ D13}} (1976) 3214;
DOI: 10.1103/PhysRevD.13.3214.

\bibitem{DZ}
S.\ Deser, and B.\ Zumino, ``Consistent Supergravity,'' 
{\bf {Phys.\ Lett.\ B 62}} (1976) 335;
DOI: 10.1016/0370-2693(76)90089-7.

\bibitem{adnkKyeoh}
S.\ J.\ Gates, Jr., and K.\ Stiffler, 
``Adinkra Color Confinement In Exemplary Off-Shell Constructions Of 4D, N = 2 Supersymmetry Representations,''
\textbf{J. High Energ. Phys. 2014, 51} (2014);
DOI: 10.1007/JHEP07(2014)051,
arXiv [hep-th: 1405.0048].

\bibitem{Hmy1}
S.\ J.\ Gates, Jr., T.\ H\" ubsch, and K.\ Stiffler,
``Adinkras and SUSY Holography: Some Explicit Examples,''
{\bf {Int.\ J.\ Mod.\ Phys.\ A 29}} (2014) 07, 1450041,
DOI: 10.1142/S0217751X14500419, arXiv [hep-th: 1208.5999].

\bibitem{Hmy2}
S.\ J.\ Gates, Jr., T.\ H\" ubsch, and K.\ Stiffler,
``On Clifford-Algebraic ÒHoloraumy,Ó Dimensional Extension, and SUSY Holography,''
{\bf {Int.\ J.\ Mod.\ Phys.\ A 30}} (2015) 09, 1550042,
 DOI: 10.1142/S0217751X15500426, arXiv [hep-th: 1409.4445].

\bibitem{Hmy3}
S.\ J.\ Gates, Jr., T.\ Grover , M.\ D.\ Miller-Dickson, B.\ A.\ 
Mondal, A.\ Oskoui, S.\ Regmi, E.\ Ross, and R.\ Shetty,
``A Lorentz Covariant Holoraumy-Induced ``GadgetÓ From Minimal Off-Shell 4D, N = 1 Supermultiplets,'' 
{\bf {J. High Energ. Phys. 11}} (2015) 113, DOI: 10.1007/JHEP11(2015)113, arXiv [hep-th: 1508.07546].

\bibitem{Hmy4}
S.\ J.\ Gates, Jr., F.\ Guyton, S.\  Harmalkar, D.\ S.\  Kessler, V.\ Korotkikh, and V.\ A.\ Meszaros, 
``Adinkras From Ordered Quartets of BC4 Coxeter Group Elements and Regarding 1,358,954,496 
Matrix Elements of the Gadget,''  {\bf {J. High Energ. Phys. 11}} (2017) 006, DOI: 10.1007/JHEP06(2017)006, 
arXiv: [hep-th: 1701.00304].

\bibitem{Hmy5}
W.\ Caldwell, A.\ N.\ Diaz, I.\ Friend, S.\ J.\ Gates, Jr., 
S.\ Harmalkar, T.\ Lambert-Brown6a, D.\ Lay, K.\
Martirosova, V.\ A.\ Meszaros, M.\ Omokanwaye, S.\ 
Rudman, D.\ Shin, and A.\ Vershov, ``On the Four Dimensional Holoraumy of the 4D, N = 1 Complex 
Linear Supermultiplet,'' {\bf Int.\ J.\ Mod.\ Phys.\ A 33} (2018) 12, 1850072, DOI:
10.1142/S0217751X18500720, arXiv [hep-th: 1702.05453].
 
\bibitem{Hmy6}
S.\ J.\ Gates, Jr., L.\ Kang, D.\ S.\  Kessler, and V.\ Korotkikh, ``Adinkras From Ordered Quartets of 
BC4 Coxeter Group Elements and Regarding Another GadgetÕs 1,358,954,496 matrix elements,'' 
{\bf Int.\ J.\ Mod.\ Phys.\ A 33} (2018) 12, 1850066, DOI: 10.1007/JHEP06(2017)006,
arXiv [hep-th: 1802.02890]

\bibitem{Hmy7}
S.\ J.\ Gates, Jr., and S.-N. H.\ Mak, ``Examples of 4D, N = 2 Holoraumy,''
{\bf Int.\ J.\ Mod.\ Phys.\ A 34} (2019) 17, 1950081, DOI: 10.1142/S0217751X19500817, 
arXiv [hep-th: 1808.07946].

\bibitem{Cosets}
S.~J.~Gates, T.~H\"ubsch, K.~Iga and S.~Mendez-Diez,
``N=4 and N=8 SUSY Quantum Mechanics and Klein's Vierergruppe,''
arXiv [hep-th: 1608.07864].

\bibitem{EFThedron2}
LY.\ Chiang, YT.\ Huang, W.\ Li, et al., 
``Into the EFThedron and UV constraints from IR consistency,''
\textbf{J. High Energ. Phys. 2022, 63} (2022);
DOI: 10.1007/JHEP03(2022)063.

\bibitem{Adnk1}
M.\ Faux, and S.\ J.\ Gates Jr., 
``Adinkras: A Graphical technology for supersymmetric 
representation theory,''
{\bf {Phys.\ Rev.\ D71}} (2005) 065002;
DOI: 10.1103/PhysRevD.71.065002, 
arXiv [hep-th:0408004].

\bibitem{pHR0n1}
P.\ Schoute, 
``Analytic treatment of the polytopes regularly derived from the regular polytopes,''
 Verhandelingen der Koninklijke Akademie van Wetenschappen Te Amsterdam, 11 (3): 87 pp.

\bibitem{pHR0n2}
G.\ Ziegler, ``Lectures on Polytopes," Springer-Verlag, Graduate Texts in Mathematics 152  (1995).

\bibitem{pHR0n3}
R.\ Thomas,  
``Chapter 9. The Permutahedron", Lectures in Geometric Combinatorics, Student Mathematical 
Library: IAS/Park City Mathematical Subseries, 33, American Mathematical Society, (2006), p. 85, 
ISBN 978-0-8218-4140-2.

\bibitem{pHR0n4}
J.\ Santmyer, ``For all possible distances look to the permutohedron," 
Mathematics Magazine, 80 (2) (2007), 120, 
DOI: 10.1080/0025570X.2007.11953465.

\bibitem{BruHT}
A.\ Bj\" orner, ``Orderings of Coxeter groups. Combinatorics and algebra,''
(Boulder, Colo., 1983), 175, 
{\bf {Contemp.\ Math.\, 34}}, Amer.\ Math.\ Soc.\ ae, Providence, RI, 1984,
05A99 (06F15 14M15 20H15 52A25).

\bibitem{holography}
S.~J.~Gates, Jr., T.~H\"ubsch and K.~Stiffler,
``Adinkras and SUSY Holography: Some Explicit Examples,''
{\bf {Int.\ J.\ Mod.\ Phys.\ A 29}}, no.07, 1450041 (2014),
DOI: 10.1142/S0217751X14500419, arXiv [hep-th: 1208.5999].

\bibitem{HYMN1}
S.\ J.\ Gates, K.\ Stiffler, Y.\ Hu,
``Adinkra Height Yielding Matrix Numbers: Eigenvalue Equivalence Classes for Minimal Four-Color Adinkras,''
{\bf {Int.\ J.\ Mod.\ Phys.\ A34}} (2019) 1950085, 
e-Print: arXiv [hep-th: 1904.01738].

\bibitem{HYMN2}
S.\ J.\ Gates, K.\ Stiffler, Y.\ Hu,
``Properties of HYMNs in Examples of Four-Color, Five-Color, and Six-Color Adinkras,''  
Brown Univ. preprint \& Univ.\ of Iowa preprint, Oct 2020,
e-Print: arXiv [hep-th :2010.14659].

\bibitem{PP}
L. Billera, and S.\ Aravamuthan, 
``The combinatorics of permutation polytopes,''
DIMACS Series in Discrete Mathematics and Theoretical Computer Science. (1996) 24. 

\bibitem{Garden}
S.\ Bellucci, S.\ J.\ Gates Jr., and E.\ Orazi,
``A Journey through garden algebras,''  
DOI: 10.1007/3-540-33314-2\_1, arXiv [hep-th: 0602259].

\bibitem{Toppan}
M.\ Gonzales, M.\ Rojas, and F.\ Toppan,
``One-Dimensional $\sigma$ -Models with N = 5, 6, 7, 8 off-shell supersymmetries,''
\textbf{International Journal of Modern Physics A 24 23} 4317 (2009);
DOI: 10.1142/S0217751X09044516.

\bibitem{groupth}
A.\ Clark, ``Elements of Abstract Algebra'', Dover (1984), 2nd ed., 63 pp., Theorem 84$\beta$ 

\end{thebibliography}
\end{document}